\documentclass[review,onefignum,onetabnum]{siamart190516}

\usepackage{subfigure}
\usepackage{amsfonts}
\usepackage{graphicx}
\usepackage{epstopdf}
\usepackage{color}
\usepackage{algorithmic}
\usepackage[title]{appendix}

\graphicspath{{figures/}}

\usepackage[normalem]{ulem}

\renewcommand\thesection{\arabic{section}}
\renewcommand\thesubsection{\thesection.\arabic{subsection}}
\renewcommand\thesubsubsection{\thesubsection.\arabic{subsubsection}}
\renewcommand\theequation{\thesection.\arabic{equation}}

\ifpdf
  \DeclareGraphicsExtensions{.eps,.pdf,.png,.jpg}
\else
  \DeclareGraphicsExtensions{.eps}
\fi

\newsiamremark{remark}{Remark}
\newsiamremark{hypothesis}{Hypothesis}
\crefname{hypothesis}{Hypothesis}{Hypotheses}
\newsiamthm{claim}{Claim}

\newcommand{\Real}{\mathbb{R}}
\renewcommand{\v}[1]{\mathbf{#1}}

\newcommand{\x}{\v{x}}

\newcommand{\y}{\v{y}}

\newcommand{\pdedomain}{\bar{\Omega}}

\newcommand{\eps}{\varepsilon}
\newcommand{\mc}{\mathcal}
\newcommand{\xiM}{\xi_{>}}
\newcommand{\xim}{\xi_{<}}

\title{Optimization of the Mean First Passage Time in Near-Disk
  and Elliptical Domains in 2-D with Small Absorbing Traps} \author{Sarafa
  Iyaniwura\thanks{Dept. of Mathematics, Univ. of British Columbia,
    Vancouver, B.C., Canada.} \and Tony Wong\footnotemark[1] \and
  Colin~B.~Macdonald\footnotemark[1]\and Michael
  J. Ward\footnotemark[1]\,\, \thanks{corresponding author,
    \texttt{ward@math.ubc.ca}}}

\headers{Optimization of MFPT in Near-Disk and Elliptical Domains}
  {S. Iyaniwura, T. Wong, C. B. Macdonald, M. J. Ward}

\ifpdf
\hypersetup{
  pdftitle={Optimization of MFPT in Near-Disk and Elliptical Domains},
  pdfauthor={S. Iyaniwura, T. Wong, C. B. Macdonald, M. Ward}}
\fi

\begin{document}
\nolinenumbers
\maketitle

\begin{abstract}
  The determination of the mean first passage time (MFPT) for a
  Brownian particle in a bounded 2-D domain containing small absorbing
  traps is a fundamental problem with biophysical applications. The
  average MFPT is the expected capture time assuming a uniform
  distribution of starting points for the random walk. We develop a
  hybrid asymptotic-numerical approach to predict optimal
  configurations of $m$ small stationary circular absorbing traps that
  minimize the average MFPT in near-disk and elliptical domains. For a
  general class of near-disk domains, we illustrate through several
  specific examples how simple, but yet highly accurate, numerical
  methods can be used to implement the asymptotic theory. From the
  derivation of a new explicit formula for the Neumann Green's
  function and its regular part for the ellipse, a numerical approach
  based on our asymptotic theory is used to investigate how the
  spatial distribution of the optimal trap locations changes as the
  aspect ratio of an ellipse of fixed area is varied. The results from
  the hybrid theory for the ellipse are compared with full PDE
  numerical results computed from the closest point method
  \cite{IWWC2019}. For long and thin ellipses, it is shown that the
  optimal trap pattern for $m=2,\ldots,5$ identical traps is collinear
  along the semi-major axis of the ellipse. For such essentially 1-D
  patterns, a thin-domain asymptotic analysis is formulated and
  implemented to accurately predict the optimal locations of collinear
  trap patterns and the corresponding optimal average MFPT.
\end{abstract}

\section{Introduction}\label{sec:intro}

The concept of first passage time arises in various applications in
biology, biochemistry, ecology, physics, and biophysics (see
\cite{grigoriev2002kinetics}, \cite{holcman2004escape}, \cite{redner},
\cite{Venu} \cite{schuss2007narrow}, \cite{ricc1985}, and the
references therein). Narrow escape or capture problems are first
passage time problems that characterize the expected time it takes for
a Brownian ``particle'' to reach some absorbing set of small
measure. These problems are of singular perturbation type as they
involve two spatial scales: the ${\mathcal O}(1)$ spatial scale of the
confining domain and the ${\mathcal O}(\eps)$ asymptotically small
scale of the absorbing set. Narrow escape and capture problems arise
in various applications, including estimating the time it takes for a
receptor to hit a certain target binding site, the time it takes for a
diffusing surface-bound molecules to reach a localized signaling
region on the cell membrane, or the time it takes for a predator to
locate its prey, among others (cf.~\cite{benichou2014first},
\cite{BL}, \cite{coombs2009}, \cite{cheviakov2010asymptotic},
\cite{Holcman2014}, \cite{LBW2017}, \cite{singer2006narrow},
\cite{PWPK2010}, \cite{Venu}).  A comprehensive overview of the
applications of narrow escape and capture problems in cellular biology
is given in \cite{HolcmanReview2014}.

In this paper, we consider a narrow capture problem that involves
determining the MFPT for a Brownian particle, confined in a bounded
two-dimensional domain, to reach one of $m$ small stationary circular
absorbing traps located inside the domain. The average MFPT for this
diffusion process is the expected time for capture given a uniform
distribution of starting points for the random walk. In the limit of
small trap radius, this narrow capture problem can be analyzed by
techniques in strong localized perturbation theory
(cf.~\cite{WHK1993}, \cite{WK1993}). For a disk-shaped domain spatial
configurations of small absorbing traps that minimize the average MFPT
domain were identified in \cite{KTW2005}. However, the problem of
identifying optimal trap configurations in other geometries is largely
open. In this direction, the specific goal of this paper is to develop
and implement a hybrid asymptotic-numerical theory to identify optimal
trap configurations in near-disk domains and in the ellipse.

In \S~\ref{sec:mfpt_asy}, we use a perturbation approach to derive a
two-term approximation for the average MFPT in a class of near-disk
domains in terms of a boundary deformation parameter $\sigma\ll
1$. In our analysis, we allow for a smooth, but otherwise arbitrary,
star-shaped perturbation of the unit disk that preserves the domain
area. At each order in $\sigma$, an approximate solution is derived
for the MFPT that is accurate to all orders in
$\nu\equiv {-1/\log\eps}$, where $\eps\ll 1$ is the common radius of
the $m$ circular absorbing traps contained in the domain. To
leading-order in $\sigma$, this small-trap singular perturbation
analysis is formulated in the unit disk and leads to a linear
algebraic system for the leading-order average MFPT involving the Neumann
Green's matrix. At order ${\mathcal O}(\sigma)$, a further
linear algebraic system that sums all logarithmic terms in $\nu$ is
derived that involves the Neumann Green's matrix and certain weighted
integrals of the boundary profile characterizing the domain
perturbation. In \S~\ref{sec:mfpt_num}, we show how to numerically
implement this asymptotic theory by using the analytical expression
for the Neumann Green's function for the unit disk together with
the trapezoidal rule to compute certain weighted integrals of the
boundary profile with high precision. From this numerical
implementation of our asymptotic theory, and combined with either a
simple gradient descent procedure or a particle swarming 
approach \cite{kennedy2010}, we can numerically identify optimal trap
configurations that minimize the average MFPT in near-disk domains. In
\S~\ref{sec:examples}, we illustrate our hybrid asymptotic-numerical
framework by determining some optimal trap configurations in various
specific near-disk domains.

For a general 2-D domain containing small absorbing traps, a singular
perturbation analysis in the limit of small trap radii, related to
that in \cite{Venu}, \cite{coombs2009}, \cite{KTW2005}, and
\cite{WHK1993}, shows that the average MFPT is closely approximated by
the solution to a linear algebraic system involving the Neumann
Green's matrix. The challenge in implementing this analytical theory
is that, for an arbitrary 2-D domain, a full PDE numerical solution of
the Neumann Green's function and its regular part is typically
required to calculate this matrix. However, for an elliptical domain,
in \eqref{cell:finz_g} and \eqref{cell:R0} below, we provide a new
explicit representation of this Neumann Green's function and its
regular part. These explicit formulae allow for a rapid numerical
evaluation of the Neumann Green's interaction matrix for a given
spatial distribution of the centers of the circular traps in the
ellipse. The linear algebraic system determining the average MFPT is
then coupled to a gradient descent numerical procedure in order to
readily identify optimal trap configurations that minimize the average
MFPT in an ellipse. Although, a similar formula for the Neumann
Green's function has been derived previously for a rectangular domain
(cf.~\cite{marshall}, \cite{MHB}, \cite{KWW2010}), and an explicit and
simple formula exists for the disk \cite{KTW2005}, to our knowledge
there has been no prior derivation of a rapidly converging infinite
series representation for the Neumann Green's function in an
ellipse. The derivation of this Neumann Green's function using
elliptic cylindrical coordinates is deferred until \S~\ref{sec:g_ell}.

With this explicit approach to determine the Neumann Green's matrix,
in \S~\ref{sec:ellipse} we develop a hybrid asymptotic-numerical
framework to approximate optimal trap configurations that minimize the
average MFPT in an ellipse of a fixed area. In \S~\ref{ell:ex} we
implement our hybrid method to investigate how the optimal trap
patterns change as the aspect ratio of the ellipse is varied. The
results from the hybrid theory for the ellipse are favorably compared
with full PDE numerical results computed from a computationally
intensive numerical procedure of using the closest point method
\cite{IWWC2019} to compute the average MFPT and a particle swarming
approach \cite{kennedy2010} to numerically identify the optimum trap
configuration. As the ellipse becomes thinner, our hybrid theory shows
that the optimal trap pattern for $m=2,\ldots,5$ identical traps
becomes collinear along the semi-major axis of the ellipse. In the
limit of a long and thin ellipse, in \S~\ref{sec:thin} a thin-domain
asymptotic analysis is formulated and implemented to accurately
predict the optimal locations of collinear trap configurations and the
corresponding optimal average MFPT.

In \S~\ref{sec:discussion}, we show that the optimal trap
configurations that minimize the average MFPT also correspond to trap
patterns that maximize the coefficient of order ${\mathcal O}(\nu^2)$
in the asymptotic expansion of the fundamental Neumann eigenvalue of
the Laplacian in the perforated domain. This fundamental eigenvalue
characterizes the rate of capture of the Brownian particle by the
traps. Eigenvalue optimization problems for the fundamental Neumann
eigenvalue in a domain with small absorbing traps have been studied in
\cite{KTW2005} for the unit disk. The results herein extend this
previous analysis to the ellipse and to near-disk domains.

\section{Asymptotics of the MFPT in Near-Disk Domains}\label{sec:mfpt_asy}

We derive an asymptotic approximation for the MFPT for a class of
near-disk 2-D domains that are defined in polar coordinates by
\begin{equation}\label{PerturbPar}
  \Omega_{\sigma} = \Big{ \{ }(r,\theta)\, \Big{|}\, 0 < r \leq 1 +
  \sigma h(\theta)\,,  \,\,  0 \leq \theta \leq 2 \pi  \Big{ \} }\,,
\end{equation}
where the boundary profile, $h(\theta)$, is assumed to be an
${\mathcal O}(1)$, $C^{\infty}$ smooth $2\pi$ periodic function with
$\int_0^{2\pi} h(\theta)\,d\theta=0$. Observe that
$\Omega_{\sigma}\to \Omega$ as $\sigma \to 0$, where $\Omega$ is the
unit disk. Since $\int_{0}^{2\pi} h(\theta)\, d\theta=0$, the domain
area $|\Omega_{\sigma}|$ for $\sigma\ll 1$ is
$|\Omega_{\sigma}|= \pi + {\mathcal O}(\sigma^2)$.

Inside the perturbed disk $\Omega_{\sigma}$, we assume that there are
$m$ circular traps of a common radius $\varepsilon\ll 1$ that are
centered at arbitrary locations $\x_1,\ldots,\x_m$ with
$|\x_i-\x_j|={\mathcal O}(1)$ and
$\mbox{dist}(\partial\Omega_{\sigma},\x_j)= {\mathcal O}(1)$ as
$\eps\to 0$. The $j$-th trap, centered at some
$\x_j\in \Omega_{\sigma}$, is labelled by
$\Omega_{\varepsilon j} = \{\x : |\x - \x_j| \leq \varepsilon \}$.
The near-disk domain with the union of the trap regions deleted is
denoted by $\pdedomain_{\sigma}$. In $\pdedomain_{\sigma}$, it is
well-known that the mean first passage time (MFPT) for a Brownian
particle starting at a point $\x \in \pdedomain_{\sigma}$ to be
absorbed by one of the traps satisfies (cf.~\cite{redner})
\begin{equation}\label{Ellip_Model}
\begin{split}
  D\, \Delta u  = -1\,, &\quad \x \in \pdedomain_{\sigma}\,; \qquad
  \pdedomain_{\sigma} \equiv \Omega_{\sigma} \setminus
  \cup_{j=1}^{m}\Omega_{\varepsilon j} \,, \\
\partial_n u = 0\,, \quad \x \in \partial \Omega_{\sigma}\,; &
\qquad u = 0\,, \quad \x \in \partial \Omega_{\varepsilon j}\,,
\quad j = 1, \ldots,m\,.
\end{split}
\end{equation}
In terms of polar coordinates, the Neumann boundary condition in
\eqref{Ellip_Model} becomes
\begin{equation}\label{PolarBC}
\begin{split}
  u_r -  &\frac{ \sigma h_{\theta}}{(1 + \sigma h)^2} u_{\theta}= 0 \quad
  \text{on} \quad r = 1 +\sigma h(\theta)\,.
\end{split}
\end{equation} 

For an arbitrary arrangement $\lbrace{\x_1,\ldots,\x_m\rbrace}$ of the
centers of the traps, and for $\sigma\to 0$ and $\varepsilon\to 0$, we
will derive a reduced problem consisting of two linear algebraic
systems that provide an asymptotic approximation to the MFPT that has
an error ${\mathcal O}(\sigma^2,\varepsilon^2)$. These
linear algebraic systems involve the Neumann Green's matrix and certain
weighted integrals of the boundary profile $h(\theta)$.

To analyze \eqref{Ellip_Model}, we use a regular perturbation series
to approximate \eqref{Ellip_Model} for the near-disk domain to
problems involving a unit disk.  We expand the MFPT $u$ as
\begin{equation}\label{U_Sigma_Expand}
\begin{split}
  u = u_0 + \sigma u_1  + \ldots \,,
\end{split}
\end{equation} 
and substitute it into \eqref{Ellip_Model} and \eqref{PolarBC}. This
yields the leading-order problem
\begin{equation}\label{LeadingOrder}
\begin{split}
  D\,  \Delta u_0  = -1\,, &\quad \x \in \pdedomain\,; \qquad
   \pdedomain \equiv \Omega \setminus \cup_{j=1}^{m}\Omega_{\varepsilon j}\,, \\
\partial_n u_0 = 0\,, \quad \text{on} \quad r = 1\,; & \qquad u_0 = 0\,,
\quad \x \in \partial \Omega_{\varepsilon j}\,, \quad j = 1, \ldots,m\,,
\end{split}
\end{equation}
together with the following problem for the next order correction $u_1$:
\begin{equation}\label{OrderSigma}
\begin{split}
  \Delta u_1 = 0 \,, &\quad \x \in \pdedomain\,; \qquad \partial_r u_1 =
  -h u_{0rr} + h_{ \theta} u_{0 \theta}\,, \quad \text{on} \quad r = 1\,;
  \\ \quad u_1 &=0\,, \quad \x \in \partial \Omega_{\varepsilon j}\,,
  \quad j = 1, \ldots,m\,.
\end{split}
\end{equation}
Observe that \eqref{LeadingOrder} and \eqref{OrderSigma} are
formulated on the unit disk and not on the perturbed disk.  Assuming
$\varepsilon^2 \ll \sigma $, we use \eqref{U_Sigma_Expand} and
$|\Omega_{\sigma}| = |\Omega| + {\mathcal O}(\sigma^2)$ to derive an
expansion for the average MFPT, defined by
$\overline{u} \equiv \frac{1}{|\pdedomain_{\sigma}|}\int_{\pdedomain_{\sigma}} u \,
\text{d}\x$, in the form
\begin{equation}\label{AveMFPT_Perturb}
\begin{split}
  \overline{u} = \frac{1}{|\Omega|} \int_{\Omega } u_0
  \,\text{d}\x + \sigma \left[ \frac{1}{|\Omega |} \int_{\Omega }
    u_1 \,\text{d}\x + \frac{1}{|\Omega|}\int_{0}^{2\pi}
    h(\theta)\,u_0|_{r=1} \, \text{d}\theta \right] +
  \mathcal{O}(\sigma^2,\varepsilon^2)\,,
\end{split}
\end{equation}
where $|\Omega|=\pi$ and $u_0|_{r=1}$ is the leading-order solution
$u_0$ evaluated on $r=1$.

Since the asymptotic calculation of the leading-order solution $u_0$
by the method of matched asymptotic expansions in the limit
$\eps\to 0$ of small trap radius was done previously in
\cite{coombs2009} (see also \cite{Venu} and \cite{WHK1993}), we only
briefly summarize the analysis here. In the inner region near the
$j$-th trap, we define the inner variables
$\v{y} = \varepsilon^{-1}(\x - \x_j)$ and
$u_0(\x) = v_j(\varepsilon \v{y}+\x_j)$ with $\rho = |\v{y}|$, for
$j = 1, \ldots, m$.  Upon writing \eqref{LeadingOrder} in terms of
these inner variables, we have for $\varepsilon \to 0$ and for each
$j=1,\ldots,m$ that
\begin{equation}\label{LeadingOrderInner}
\begin{split}
 \Delta_{\rho}\, v_j  & = 0 \,, \quad \rho > 1\,;\qquad
 v_j =0\,, \quad   \mbox{on} \,\,\, \rho = 1\,, 
\end{split}
\end{equation} 
where
$\Delta_{\rho} \equiv \partial_{\rho \rho} + \rho^{-1}
\partial_{\rho}$. This admits the radially symmetric solution 
$v_j=A_j\log\rho$, where $A_j$ is an unknown constant. From
an asymptotic matching of the inner and outer solutions we
obtain the required singularity condition for the outer
solution $u_0$ as $\x \to \x_j$ for $j = 1, \ldots, m$.
In this way, we obtain that $u_0$ satisfies
\begin{subequations}\label{LeadingOrder_CompleteOuter}
\begin{align}
  \Delta u_0  = -{1/D}\,, \quad \x & \in \Omega\setminus
      \lbrace{\x_1,\ldots,\x_{m}\rbrace}\,;\quad \partial_r u_{0} =0\,,
    \,\,\, \x \in \partial \Omega\,; \label{LeadingOrder_CompleteOuterA}\\
  u_0 \sim A_j  \log |\x - \x_j| &+ A_j/\nu \quad \text{as} \quad\x \to \x_j\,,
       \qquad j = 1, \ldots, m\,,\label{LeadingOrder_CompleteOuterB}
\end{align}
\end{subequations} 
where $\nu \equiv -1/\log\varepsilon$. In terms of the Delta distribution,
\eqref{LeadingOrder_CompleteOuter} implies that
\begin{equation}\label{LeadingOuter_delta}
  \Delta u_0  = -\frac{1}{D} + 2 \pi \sum_{j = 1}^{m} A_j \delta(\x - \x_j)\,,
  \quad \x \in \Omega \,; \qquad \partial_r u_{0} =0\,,  \,\,\,
  \x \in \partial \Omega\,.
\end{equation}
By applying the divergence theorem to \eqref{LeadingOuter_delta} over
the unit disk we obtain that $\sum_{j=1}^{m} A_j = {|\Omega|/(2\pi D)}$.
The solution to \eqref{LeadingOuter_delta} is represented as
\begin{align}\label{SolutionOuterLead}
  u_0 = -2 \pi \sum_{k=1}^{m}  A_k G(\x ; \x_k)  + \overline{u}_0 \,; \qquad
 \overline{u}_0 = \frac{1}{|\Omega|}  \int_{\Omega} u_0 \, \text{d}\x\,,
\end{align}
where $G(\x ; \x_j)$ is the Neumann Green's function for the unit disk,
which satisfies
\begin{subequations}\label{GreenFunctionProb}
\begin{gather}
  \Delta G  = \frac{1}{|\Omega|} - \delta(\x - \x_j)\,,\quad \x \in \Omega\,;
              \quad \partial_n G =0\,, \,\,\, \x \in \partial \Omega\,; \quad
              \int_{\Omega} G \,\text{d}\x=0\,, \label{GreenFunctionProb_A}\\
  G  \sim -\frac{1}{2\pi} \log{|\x - \x_j|} + R_j + \nabla_{\x}R_j\cdot(\x-\x_j)
  \quad \text{as} \quad\x \to \x_j\,.     \label{GreenFunctionProb_B}
\end{gather}
\end{subequations}
Here, $R_j \equiv R(\x_j)$ is the regular part of the Green's function
at $\x = \x_j$. Expanding \eqref{SolutionOuterLead} as $\x \to \x_j$,
and using the singularity behaviour of $G(\x ; \x_j)$ given in
\eqref{GreenFunctionProb_B}, together with the far-field behavior
\eqref{LeadingOrder_CompleteOuterB} for $u_0$, we obtain the
matching conditon:
\begin{equation}\label{SolutionOuterLead_Expand}
   -2 \pi A_j\,R_j  -2 \pi
  \sum_{i \neq j}^{m} A_i \, G(\x_j ; \x_i) + \overline{u}_0 \sim
    {A_j/\nu} \,, \qquad \mbox{for} \quad j=1,\ldots, m\,.
\end{equation}
This yields a linear algebraic system for $\overline{u}_0$ and
$\mathcal{A} \equiv (A_1, \ldots, A_{m})^T$, given by
\begin{align}\label{Alg_Matrix}
  ( I + 2\pi \nu \, \mathcal{G}) \mathcal{A} = \nu\, \overline{u}_0 \,\v{e}\,,
  \qquad \v{e}^T \mathcal{A} = \frac{|\Omega|}{2\pi D} \,.
\end{align} 
Here, $\v{e} \equiv (1,\ldots,1)^T$, $\nu = -1/\log\varepsilon$, $I$
is the $m\times m$ identity matrix, and $\mathcal{G}$ is the symmetric
Green's matrix with matrix entries given by
\begin{align}\label{GreenMAtrix}
  (\mathcal{G})_{jj} = R_j \,\,\, \text{for} \,\,\, i = j
  \quad \text{and} \quad (\mathcal{G})_{ij} = (\mathcal{G})_{ji}  =
  G(\x_i ; \x_j) \,\,\, \text{for} \,\,\, i \neq j \,.
\end{align}
We left-multiply the equation for $\mathcal{A}$ in \eqref{Alg_Matrix}
by $\v{e}^T$, which isolates $\overline{u}_0$. By using this
expression in \eqref{Alg_Matrix}, and defining the matrix $E$ by
$E={\v{e}\v{e}^T/m}$, we get
\begin{equation}\label{u0_bar}
  \Big{[} I + 2 \pi \nu (I - E)\mc{G}  \Big{]} \mathcal{A} =
  \frac{|\Omega|}{2\pi D m} \v{e} \,, \quad \text{and}  \quad
  \overline{u}_0 = \frac{|\Omega|}{2\pi D \nu m}
 + \frac{2 \pi}{m} \v{e}^T \mc{G} \mathcal{A}  \,.
\end{equation}

\begin{remark} {\em The result \eqref{u0_bar} effectively sums all the
  logarithmic terms in powers of $\nu={-1/\log\eps}$. To estimate the
  error with this approximation with regards to the leading-order
  in $\sigma$ problem \eqref{LeadingOrder}, we calculate using
  \eqref{SolutionOuterLead} the refined local behavior
\begin{equation}\label{err:1}
  u_0\sim    -2 \pi \left(A_j\,R_j  +\sum_{i \neq j}^{m} A_i \, G(\x_j ; \x_i)
  \right) + \overline{u}_0 + \v{f}_j \cdot (\x-\x_j) \,, \quad
  \mbox{as} \quad \x\to \x_j,
\end{equation}
where
$\v{f}_j\equiv -2\pi\left(A_j\nabla_{\x} R_{j} + \sum_{i \neq j}^{m}
  A_i \, \nabla_{\x}G(\x ; \x_i)\vert_{\x=\x_j}\right)$. To account for this
  gradient term, near the $j$-th trap we must modify the inner
  expansion as $v_j\sim A_j\log\rho + \eps v_{j1}$. Here $\Delta_{\y}v_{j1}=0$
  in $|\y|\geq 1$, with $v_{j1}=0$ on $|\y|=1$ and
  $v_{j1}\sim \v{f}_j\cdot \y$ as $|\y|\to\infty$. The solution is
  $v_{j1}=\v{f}_j\cdot\left(\y - {\y/|\y|^2}\right)$. The far field behavior
  for $v_{j1}$ implies that in the outer region we must have that $u\sim u_0
  +\eps^2 w_0+\cdots$, where $w_0\sim -\v{f}_j\cdot{(\x-\x_j)/|\x-\x_j|^2}$
  as $\x\to \x_j$. This shows that the $\eps$-error estimate for $u_0$ is
  ${\mathcal O}(\eps^2)$, as claimed in \eqref{AveMFPT_Perturb}.}
\end{remark}

Next, we study the $\mathcal{O}(\sigma)$ problem for $u_1$ given in
\eqref{OrderSigma}. We construct an inner region near each of the
traps by introducing the inner variables
$\v{y} = \varepsilon^{-1}(\x - \x_j)$ and
$u_1(\x) = V_j(\varepsilon \v{y}+\x_j)$ with $ \rho = |\v{y}| $. From
\eqref{OrderSigma}, this yields the same leading-order inner problem
\eqref{LeadingOrderInner} with $v_j$ replaced by $V_j$. The radially
symmetric solution is $V_j = B_j \log\rho$, where $B_j$ is a constant
to be found. By matching this far-field behavior of the inner solution
to the outer solution we obtain the singularity behavior for
$u_1$ as $\x \to \x_j$ for $j = 1, \ldots, m$.  In this way, we find
from \eqref{OrderSigma} that $u_1$ satisfies
\begin{subequations}\label{OrderSigma_CompleteOuter}
\begin{align}
\Delta u_1  & = 0\,, \quad \x\in \Omega \setminus \{\x_1,\ldots,\x_{m}  \}\,;
 \quad \partial_r u_1  = F(\theta) \,, \quad \text{on}
                  \quad r = 1; \label{OrderSigma_CompleteOuterA}\\
  u_1 &\sim  B_j  \log{|\x - \x_j|} + B_j/\nu \quad \text{as} \quad
   \x \to \x_j \quad  j = 1, \ldots, m\,,\label{OrderSigma_CompleteOuterB}
\end{align}
where $\nu = -1/ \log\varepsilon$ and $F(\theta)$ is defined by
\begin{equation} \label{ftheta}
  F(\theta) \equiv  -h u_{0rr}\vert_{r=1} + h_{ \theta} u_{0 \theta}\vert_{r=1} =
\left( h u_{0\theta} \right)_{\theta} + \frac{h}{D} \,.
\end{equation}
In deriving \eqref{ftheta} we used
$u_{0rr}=-u_{0\theta\theta}+{1/D}$ at $r=1$, as obtained from
\eqref{LeadingOrder}.
\end{subequations}

Next, we introduce the Dirac distribution and write the problem
\eqref{OrderSigma_CompleteOuter} for $u_1$ as
\begin{align}\label{OrderSigma_CompleteOuter2}
  \Delta u_1   = 2\pi \sum_{i=1}^{m} B_i \,\,\delta(\x - \x_i)\,,
  \quad \x \in \Omega \,; \qquad u_{1r}  = F(\theta)\,, \quad \text{on}
  \quad r = 1\,.
\end{align}
Since $\int_0^{2 \pi} F(\theta) \, \text{d}\theta = 0$, the divergence
theorem yields $\sum_{j=1}^{m} B_j = 0$.  We decompose
\begin{equation}\label{u1_Sol}
 u_1 = -2 \pi \sum_{i=1}^{m}  B_i  G(\x ; \x_i) + u_{1p}+ \overline{u}_1\,,
\end{equation}
where $\overline{u}_1$ is the unknown average of $u_1$ over the unit
disk, and $G(\x;\x_i)$ is the Neumann Green's function satisfying
\eqref{GreenFunctionProb}. Here, $u_{1p}$ is taken to be the unique
solution to
\begin{align}\label{u1P_Prob}
 \Delta u_{1p} &= 0, \quad \x  \in \Omega ; \quad
  \partial_r u_{1p} =  F(\theta) \, \quad \text{on} \quad r = 1; \quad
       \int_{\Omega} u_{1p} \, \text{d}\x = 0 \,.
\end{align}

Next, by expanding \eqref{u1_Sol} as $\x \to \x_j$, we use the
singularity behaviour of $G(\x ; \x_j)$ as given in
\eqref{GreenFunctionProb_B} to obtain the local behavior of $u_1$ as
$\x \to \x_j$, for each $j=1,\ldots,m$. The asymptotic matching
condition is that this behavior must agree with that given in
\eqref{OrderSigma_CompleteOuterB}. In this way, we obtain a linear
algebraic system for the constant $\overline{u}_1$ and the vector
$\v{B} = (B_1,\ldots,B_{m})^T$, which is given in matrix form by
\begin{equation}\label{System_BNu_Mat}
  (I + 2 \pi \nu \mc{G} )\v{B} = \nu \overline{u}_1 \v{e} + \nu \v{u}_{1p}\,,
  \qquad  \v{e}^T \v{B} = 0 \,.
\end{equation}
Here, $I$ is the identity, $\v{e} = (1,\ldots,1)^T$, and
$\v{u}_{1p} = (u_{1p}(\x_1), \ldots, u_{1p}(\x_{m}))^T$.  Next, we left
multiply the equation for $\v{B}$ by $\v{e}^T$. This determines
$\overline{u}_1$, which is then re-substituted into
\eqref{System_BNu_Mat} to obtain the uncoupled problem
\begin{equation}\label{u1_bar}
\Big{[} I + 2 \pi \nu (I - E)\mc{G}  \Big{]} \v{B} = \nu (I - E)\v{u}_{1p}\,,
\quad \text{and}  \quad \overline{u}_1 = - \frac{1}{m}
\v{e}^T \v{u}_{1p} + \frac{ 2 \pi}{m} \v{e}^T \mc{G} \v{B} \,,
\end{equation}
where $E\equiv {\v{e}\v{e}^T/m}$. Since $\v{e}^T(I-E)=0$, we observe from
\eqref{u1_bar} that $\v{e}^T\v{B}=0$, as required. Equation
\eqref{u1_bar} gives a linear system for the $\mc{O}(\sigma)$ average
MFPT $\overline{u}_1$ in terms of the Neumann Green's matrix $\mc{G}$,
and the vector $\v{u}_{1p}$.

To determine $u_{1p}(\x_j)$, we use Green's second identity on
\eqref{u1P_Prob} and \eqref{GreenFunctionProb} to obtain a
line integral over the boundary $\x\in \partial\Omega$ of the unit
disk. Then, by using \eqref{ftheta} for $F(\theta)$, integrating
by parts and using $2\pi$ periodicity we get
\begin{equation}\label{u1p:bnd_2}
  u_{1p}(\x_j) = \int_{0}^{2\pi}  G(\x;\x_j) F(\theta) \, d\theta
   = \int_{0}^{2\pi} G(\x;\x_j) \frac{h(\theta)}{D}\,
  d\theta - \int_{0}^{2\pi} h(\theta) u_{0\theta} \partial_{\theta}
  G(\x;\x_j) \, d\theta \,.
\end{equation}
Then, by setting \eqref{SolutionOuterLead} for $u_0$ into
\eqref{u1p:bnd_2}, we obtain in terms of the $A_k$ of \eqref{u0_bar}
that%
\begin{subequations}\label{u1p:all}
\begin{equation}\label{u1p:bnd}
  u_{1p}(\x_j) = \frac{1}{D} \int_{0}^{2\pi} G(\x;\x_j) h(\theta) \,
  d\theta + 2\pi \sum_{k=1}^{m} A_k J_{jk}\,.
\end{equation}
Here, $J_{jk}$ is defined by the following
boundary integral with $\x=(\cos(\theta), \sin(\theta))^T$:
\begin{equation}\label{u1p:Jjk}
  J_{jk} \equiv \int_{0}^{2\pi}  h(\theta) \left(\partial_{\theta} G(\x;\x_j)\right)
  \left(\partial_{\theta} G(\x;\x_k)\right)\, d\theta \,.
\end{equation}
\end{subequations}

>From a numerical evaluation of the boundary integrals in
\eqref{u1p:all}, we can calculate
$\v{u}_{1p}= (u_{1p}(\x_1), \ldots, u_{1p}(\x_{m}))^T$, which
specifies the right-hand side of the linear system \eqref{u1_bar} for
$\v{B}$.  After determining $\v{B}$, we obtain $\overline{u}_1$ from
the second relation in \eqref{u1_bar}.  Finally, by substituting
\eqref{SolutionOuterLead} for $u_0$ into \eqref{AveMFPT_Perturb}, and
recalling that $\int_{0}^{2\pi} h(\theta)\,d\theta=0$, we obtain a
two-term expansion for the average MFPT given by
\begin{equation}\label{final:avemfpt}
  \overline{u} \sim \overline{u}_0 + \sigma
  \left( \overline{u}_1 - 2 \sum_{k=1}^{m} A_k \int_{0}^{2\pi}
    G(\x;\x_k) h(\theta) \, d\theta \right) \,.
\end{equation}
Here, $\x\in \partial\Omega$ and $\overline{u}_0$ is determined
from \eqref{u0_bar}.

\section{Optimizing Trap Configurations for the MFPT
in the Near-Disk}\label{sec:mfpt_num}

To numerically evaluate the boundary integrals in \eqref{u1p:all} and
\eqref{final:avemfpt}, we need explicit formulae for $G(\x;\x_j)$ and
$\partial_\theta G(\x;\x_j)$ on the boundary of the unit disk where
$\x=(\cos\theta,\sin\theta)^T$. For the unit disk, we obtain from
equation (4.3) of \cite{KTW2005} that 
\begin{subequations}\label{gr:gmrm}
\begin{gather}
  G(\x;\x_j) = -\frac{1}{2\pi}\log|\x-\x_j| - \frac{1}{4\pi}
  \log\left( |\x|^2|\x_j|^2 + 1 - 2\x \cdot \x_j\right)
   + \frac{ (|\x|^2 + |\x_j|^2 )}{4\pi} - \frac{3}{8\pi} ,
               \label{gr:gm} \\
  R(\x_j;\x_j)  = -\frac{1}{2\pi}\log\left(1 - |\x_j|^2\right) +
     \frac{|\x_j|^2}{2\pi} - \frac{3}{8\pi} \,.
    \label{gr:rm}
\end{gather}
\end{subequations}
For an arbitrary configuration $\lbrace{\x_1,\ldots,\x_m\rbrace}$ of
traps, these expressions can be used to evaluate the Neumann Green's
matrix $\mc{G}$ of \eqref{GreenMAtrix} as needed in \eqref{u0_bar} and
\eqref{u1_bar}.

Next, by setting $\x=(\cos\theta,\sin\theta)^T$ we can evaluate
$G(\x;\x_j)$ on $\partial\Omega$, and then calculate its tangential
boundary derivative $\partial_\theta G(\x;\x_j)$. By using
\eqref{gr:gm}, we obtain
\begin{subequations}\label{disk:gall}
\begin{align}
  G(\x;\x_j) & = -\frac{1}{2\pi} \log\left(1 + r_j^2 - 2 r_j
               \cos(\theta-\theta_j)\right) + \frac{1}{4\pi}(1+r_j^2) -
               \frac{3}{8\pi} \,, \label{disk:g} \\
  \partial_\theta G(\x;\x_j) & = - \frac{r_j}{\pi}
 \frac{\sin(\theta-\theta_j)}{\left[r_j^2+1-2r_j\cos(\theta-\theta_j)\right]}
                               \,, \label{disk:gtheta}
\end{align}
\end{subequations}
where $r_j\equiv |\x_j|$ and
$\x_j=r_j(\cos\theta_j,\sin\theta_j)^T$. Then, since
$\int_{0}^{2\pi} h(\theta)\, d\theta=0$, we can write the two boundary
integrals appearing in \eqref{u1p:all} and \eqref{final:avemfpt} explicitly
as
\begin{subequations}\label{disk:int_all}
\begin{gather}
  \int_{0}^{2\pi} G(\x;\x_j) h(\theta)\, d\theta  = -\frac{1}{2\pi}
  \int_{0}^{2\pi} h(\theta) \log\left(1+r_j^2 - 2r_j \cos(\theta-\theta_j)\right)
                                                    \,d\theta\,,\\
   J_{jk} = \frac{r_j r_k}{\pi^2} \int_{0}^{2\pi} 
   \frac{h(\theta) \sin(\theta-\theta_j)\sin(\theta-\theta_k)}{
   \left[ r_j^2+1-2r_j\cos(\theta-\theta_j)\right]
   \left[ r_k^2+1-2r_k\cos(\theta-\theta_k)\right]} \, d\theta \,.
\end{gather}                                                    
\end{subequations}

Although for an arbitrary $h(\theta)$ the integrals in
\eqref{disk:int_all} cannot be evaluated in closed form, they can be
computed to a high degree of accuracy with relatively few grid points
using the trapezoidal rule since this quadrature rule is exponentially
convergent for $C^{\infty}$ smooth periodic functions
\cite{tref_trap}. When $|x_j|<1$, the logarithmic singularities off of
the axis of integration for $J_{jk}$ in \eqref{disk:int_all} are mild
and pose no particular problem. In this way, we can numerically
calculate the two-term expansion \eqref{final:avemfpt} for the average
MFPT with high precision.

Then, to determine the optimal trap configuration we can either use the
particle swarming approach \cite{kennedy2010}, or the ODE relaxation
dynamics scheme
\begin{equation}
  \frac{d\v{z}}{dt} = -\nabla_{\v{z}} \overline{u} \,, \qquad
  \mbox{where} \quad \v{z}\equiv (x_1,y_1,\ldots,x_m,y_m)^T \,,
  \label{near_disk:relax}
\end{equation}
and $\overline{u}$ is given in \eqref{final:avemfpt}. Starting from an
admissible initial state $\v{z}\vert_{t=0}$, where
$\x_j=(x_j,y_j)\in\Omega_0$ at $t=0$ for $j=1,\ldots,m$, the gradient
flow dynamics \eqref{near_disk:relax} converges to a local minimum of
$\overline{u}$. Because of our high precision in calculating
$\overline{u}$, a centered difference scheme with mesh spacing
$10^{-4}$ was used to estimate the gradient in
\eqref{near_disk:relax}.

\subsection{Examples of the Theory}\label{sec:examples}

\begin{figure}[htbp]
\begin{center}
{\includegraphics[height=3.5cm,width=0.24\textwidth]{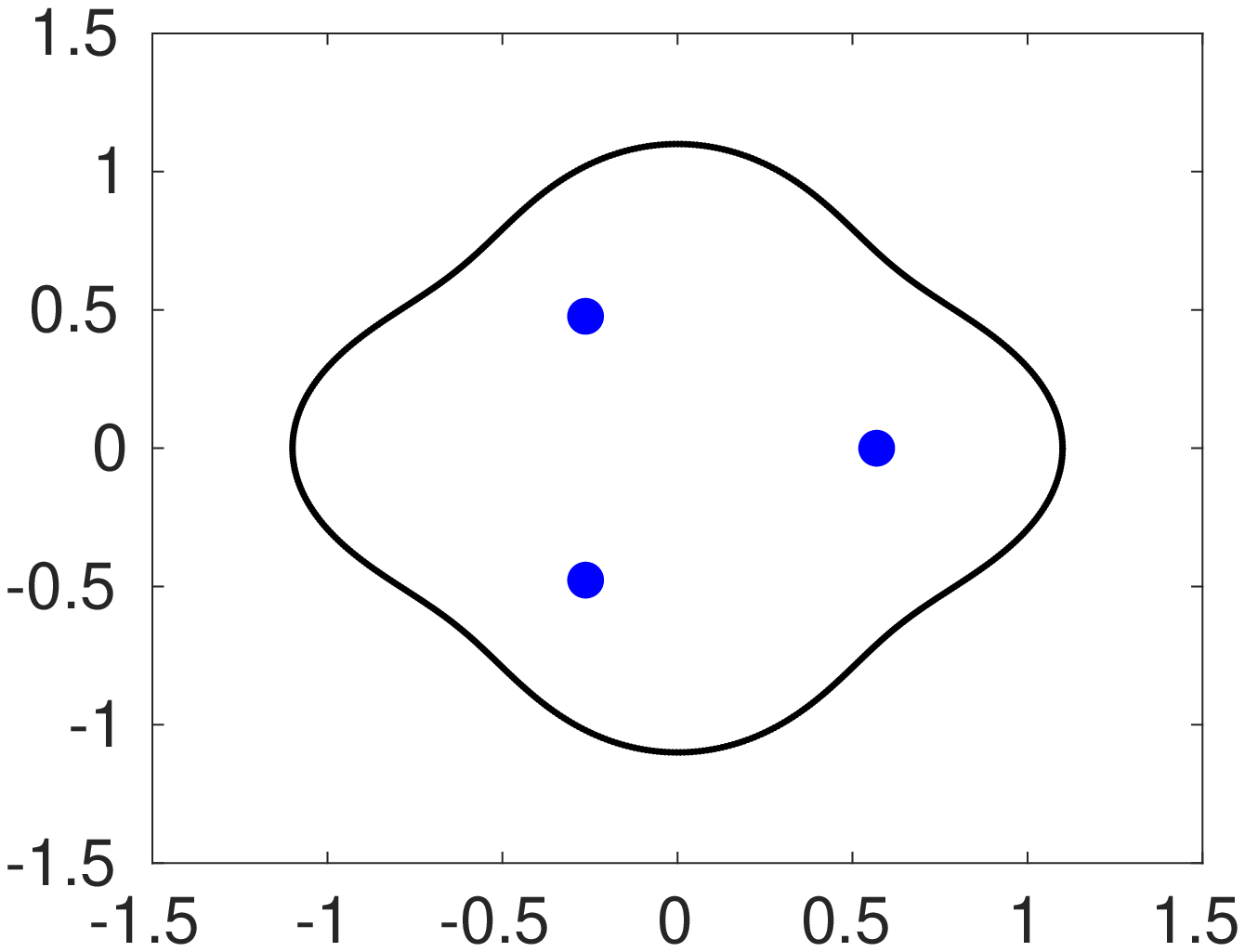}\label{fig:cos4_3}}
{\includegraphics[height=3.5cm,width=0.24\textwidth]{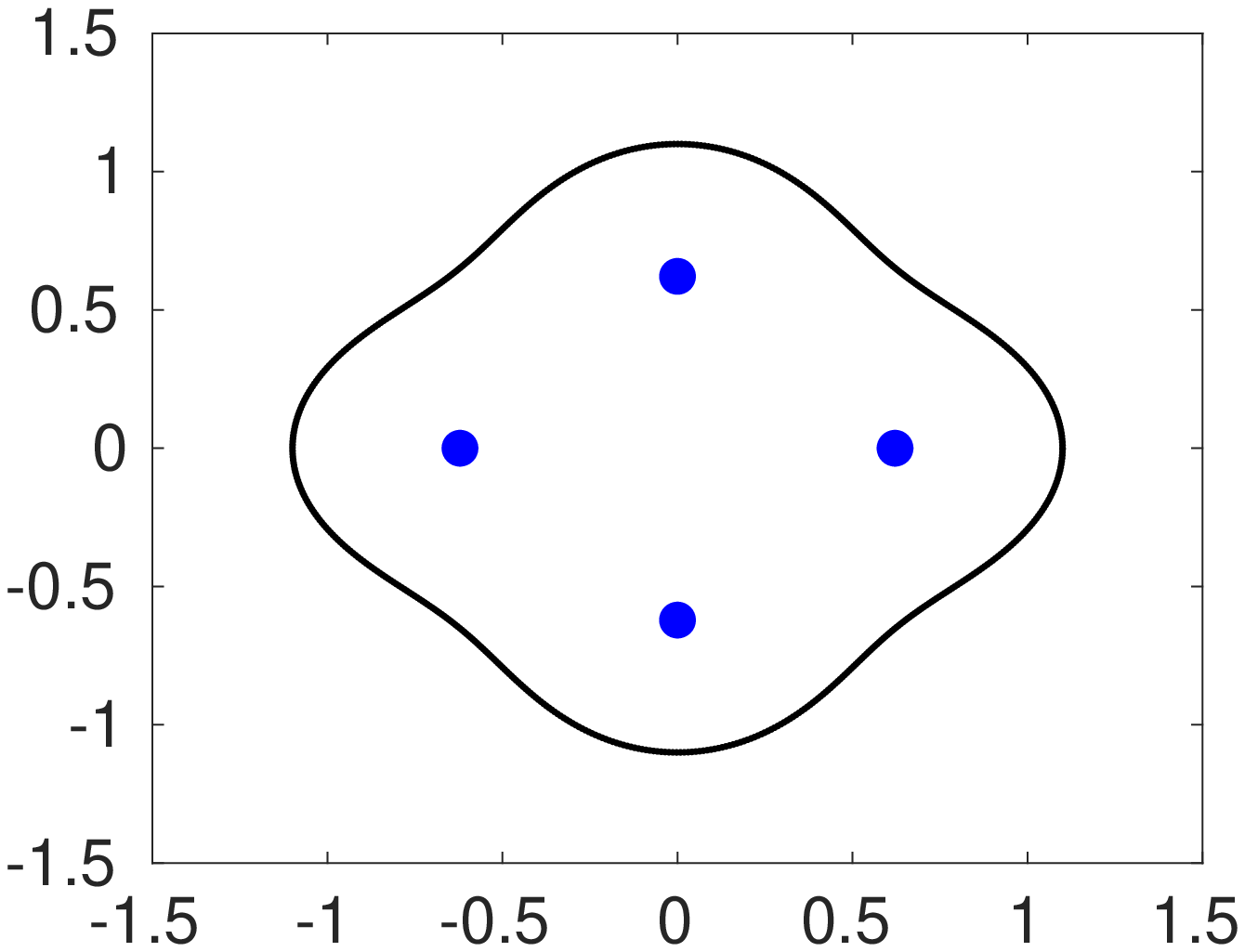}\label{fig:cos4_4}}
{\includegraphics[height=3.5cm,width=0.24\textwidth]{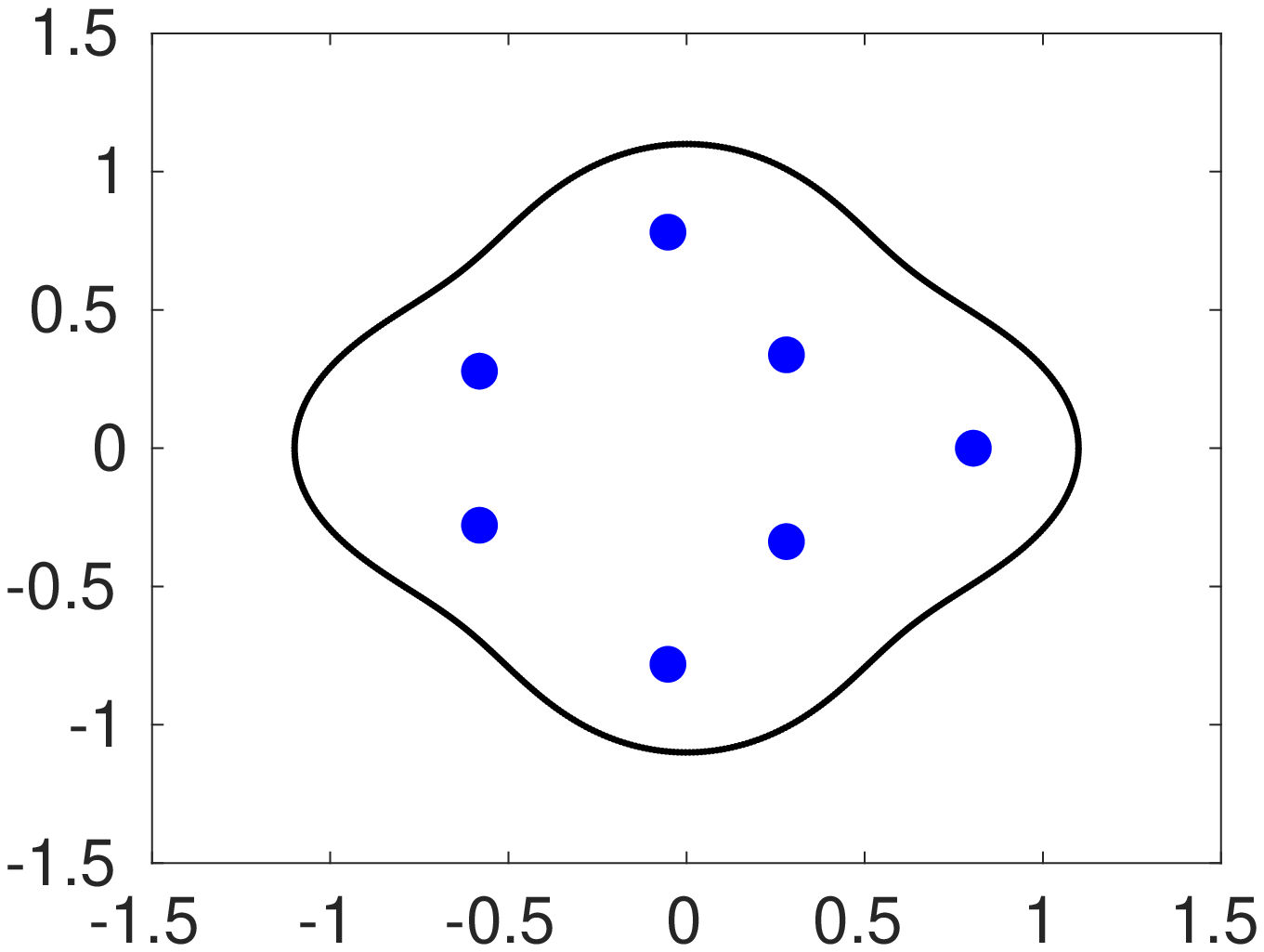}\label{fig:cos4_7}}
{\includegraphics[height=3.5cm,width=0.24\textwidth]{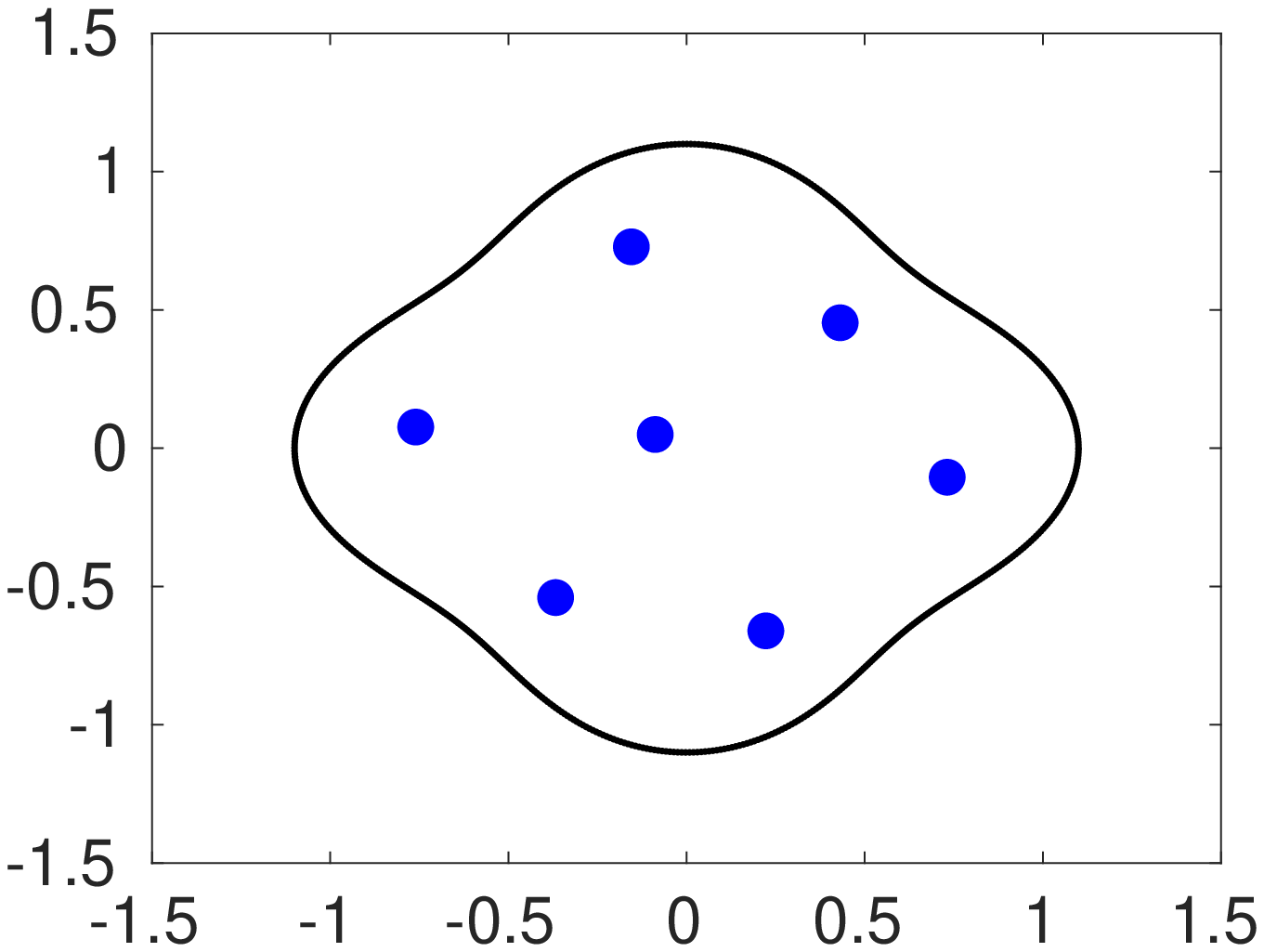}\label{fig:cos4_7_better}}
\caption{Optimal trap patterns for $D=1$ in a near-disk domain with
  boundary $r=1+\sigma \cos(4\theta)$, with $\sigma=0.1$, that
  contains $m$ traps of a common radius $\eps=0.05$. Computed from
  minimizing \eqref{final:avemfpt} using the ODE relaxation scheme
  \eqref{near_disk:relax}.  Left: $m=3$, $\overline{u}\approx
  0.2962$. Inter-trap computed distances are $0.9588$, $0.9588$, and
  $0.9540$. This result is close to the full PDE simulation results of
  Fig.~\ref{fig:full_near}. Left middle: $m=4$, $\overline{u}\approx
  0.1927$. This is a ring pattern of traps with ring radius
  $r_c\approx 0.6215$.  Right Middle: $m=7$,
  $\overline{u}\approx 0.0925$. Right: $m=7$,
  $\overline{u}\approx 0.0912$. The two patterns for $m=7$ give nearly
  the same values for $\overline{u}$, with the rightmost pattern
  giving a slightly lower value.}
\end{center}
\label{fig:neardisk_cos4}
\end{figure}

\begin{figure}[htbp]
  \begin{center}
\includegraphics[height=4.05cm,width=0.45\textwidth]{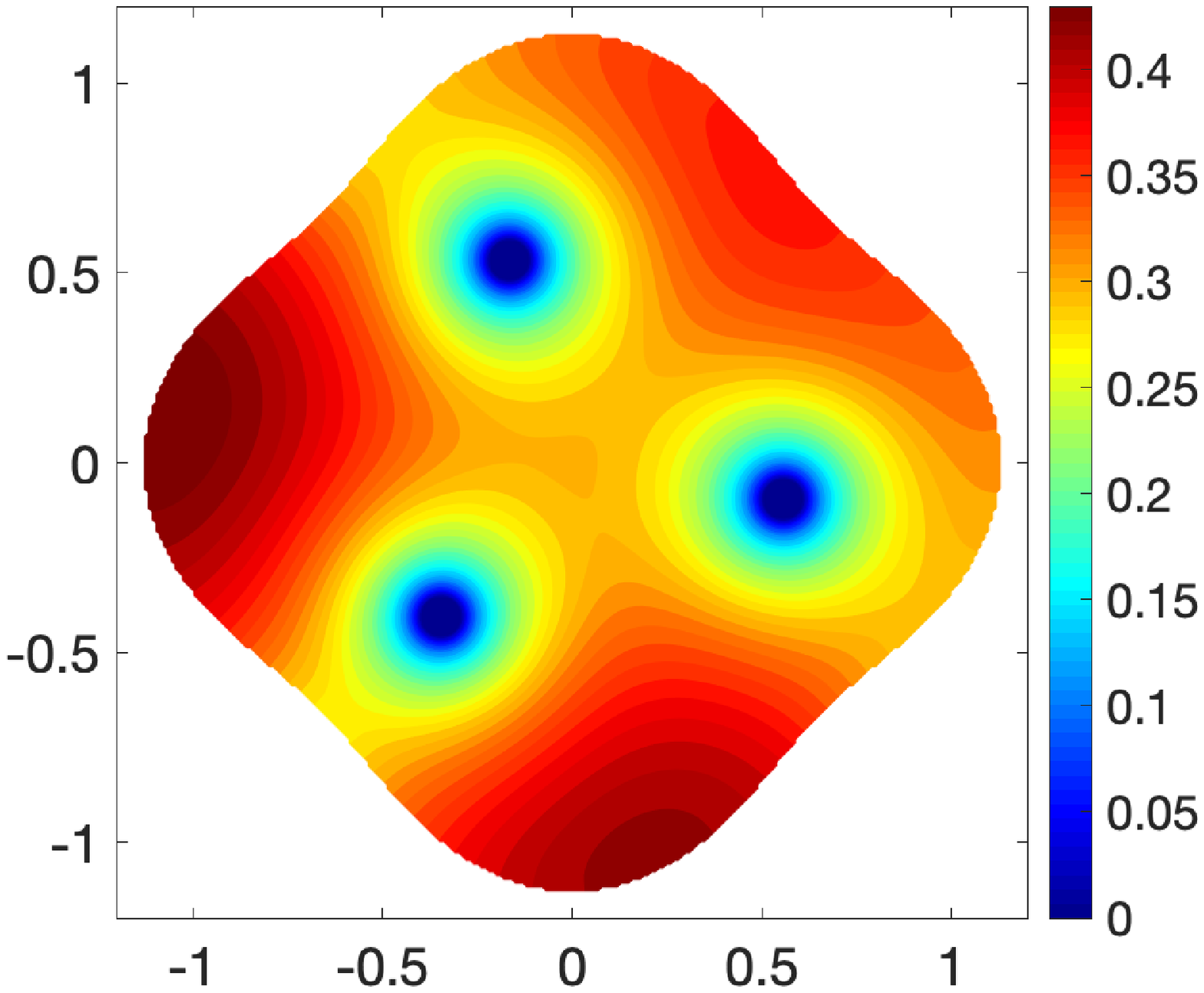}
\includegraphics[height=3.8cm,width=0.29\textwidth]{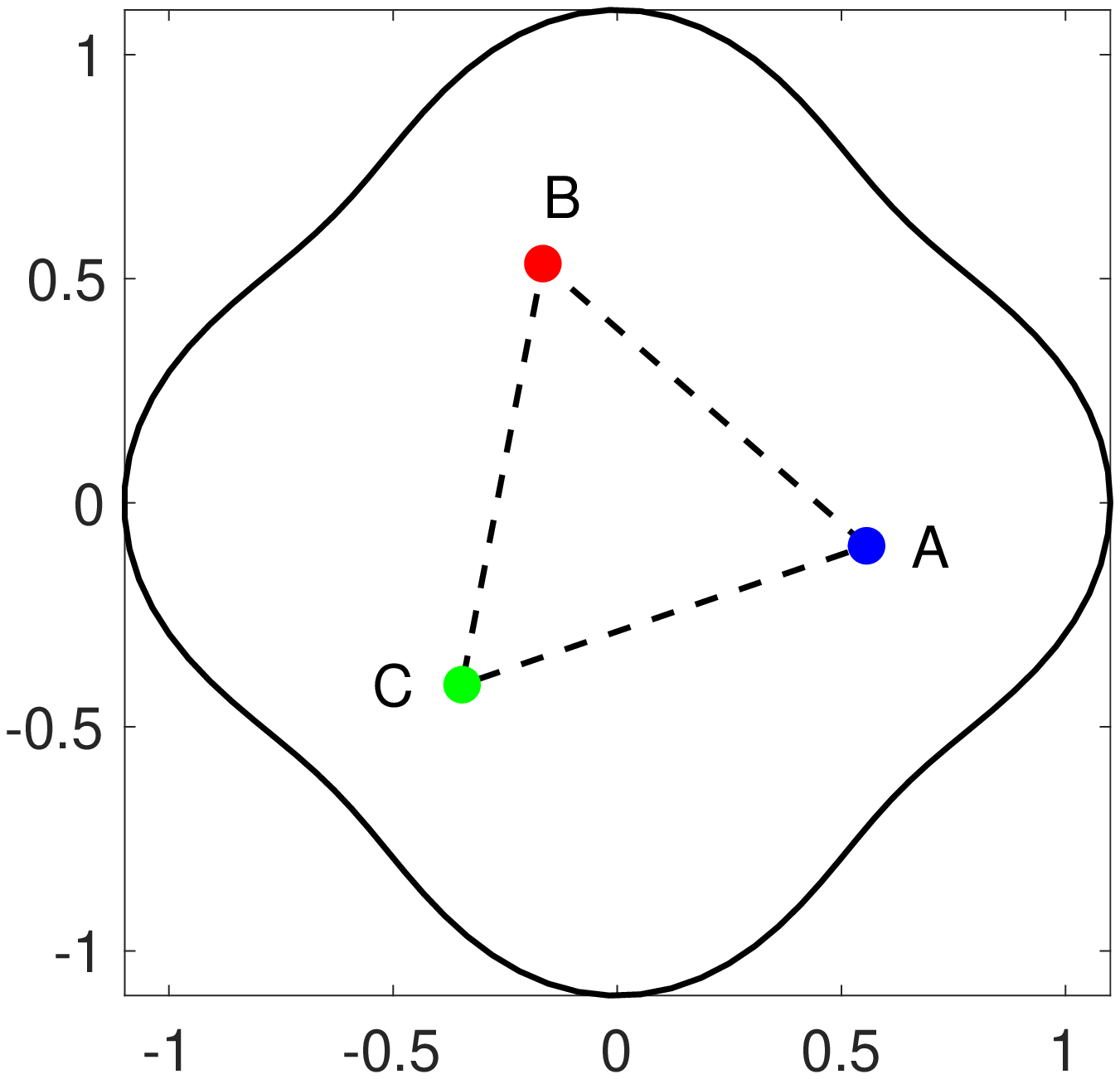}
\end{center}
\caption{Optimizing a three-trap pattern, with a common trap radius
  $\eps=0.05$, in a four-fold star-shaped domain (4-star) with boundary
  profile $h(\theta)=\cos(4\theta)$ and $\sigma=0.1$. Left panel:
  contour plot of the optimal PDE solution computed with closest point
  method. Right panel: optimal traps locations in the 4-star domain with
  computed side-lengths: $\mathbf{AB}\approx 0.9581$,
  $\mathbf{BC}\approx 0.9569$, and $\mathbf{CA}\approx 0.9541$. All of
  the computed interior angles are ${\pi/3}\pm \delta$, where
  $|\delta|\leq 0.0015$.}\label{fig:full_near}
\end{figure}

We first set $\sigma=0.1$ and consider the boundary profile
$h(\theta)=\cos(N\theta)$, where $N$ is a positive integer
representing the number of boundary folds.  In \cite{IWWC2019}, an
explicit two-term expansion for the average MFPT $\overline{u}$ was
derived for the special case where $m$ traps are equidistantly spaced
on a ring of radius $r_c$, concentric within the unperturbed disk. For
such a ring pattern, in Proposition 1 of \cite{IWWC2019} it was proved
that when ${N/m}\notin \mathbb{Z}^{+}$, then
$\overline{u}\sim \overline{u}_0 + {\mathcal O}(\sigma^2)$, as the
correction at order ${\mathcal O}(\sigma)$ vanishes
identically. Therefore, in order to determine the optimal trap pattern
when ${N/m}\notin \mathbb{Z}^{+}$ we must consider arbitrary trap
configurations, and not just ring patterns of traps. By minimizing
\eqref{final:avemfpt} using the ODE relaxation scheme
\eqref{near_disk:relax}, in the left panel of
Fig.~\ref{fig:neardisk_cos4} we show our asymptotic prediction for the
optimal trap configuration for $N=4$ folds and $m=3$ traps of a common
radius $\eps=0.05$. The optimal pattern is not of ring-type.  The
corresponding results computed from the closest point method of
\cite{IWWC2019}, shown in Fig.~\ref{fig:full_near}, are very close to
the asymptotic result.

In the left-middle panel of Fig.~\ref{fig:neardisk_cos4}, we show the
optimal trap pattern computed from our asymptotic theory
\eqref{final:avemfpt} and \eqref{near_disk:relax} for the boundary
profile $h(\theta)=\cos(4\theta)$ with $m=4$ traps and
$\sigma=0.1$. The optimal pattern is now a ring pattern of traps. In
this case, as predicted by Proposition 1 of \cite{IWWC2019}, the
optimal pattern has traps on the rays through the origin that coincide
with the maxima of the domain boundary. By applying Proposition 2 of
\cite{IWWC2019}, the optimal perturbed ring radius has the expansion
$r_{c,opt}\sim 0.5985+0.1985\sigma$. When $\sigma=0.1$, this
gives $r_{c,opt}\approx 0.6184$, and compares well with the
value $r_c\approx 0.6215$ calculated from \eqref{final:avemfpt} and
\eqref{near_disk:relax}.

In the two rightmost panels of Fig.~\ref{fig:neardisk_cos4}, we show 
for $h(\theta)=\cos(4\theta)$ and $\sigma=0.1$, that there are
two seven-trap patterns that give local minima for the average MFPT
$\bar{u}_0$. The minimum values of $\bar{u}_0$ for these patterns are
very similar.

Next, we construct a boundary profile with a localized protrusion,
or bulge, near $\theta=0$. To this end, we define
$f(\theta)\equiv -1 + \beta e^{-\chi \sin^{2}\left({\theta/2}\right)}$. By
using the Taylor expansion of $e^z$, combined with a simple
identity for $\int_{0}^{2\pi} \sin^{2n}(\psi)\,d\psi$, we conclude
that $\int_{0}^{2\pi} f(\theta)\,d\theta=0$ when $\beta$ is related to
$\chi$ by
\begin{equation}\label{near:bulge_prof}
  \frac{1}{\beta} = \frac{1}{2\pi}\int_{0}^{2\pi} e^{-\chi \sin^{2}\left({\theta/2}\right)}\,
  d\theta = \sum_{n=0}^{\infty} \frac{(-1)^n \chi^n}{2\pi n!}
  \int_{0}^{2\pi} \sin^{2n}\left(\frac{\theta}{2}\right)\, d\theta =
  \sum_{n=0}^{\infty} (-1)^n \frac{\chi^n (2n)!}{4^n\left(n!\right)^3} \,.
\end{equation}
As $\chi$ increases, the boundary deformation becomes increasingly
localized near $\theta=0$.

\begin{figure}[htbp]
  \begin{center}
{\includegraphics[height=3.5cm,width=0.24\textwidth]{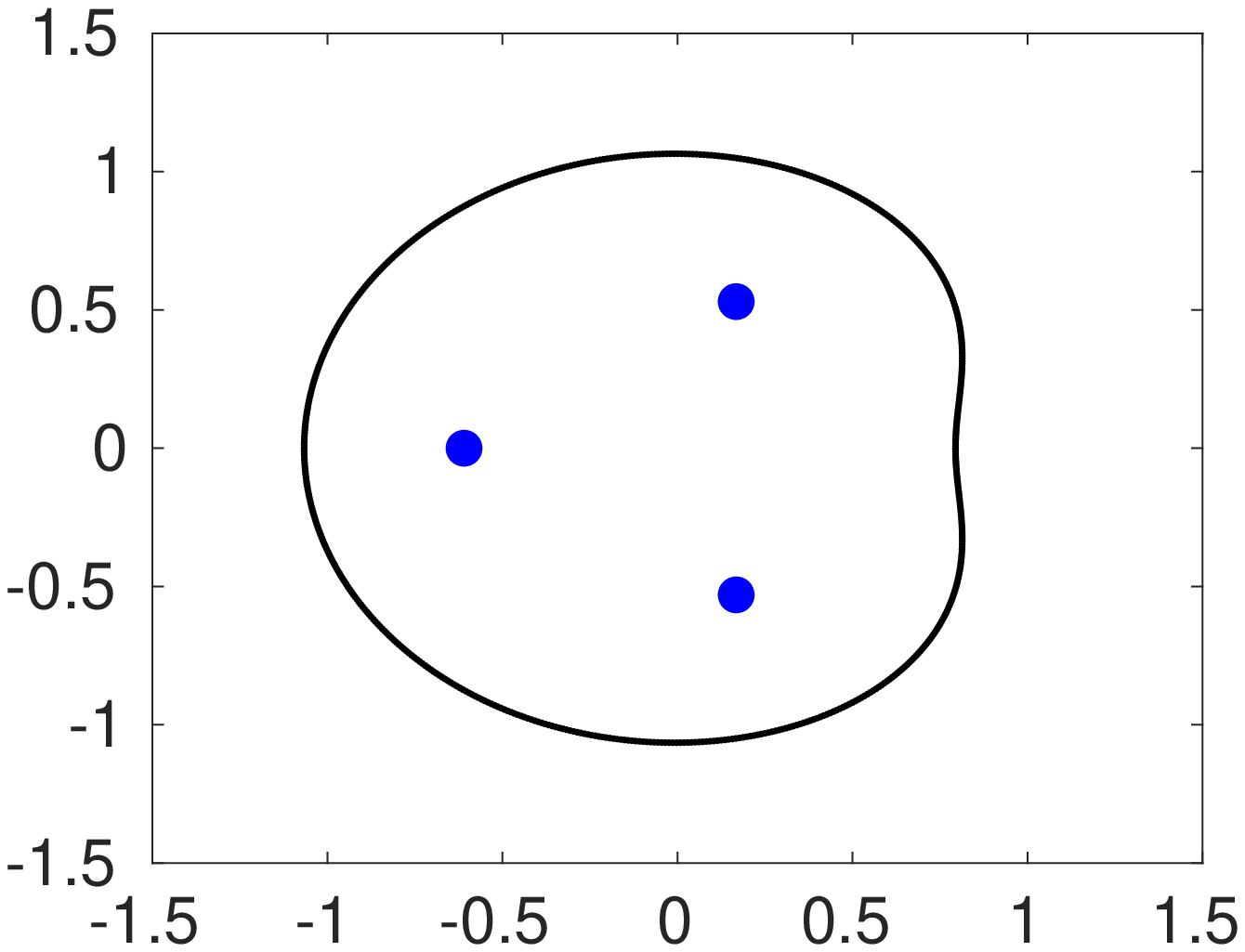}\label{fig:3bule_in}}
{\includegraphics[height=3.5cm,width=0.24\textwidth]{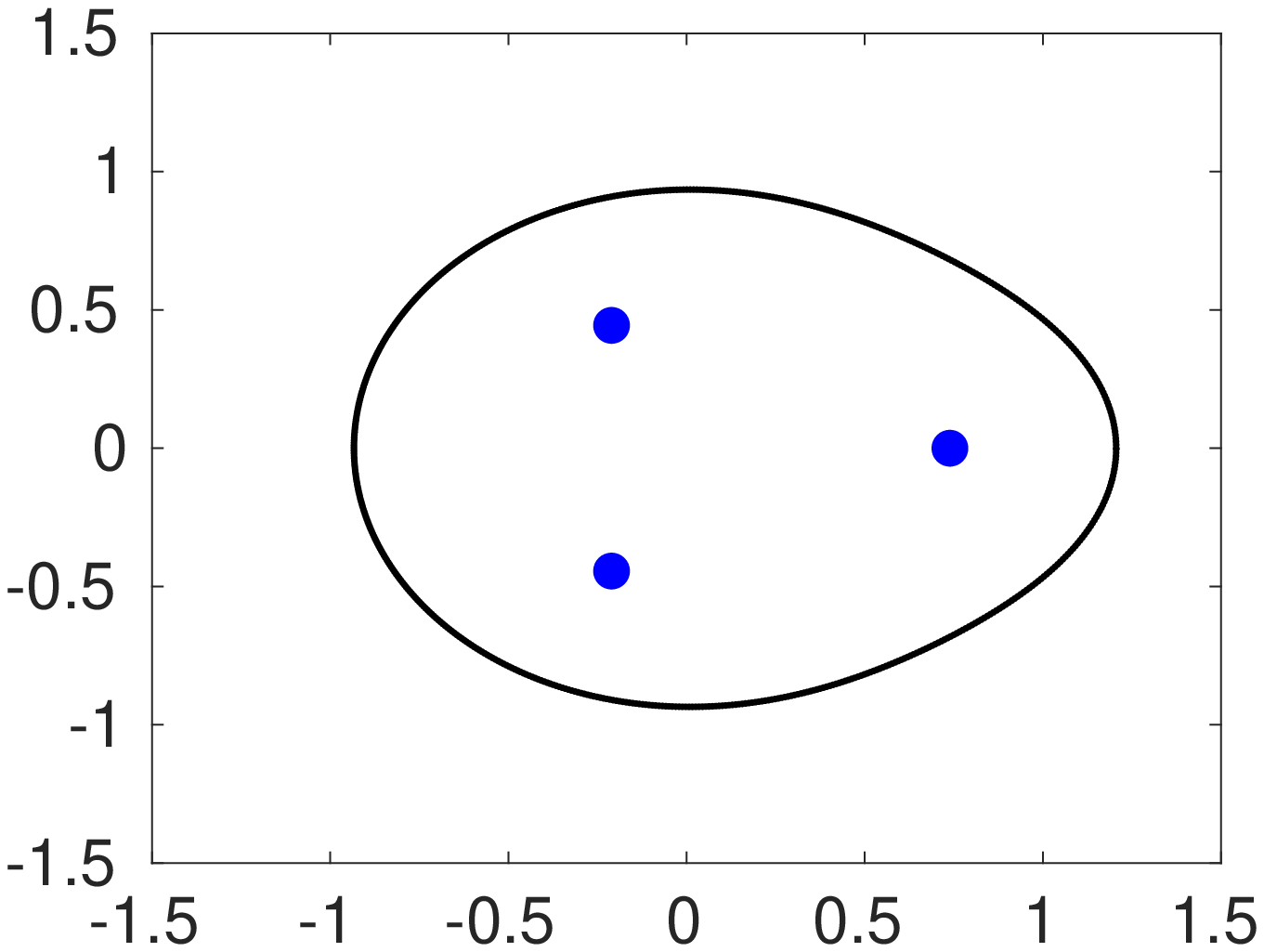}\label{fig:3bulge_out}}
{\includegraphics[height=3.5cm,width=0.24\textwidth]{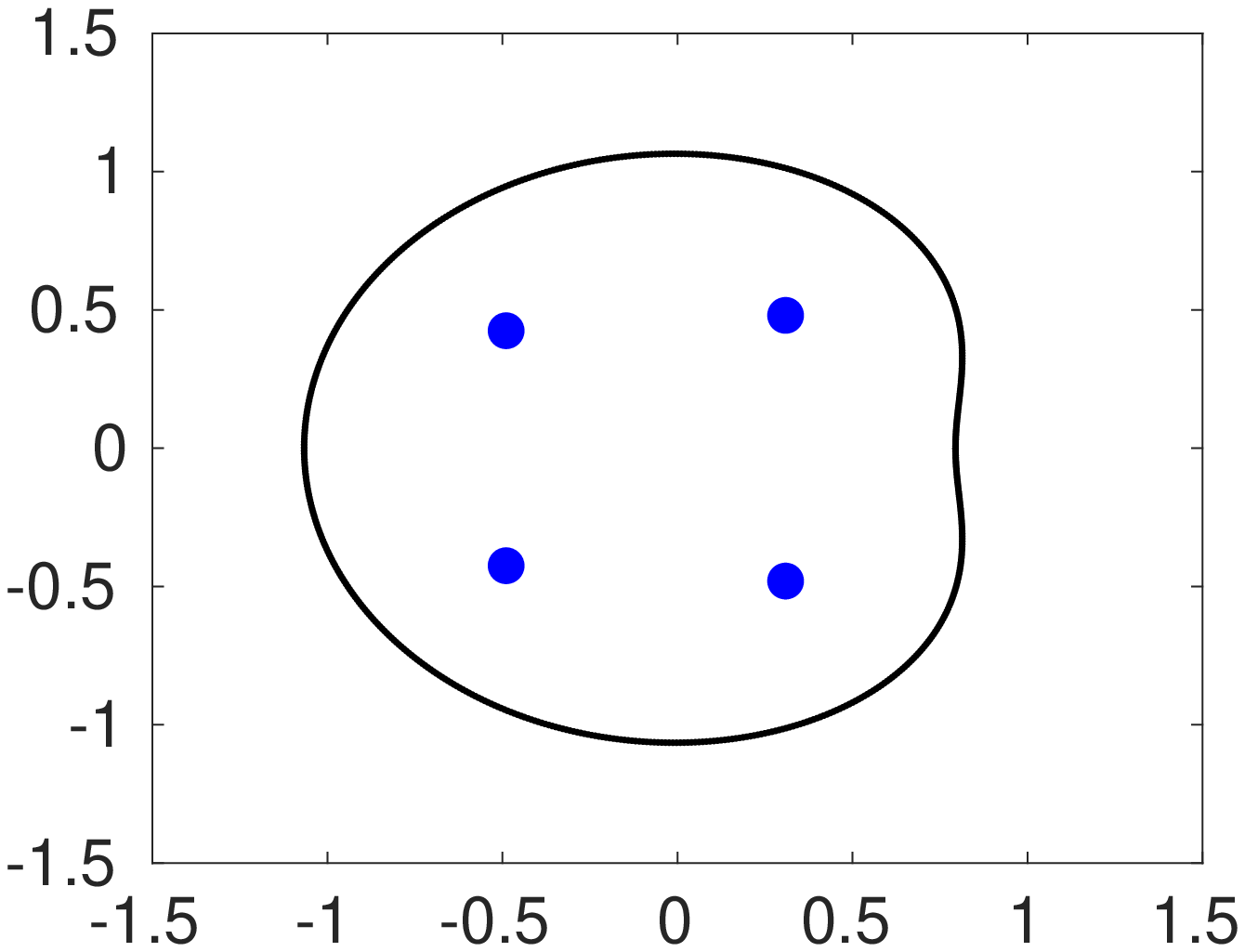}\label{fig:4bulge_in}}
{\includegraphics[height=3.5cm,width=0.24\textwidth]{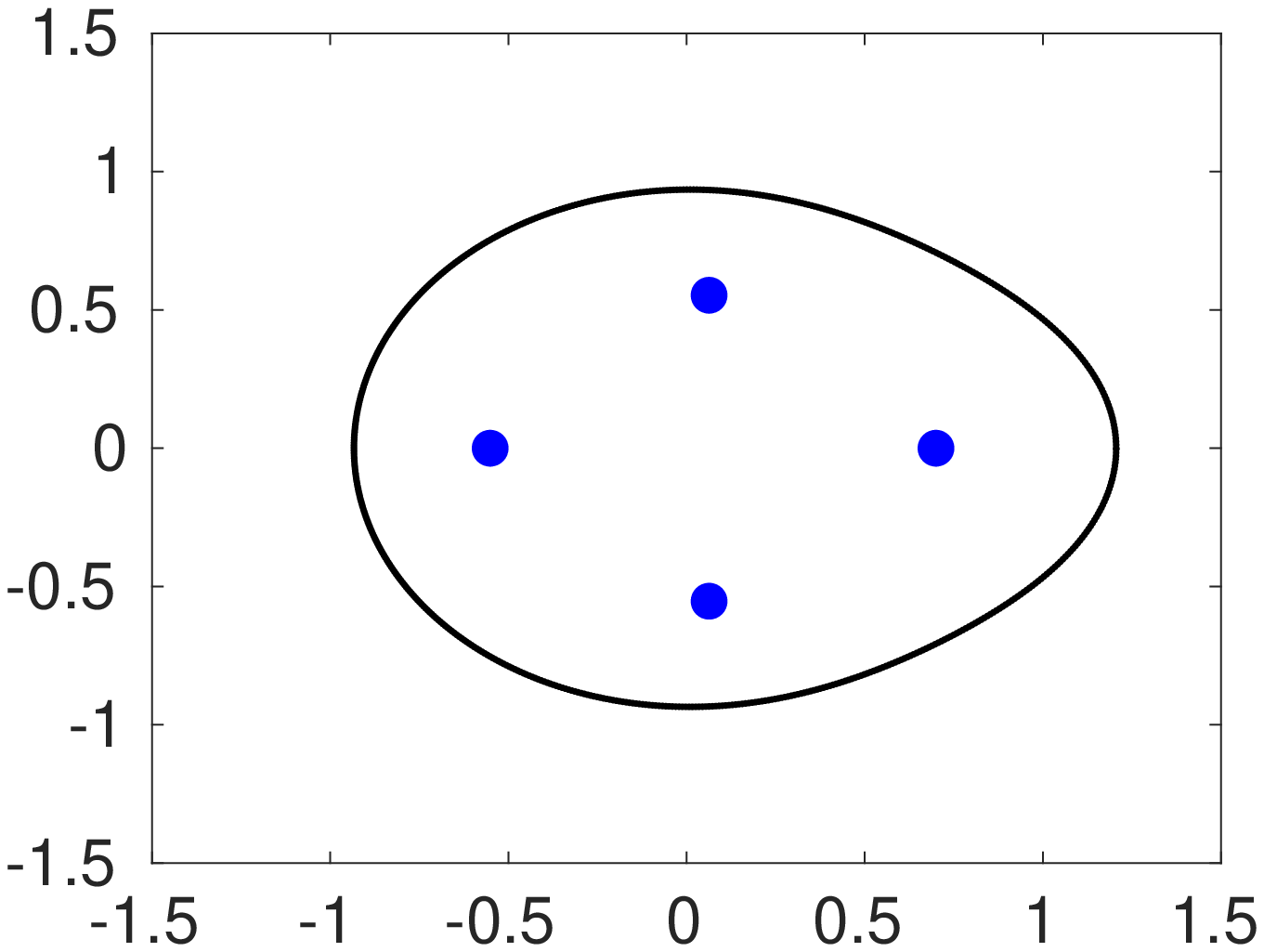}\label{fig:4bulge_out}}
\caption{Optimal trap patterns for $D=1$ with $m$ traps each of radius
  $\eps=0.05$ in a near-disk domain with boundary
  $r=1\pm \sigma f(\theta)$, where $\sigma=0.05$ and
  $f(\theta)=-1+\beta e^{-10\sin^{2}\left( {\theta/2}\right)}$, with
  $\beta=5.4484$. Computed from minimizing \eqref{final:avemfpt} using
  the ODE relaxation scheme \eqref{near_disk:relax}.  Left: $m=3$ and
  inward domain bulge $r=1 - \sigma f(\theta)$. Centroid of trap
  pattern is at $(-0.0886,0.0)$ and $\overline{u}\approx 0.2842$. Left
  Middle: $m=3$ and outward bulge $r=1 + \sigma f(\theta)$. Centroid
  is at $(0.1061,0.0)$, and $\overline{u}\approx 0.2825$. Right
  Middle: $m=4$ and inward bulge $r=1 - \sigma f(\theta)$,
  $\overline{u}\approx 0.1918$. Right: $m=4$ and outward bulge
  $r=1 + \sigma f(\theta)$, $\overline{u}\approx 0.1916$.}
\end{center}
\label{fig:neardisk_bulge}
\end{figure}

For $\chi=10$, for which $\beta=5.4484$, in
Fig.~\ref{fig:neardisk_bulge} we show optimal trap patterns for $m=3$
and $m=4$ traps for both an outward domain bulge, where
$r=1+\sigma f(\theta)$, and an inward domain bulge, were
$r=1-\sigma f(\theta)$, with $\sigma=0.05$. For the three-trap case,
by comparing the two leftmost plots in Fig.~\ref{fig:neardisk_bulge},
we observe that an inward domain bulge will displace the trap
locations to the left, as expected intuitively. Alternatively, for an
outward bulge, the location of the optimal trap on the line of
symmetry becomes closer to the domain protrusion. An intuitive, but as
we will see below in Fig.~\ref{fig:odd_5_sar}, na\"ive interpretation
of the qualitative effect of this domain bulge is that it acts to
confine or pin a Brownian particle in this region, and so in order to
reduce the mean capture time of such a pinned particle, the best
location for a trap is to move closer to the region of protrusion.
For the case of four traps, a similar qualitative comparison of the
optimal trap configuration for an inward and outward domain bulge is
seen in the two rightmost plots in Fig.~\ref{fig:neardisk_bulge}.

In Fig.~\ref{fig:neardisk_odd}, we show optimal trap patterns from our
hybrid theory for $3\leq m\leq 5$ circular traps of radius $\eps=0.05$
in a domain with boundary profile $r=1+\sigma h(\theta)$, where
$h(\theta)=\cos(3\theta)-\cos(\theta)-\cos(2\theta)$ and
$\sigma=0.075$. This boundary profile perturbs the unit disk inwards
near $\theta=\pi$ and outwards near $\theta=0$. For $m=3$, in
Fig.~\ref{fig:neardisk_full_PDE} we show a favorable comparison
between the full numerical PDE results and the hybrid results for the
optimal average MFPT and trap locations.  Moreover, from the two
rightmost plots in Fig.~\ref{fig:neardisk_odd}, we observe that there
are two five-trap patterns that give local minima for $\bar{u}_0$. The
pattern that has a trap on the line of symmetry near the outward
bulge at $\theta=0$ is, in this case, not a global minimum of
the average MFPT. This indicates that hard-to-assess global
effects, rather than simply the local geometry near a protrusion, play
a central role for characterizing the optimal trap pattern.

\begin{figure}[htbp]
\begin{center}
{\includegraphics[height=3.5cm,width=0.24\textwidth]{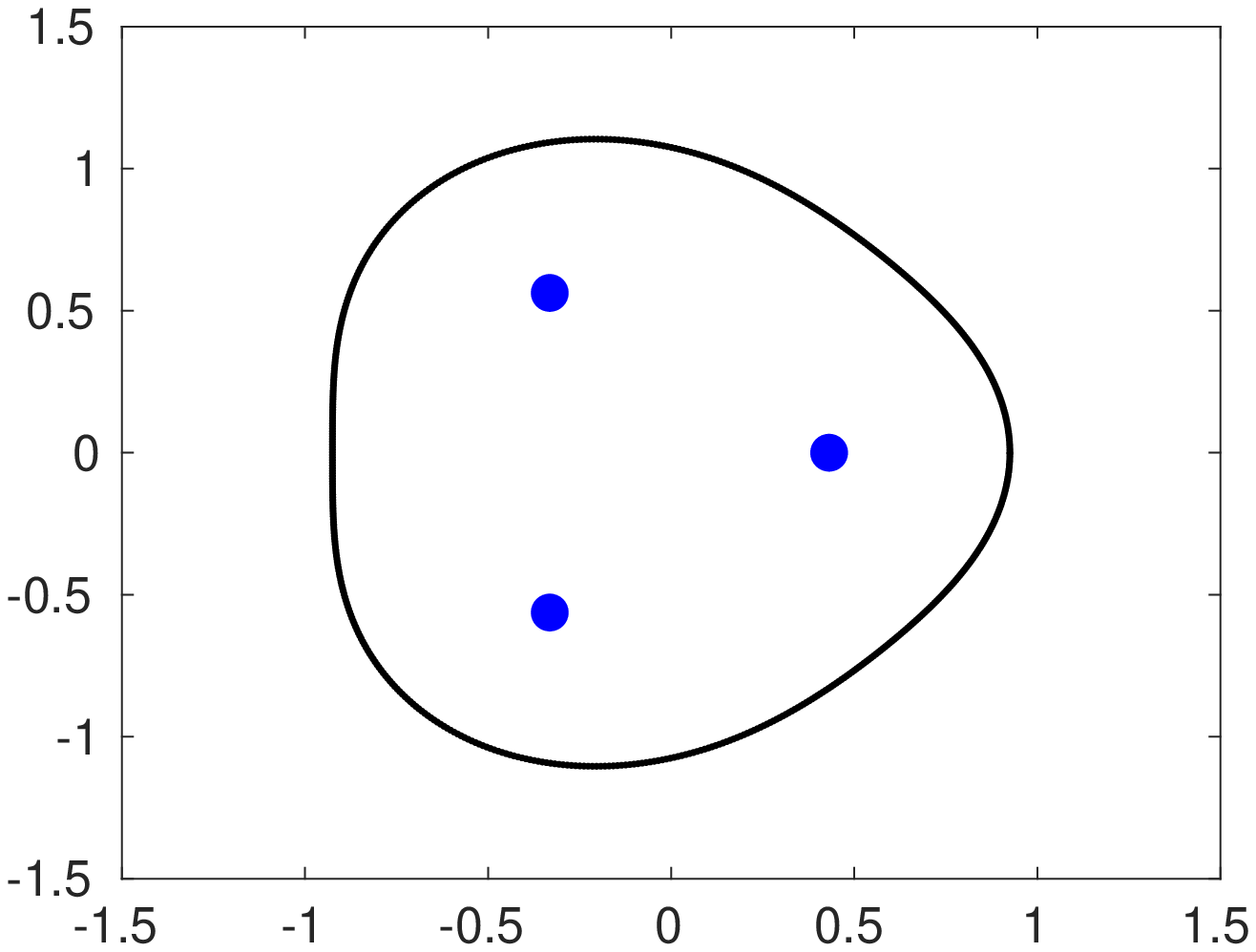}\label{fig:odd_3}}
{\includegraphics[height=3.5cm,width=0.24\textwidth]{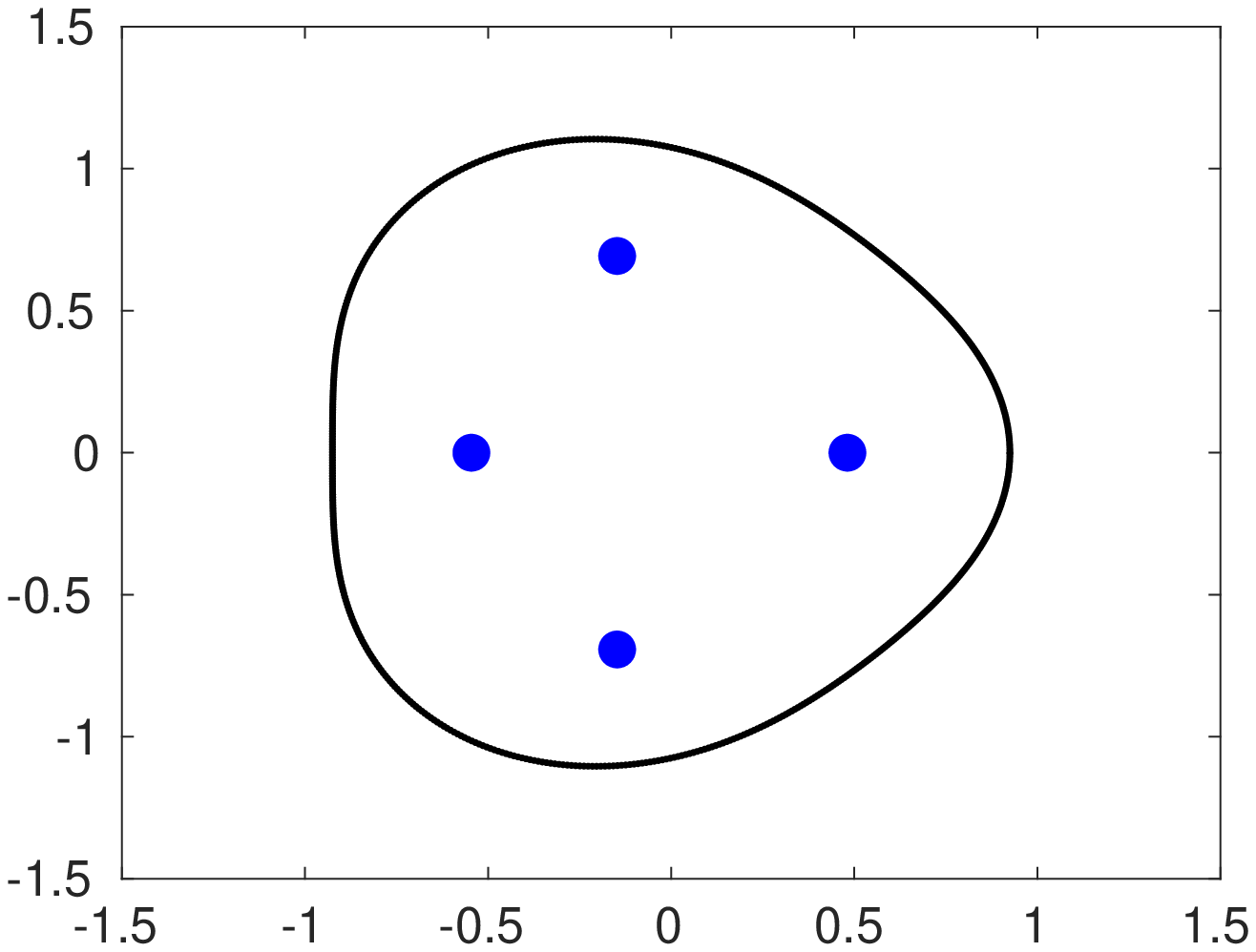}\label{fig:odd_4}}
{\includegraphics[height=3.5cm,width=0.24\textwidth]{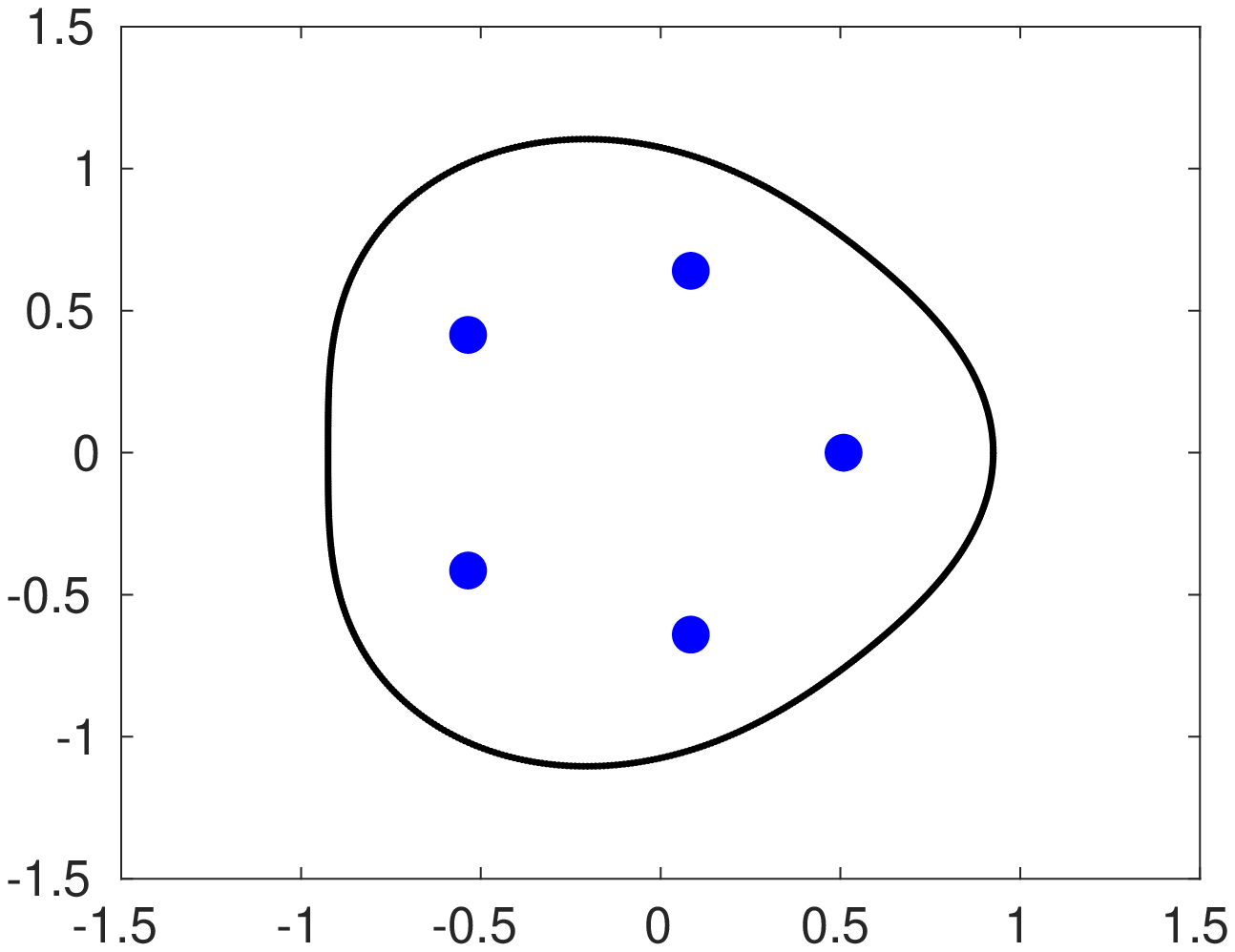}\label{fig:odd_5}}
{\includegraphics[height=3.5cm,width=0.24\textwidth]{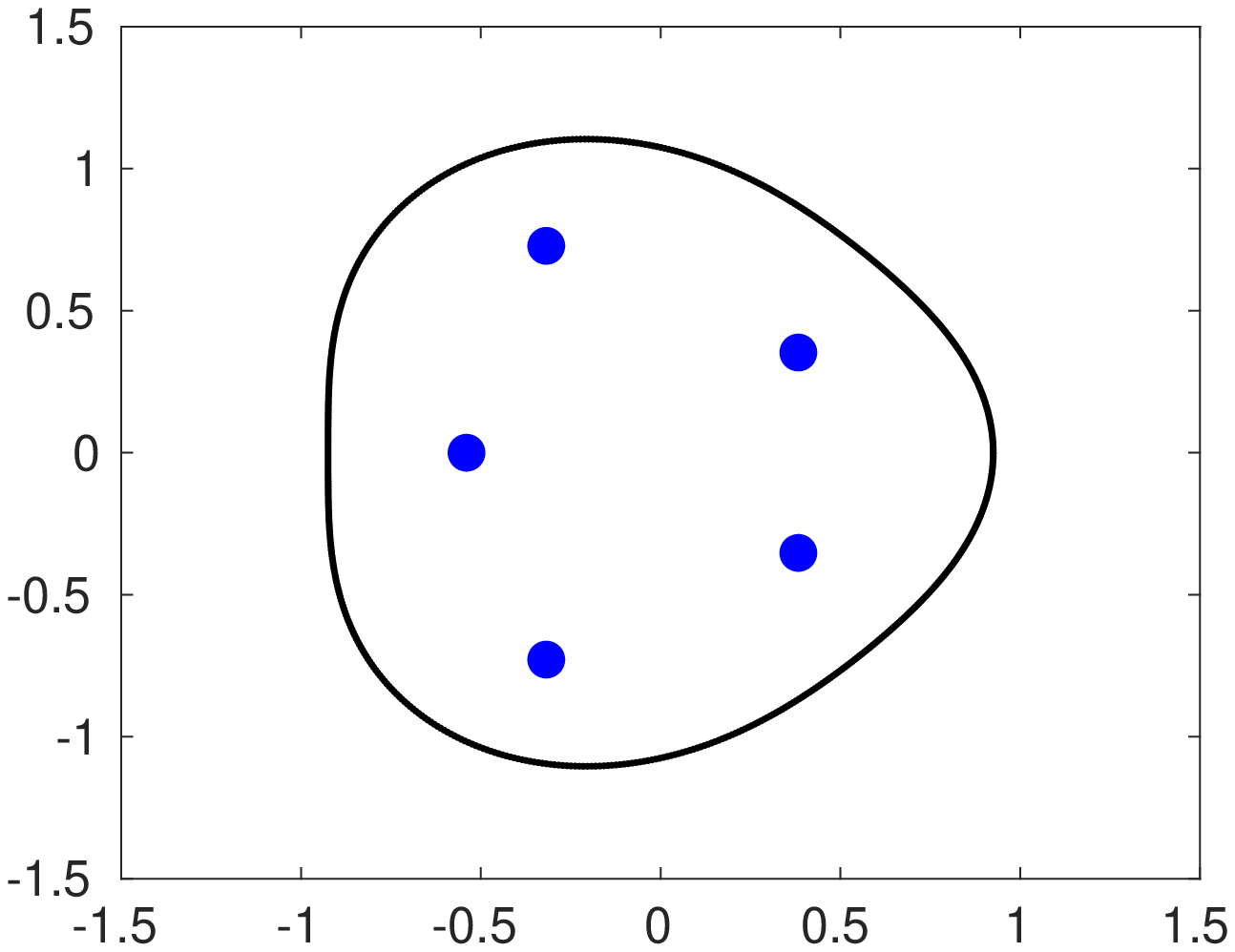}\label{fig:odd_5_sar}}
\caption{Optimal trap patterns for $D=1$ in a near-disk domain with
  boundary $r=1+\sigma h(\theta)$, $\sigma=0.075$ and
  $h(\theta)=\cos(3\theta)-\cos(\theta)-\cos(2\theta)$, that contains
  $m$ traps of a common radius $\eps=0.05$. Computed from minimizing
  \eqref{final:avemfpt} using the ODE relaxation scheme
  \eqref{near_disk:relax}.  Left: $m=3$ and
  $\overline{u}\approx 0.2794$. Left-Middle: $m=4$ and
  $\overline{u}\approx 0.19055$. Right-Middle: $m=5$ and
  $\overline{u}\approx 0.1418$. Right: $m=5$,
  $\overline{u}\approx 0.1383$.  The two patterns for $m=5$ are local
  minimizers, with rather close values for $\overline{u}$. The global
  minimum is achieved for the rightmost pattern.}
\end{center}
\label{fig:neardisk_odd}
\end{figure}

\begin{figure}[htbp]
\begin{center}
  \includegraphics[height=4.0cm,width=0.60\textwidth]{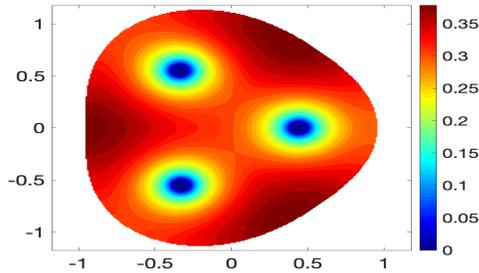}
  \caption{Contour plot of the PDE numerical solution for the optimal
    average MFPT and trap locations computed from the closest point
    method corresponding to the parameter values in the left panel of
    Fig.~\ref{fig:neardisk_odd}. Full PDE results for optimal
    locations: $(-0.3382, 0.5512)$, $(-0.3288,-0.5510)$,
    $(0.4410, 0.0012)$, and $\overline{u}=0.2996$.  Hybrid results:
    $(-0.3316, 0.5626)$, $(-0.3316, 0.5626)$, $(0.4314,0.000)$, and
    $\overline{u}_0=0.2794$.}
  \label{fig:neardisk_full_PDE}
\end{center}
\end{figure}

\section{Optimizing Trap Configurations for the MFPT
in an Ellipse}\label{sec:ellipse}

Next, we consider the trap optimization problem in an ellipse of
arbitrary aspect ratio, but with fixed area $\pi$. Our analysis uses
a new explicit analytical formula, as derived in
\S~\ref{sec:g_ell}, for the Neumann Green's function $G(\x;\x_0)$ and
its regular part $R_e$ of \eqref{ell:g}.

For $m$ circular traps each of radius $\eps$, the average MFPT
$\overline{u}_0$ satisfies (see \eqref{u0_bar})
\begin{equation}\label{e:u0_bar}
  \overline{u}_0 = \frac{|\Omega|}{2\pi D \nu m}
  + \frac{2 \pi}{m} \v{e}^T \mc{G} \mathcal{A}  \,,
  \quad \mbox{where} \quad 
  \Big{[} I + 2 \pi \nu (I - E)\mc{G}  \Big{]} \mathcal{A} =
  \frac{|\Omega|}{2\pi D m} \v{e} \,.
\end{equation}
Here $E\equiv {\v{e}\v{e}^T/m}$, $\v{e}=(1,\ldots,1)^T$,
$\nu\equiv {-1/\log\eps}$, and the Green's matrix $\mc{G}$ depends on
the trap locations $\lbrace{\x_1,\ldots,\x_m\rbrace}$. To determine
optimal trap configurations that are minimizers of the average MFPT, given
in \eqref{e:u0_bar}, we use the ODE relaxation scheme
\begin{equation}
  \frac{d\v{z}}{dt} = -\nabla_{\v{z}} \overline{u}_0 \,, \qquad
  \mbox{where} \quad \v{z}\equiv (x_1,y_1,\ldots,x_m,y_m) \,.
  \label{ode:relax}
\end{equation}
In our implementation of \eqref{ode:relax}, the gradient was
approximated using a centered difference scheme with mesh spacing
$10^{-4}$. The results shown below for the optimal trap patterns are
confirmed from using a particle swarm approach \cite{kennedy2010}.

The derivation of the Neumann Green's function and its regular part in
\S~\ref{sec:g_ell} is based on mapping the elliptical domain to a
rectangular domain using 
\begin{subequations}\label{ell:coord}
\begin{equation}\label{ell:coord_1}
  x = f \cosh\xi \cos\eta \,, \quad y = f \sinh\xi \sin\eta \,, \qquad
  f=\sqrt{a^2 - b^2} \,.
\end{equation}
With these elliptic cylindrical coordinates, the ellipse is mapped to the
rectangle $0\leq \xi\leq \xi_b$ and $0\leq\eta\leq 2\pi$, where
$a=f\cosh\xi_b$ and $b=f\sinh\xi_b$, so that
\begin{equation}
  f = \sqrt{a^2 - b^2} \,, \qquad \xi_b = \tanh^{-1}\left(\frac{b}{a}\right)
  = -\frac{1}{2} \log\beta\,, \qquad \beta 
 \equiv \left(\frac{a-b}{a+b}\right)\,. \label{ell:coord_2}
\end{equation}
\end{subequations}
To determine $(\xi,\eta)$, given a pair $(x,y)$, we invert the
transformation \eqref{ell:coord_1} using
\begin{subequations}\label{ell:inverse_mapping}
\begin{equation}\label{ell:xy_to_xi}
  \xi = \frac{1}{2} \log\left( 1 - 2s + 2 \sqrt{s^2-s}\right) \,, \quad
s \equiv \frac{-\mu - \sqrt{\mu^2 + 4 f^2 y^2}}{2f^2} \,, \quad \mu\equiv
x^2+y^2-f^2 \,.
\end{equation}
To recover $\eta$, we define $\eta_{\star}\equiv \sin^{-1}(\sqrt{p})$ and use
\begin{equation}\label{ell:xy_to_eta}
  \eta = \begin{cases}
      \eta_{\star}, & \text{if } x\geq 0\,, \,\, y\geq 0\\
 \pi -\eta_{\star}, & \text{if } x<0\,, \,\, y\geq 0\\
 \pi +\eta_{\star}, & \text{if } x\leq 0\,, \,\, y< 0\\
2\pi- \eta_{\star}, & \text{if }  x>0\,, \,\, y< 0
                \end{cases}\,,\quad \mbox{where} \quad
   p \equiv  \frac{-\mu + \sqrt{\mu^2 + 4 f^2 y^2}}{2f^2} \,.
\end{equation}
\end{subequations}
As derived in \S~\ref{sec:g_ell}, the matrix entries in $\mc{G}$ are
obtained from the explicit result
\begin{subequations}\label{cell:finz_g}
\begin{equation}\label{cell:finz_g1}
\begin{split}
    G(\x;\x_0) &= \frac{1}{4|\Omega|} \left(|\x|^2 + |\x_0|^2\right) -
    \frac{3}{16|\Omega|}(a^2 + b^2) - \frac{1}{4\pi}\log\beta -
    \frac{1}{2\pi} \xiM 
    \\
   & \qquad  -\frac{1}{2\pi} \sum_{n=0}^{\infty} \log\left(
    \displaystyle \prod_{j=1}^{8} |1-\beta^{2n} z_j| \right)  \,, \quad
  \mbox{for} \quad \x\neq \x_0 \,,
\end{split}
\end{equation}
where $|\Omega|=\pi ab$, $\xiM\equiv \max(\xi,\xi_0)$, and the complex
constants $z_1,\ldots,z_8$ are defined in terms of $(\xi,\eta)$,
$(\xi_0,\eta_0)$ and $\xi_b$ by
\begin{equation}\label{cell:def_z}
\begin{split}
  z_1 &\equiv e^{-|\xi-\xi_0| + i(\eta-\eta_0)} \,, \quad
  z_2 \equiv e^{|\xi-\xi_0|-4\xi_b + i(\eta-\eta_0)}\,, \quad
  z_3 \equiv e^{-(\xi+\xi_0) - 2\xi_b + i(\eta-\eta_0)}\,, \\
  z_4 &\equiv e^{\xi+\xi_0-2\xi_b + i(\eta-\eta_0)} \,, \quad
  z_5 \equiv e^{\xi+\xi_0 -4\xi_b + i(\eta+\eta_0)} \,, \quad
  z_6 \equiv e^{-(\xi+\xi_0) + i(\eta+\eta_0)} \,, \\
  z_7 &\equiv e^{|\xi-\xi_0| -2\xi_b + i(\eta+\eta_0)} \,, \quad
  z_8 \equiv  e^{-|\xi-\xi_0| -2\xi_b + i(\eta+\eta_0)} \,.
\end{split}
\end{equation}
\end{subequations}
Observe that the Dirac point at $\x_0=(x_0,y_0)$ is mapped to
$(\xi_0,\eta_0)$. The transformation \eqref{ell:coord} and its inverse
\eqref{ell:inverse_mapping}, determines $G(\x;\x_0)$ explicitly in
terms of $\x \in \Omega$.

Moreover, as shown in \S~\ref{sec:g_ell}, the regular part of the
Neumann Green's function, $R_e$, satisfying
$G(\x;\x_0)\sim -(2\pi)^{-1}\log|\x-\x_0|+R_e$ as $\x\to \x_0$, is
given by
\begin{subequations}\label{cell:R0}
\begin{equation}
  \begin{split}
  R_{e} &= \frac{|\x_0|^2}{2|\Omega|} - \frac{3}{16|\Omega|} (a^2+b^2)
  + \frac{1}{2\pi}\log(a+b) - \frac{\xi_0}{2\pi} + \frac{1}{4\pi}
  \log\left(\cosh^2\xi_0 - \cos^2\eta_0\right) \\
  &\quad -\frac{1}{2\pi} \sum_{n=1}^{\infty} \log(1-\beta^{2n}) -
  \frac{1}{2\pi} \sum_{n=0}^{\infty}\log\left(
    \displaystyle \prod_{j=2}^{8} |1-\beta^{2n} z_j^0| \right)  \,.
\end{split}
\end{equation}
Here, $z_j^{0}$ is the limiting value of $z_j$, defined in
\eqref{cell:def_z}, as $(\xi,\eta)\to(\xi_0,\eta_0)$, given
by
\begin{equation}
\begin{split}
  z_2^{0}&=\beta^2\,, \quad z_3^{0}=\beta e^{-2\xi_0}\,, \quad
  z_4^{0}=\beta e^{2\xi_0}\,, \quad z_5^{0}=\beta^2 e^{2\xi_0+2i\eta_0} \,,\\
  z_6^{0}&=e^{-2\xi_0+2i\eta_0} \,, \quad z_7^{0}=\beta e^{2i\eta_0} \,, \quad
  z_8^{0}=\beta e^{2i\eta_0}\,, \quad \mbox{where} \quad
  \beta \equiv \frac{a-b}{a+b} \,.
\end{split}
\end{equation}
\end{subequations}

\subsection{Examples of the Theory}\label{ell:ex}

In this subsection, we will apply our hybrid analytical-numerical
approach based on \eqref{e:u0_bar}, \eqref{cell:finz_g},
\eqref{cell:R0} and the ODE relaxation scheme \eqref{ode:relax}, to
compute optimal trap configurations in an elliptical domain of area
$\pi$ with either $m=2,\ldots,5$ circular traps of a common radius
$\eps=0.05$. In our examples below, we set $D=1$ and we study how the
optimal pattern of traps changes as the aspect ratio of the ellipse is
varied. We will compare our results from this hybrid theory with the
near-disk asymptotic results of \eqref{final:avemfpt}, with full PDE
numerical results computed from the closest point method
\cite{IWWC2019}, and with the asymptotic approximations derived below
in \S~\ref{sec:thin}, which are valid for a long and thin ellipse.

\begin{figure}[htbp]
\begin{center}
\includegraphics[height=4.1cm,width=0.49\textwidth]{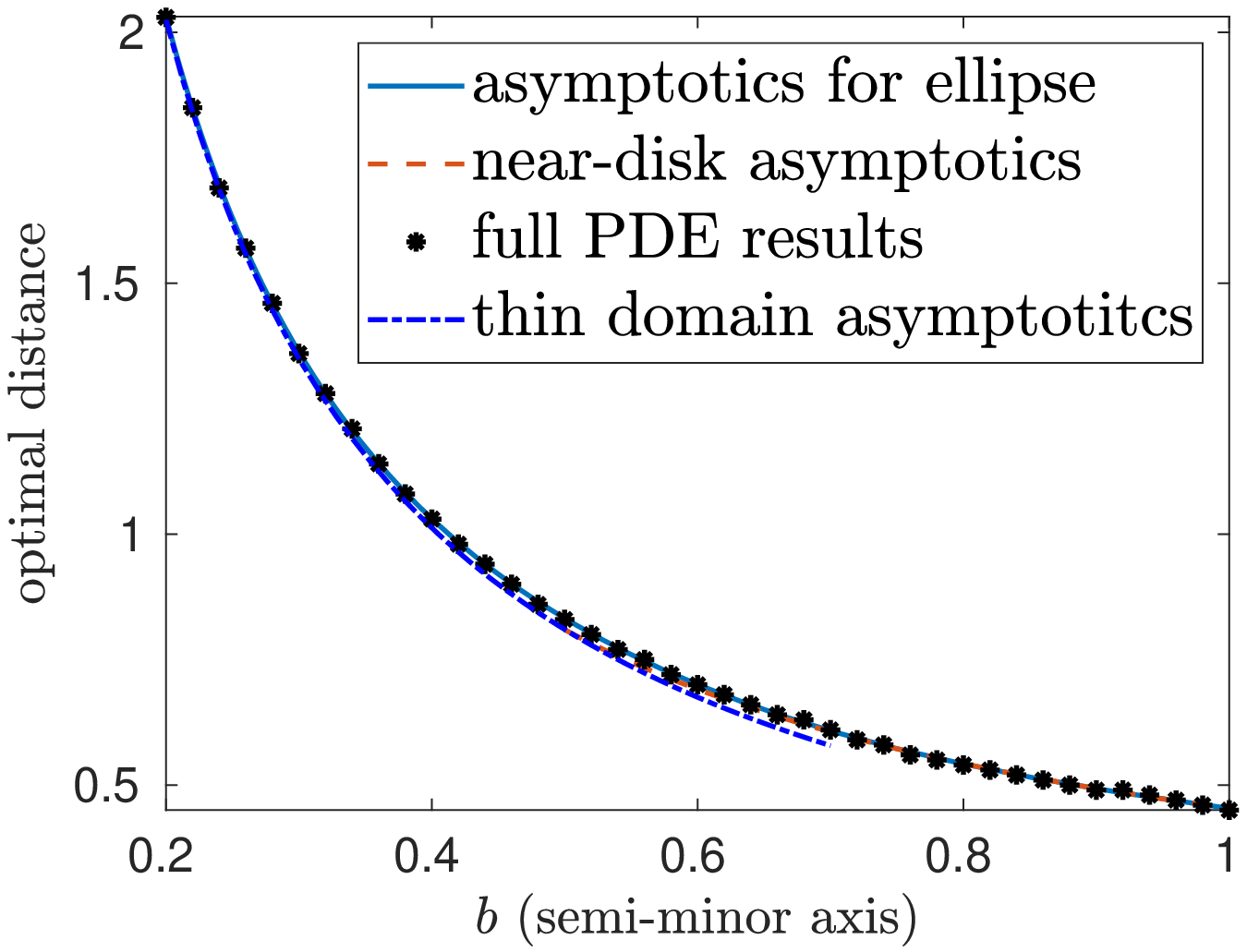}
\includegraphics[height=4.1cm,width=0.49\textwidth]{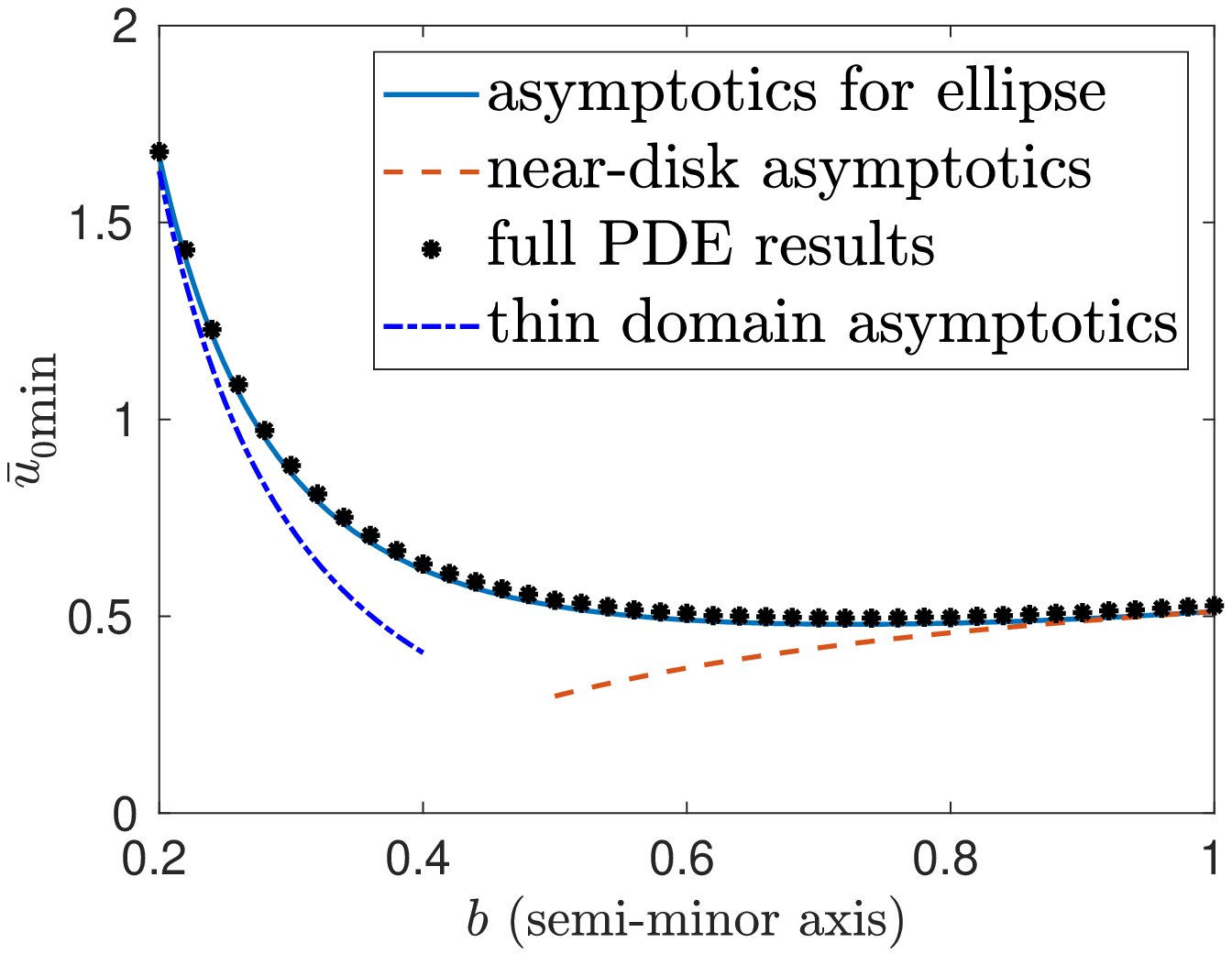}
\caption{The optimal trap distance from the origin (left panel) and
  optimal average MFPT $\overline{u}_{0\mbox{min}}$ (right panel) versus
  the semi-minor axis $b$ of an elliptical domain of area $\pi$ that
  contains two traps of a common radius $\eps=0.05$ and $D=1$. The
  optimum trap locations are on the semi-major axis, equidistant from
  the origin. Solid curves: hybrid asymptotic theory \eqref{e:u0_bar}
  for the ellipse coupled to the ODE relaxation scheme
  \eqref{ode:relax} to find the minimum.  Dashed line (red):
  near-disk asymptotics of \eqref{final:avemfpt}. Discrete
  points: full numerical PDE results computed from the closest point
  method. Dashed-dotted line (blue): thin-domain asymptotics \eqref{thin:m2}.
  These curves essentially overlap with those from the hybrid theory for the
  optimal trap distance.}
\end{center}
\label{fig:two_ellipse}
\end{figure}

For $m=2$ traps, in the right panel of Fig.~\ref{fig:two_ellipse} we
show results for the optimal average MFPT versus the semi-minor axis
$b$ of the ellipse. The hybrid theory is seen to compare very
favorably with full numerical PDE results for all $b\leq 1$. For $b$
near unity and for $b$ small, the near-disk theory of
\eqref{final:avemfpt} and \eqref{near_disk:relax}, and the thin-domain
asymptotic result in \eqref{thin:m2} are seen to provide,
respectively, good predictions for the optimal MFPT. Our hybrid theory
shows that the optimal trap locations are on the semi-major axis for
all $b<1$. In the left panel of Fig.~\ref{fig:two_ellipse}, the optimal
trap locations found from the steady-state of our ODE relaxation
\eqref{ode:relax} are seen to compare very favorably with full PDE
results. Remarkably, we observe that the thin-domain asymptotics
prediction in \eqref{thin:m2} agrees well with the optimal locations
from our hybrid theory for $b<0.7$.

\begin{figure}[htbp]
\begin{center}
\includegraphics[height=4.5cm,width=0.60\textwidth]{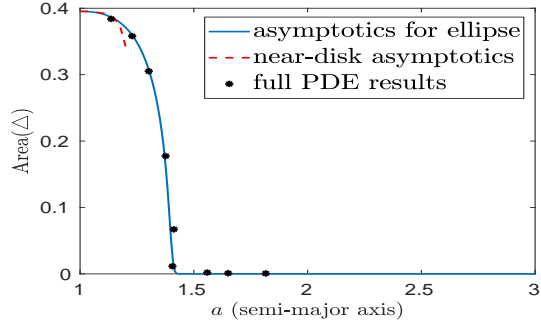}
\caption{Area of the triangle formed by the three optimally located
  traps of a common radius $\eps=0.05$ with $D=1$ in a deforming
  ellipse of area $\pi$ versus versus the semi-major axis $a$. The
  optimal traps become collinear as $a$ increases. Solid curve: hybrid
  asymptotic theory \eqref{e:u0_bar} for the ellipse coupled to the
  ODE relaxation scheme \eqref{ode:relax} to find the minimum.  Dashed
  line: near-disk asymptotics of \eqref{final:avemfpt}. Discrete
  points: full numerical PDE results computed from the closest point
  method.}
\end{center}
\label{fig:three_area}
\end{figure}

\begin{figure}[htbp]
\begin{center}
{\includegraphics[height=3.5cm,width=0.24\textwidth]{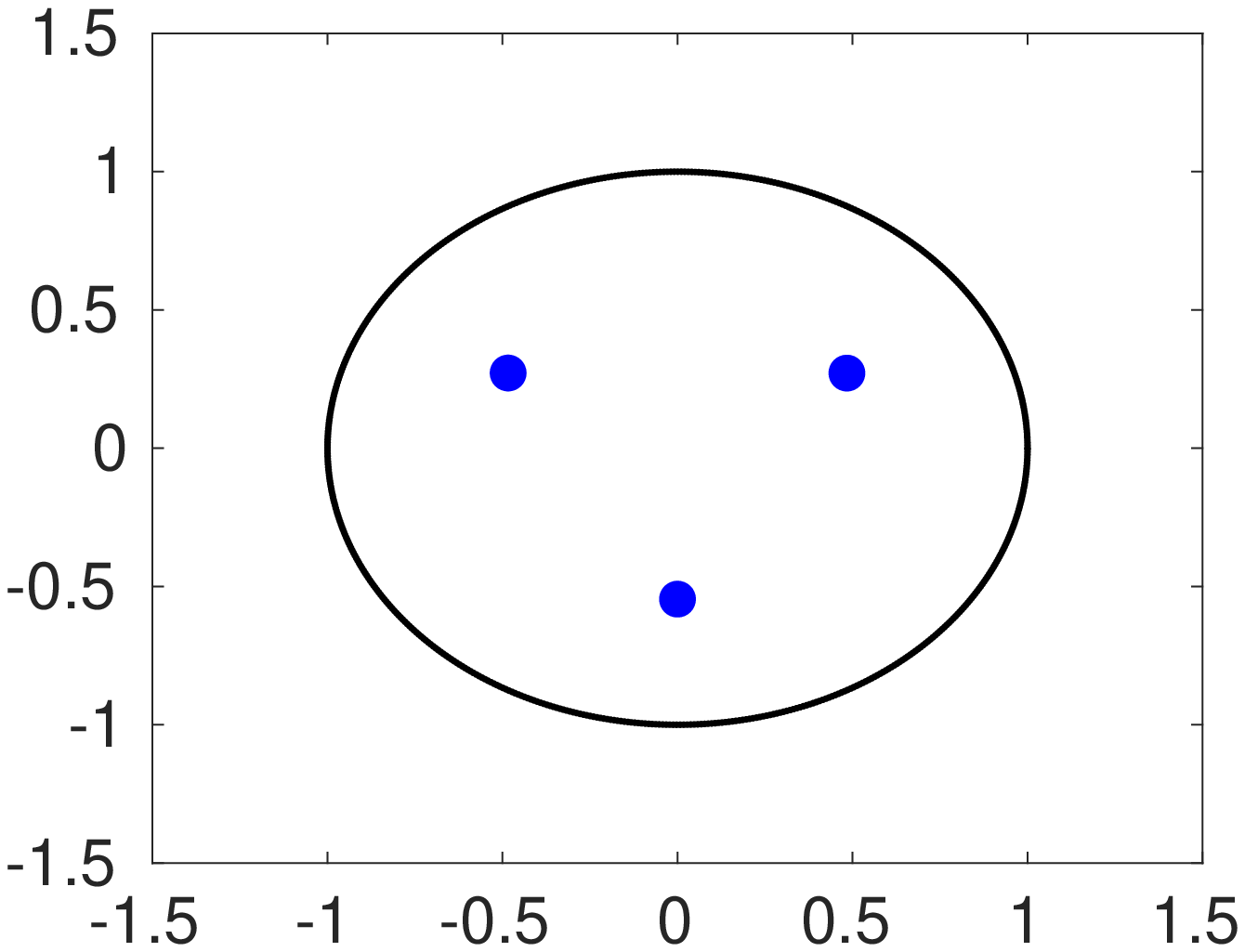}\label{fig:three_snap1}}
{\includegraphics[height=3.5cm,width=0.24\textwidth]{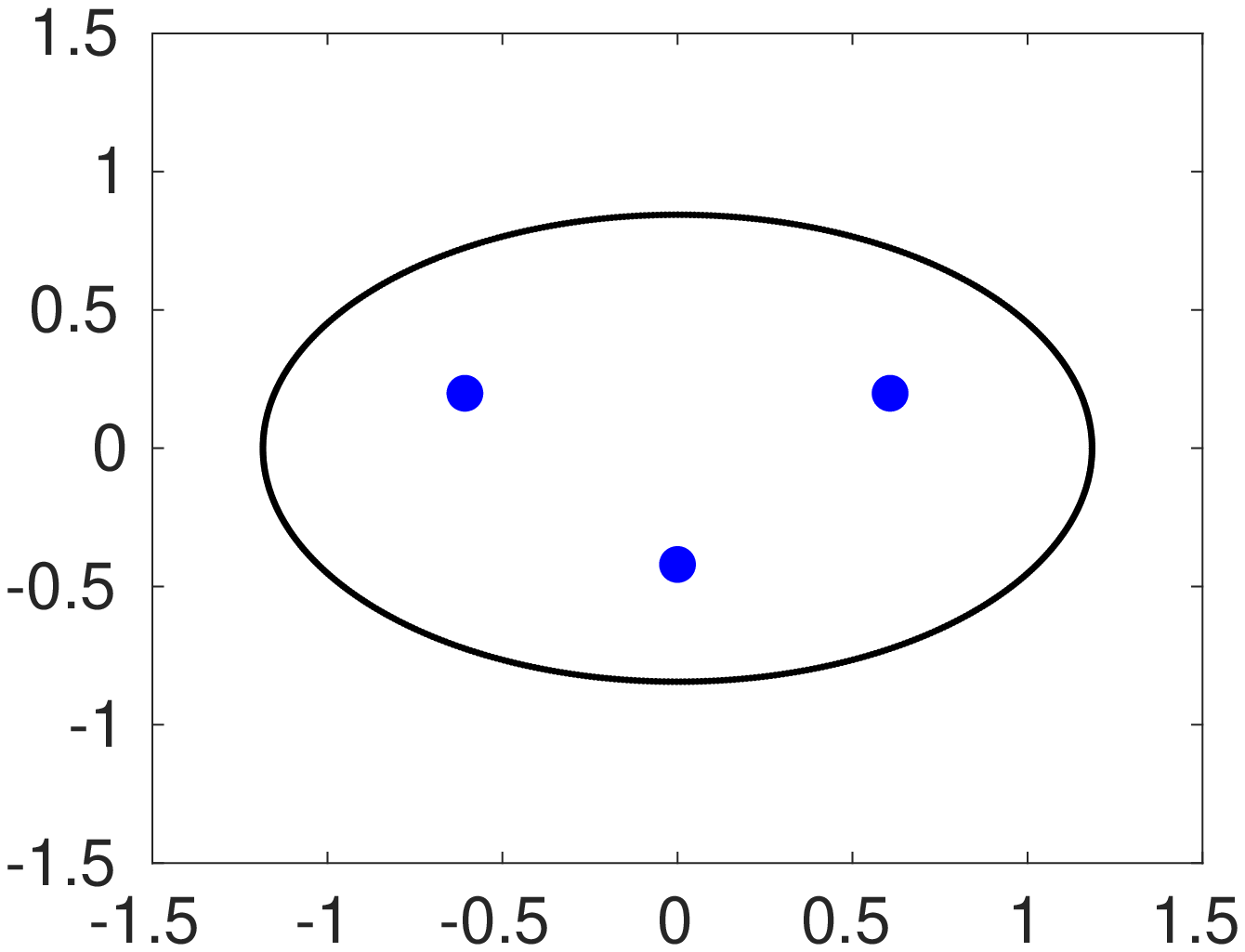}\label{fig:three_snap2}}
{\includegraphics[height=3.5cm,width=0.24\textwidth]{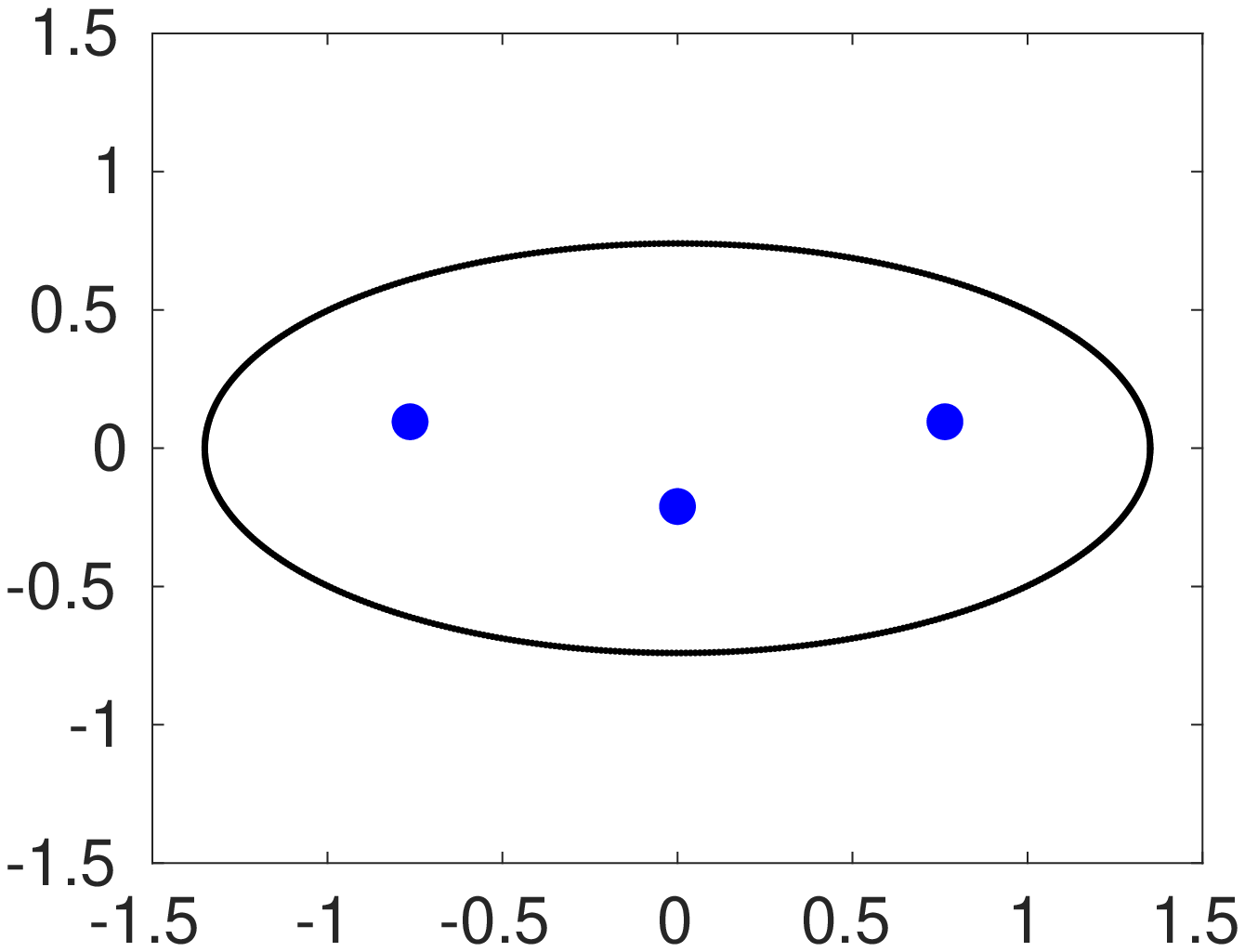}\label{fig:three_snap3}}
{\includegraphics[height=3.5cm,width=0.24\textwidth]{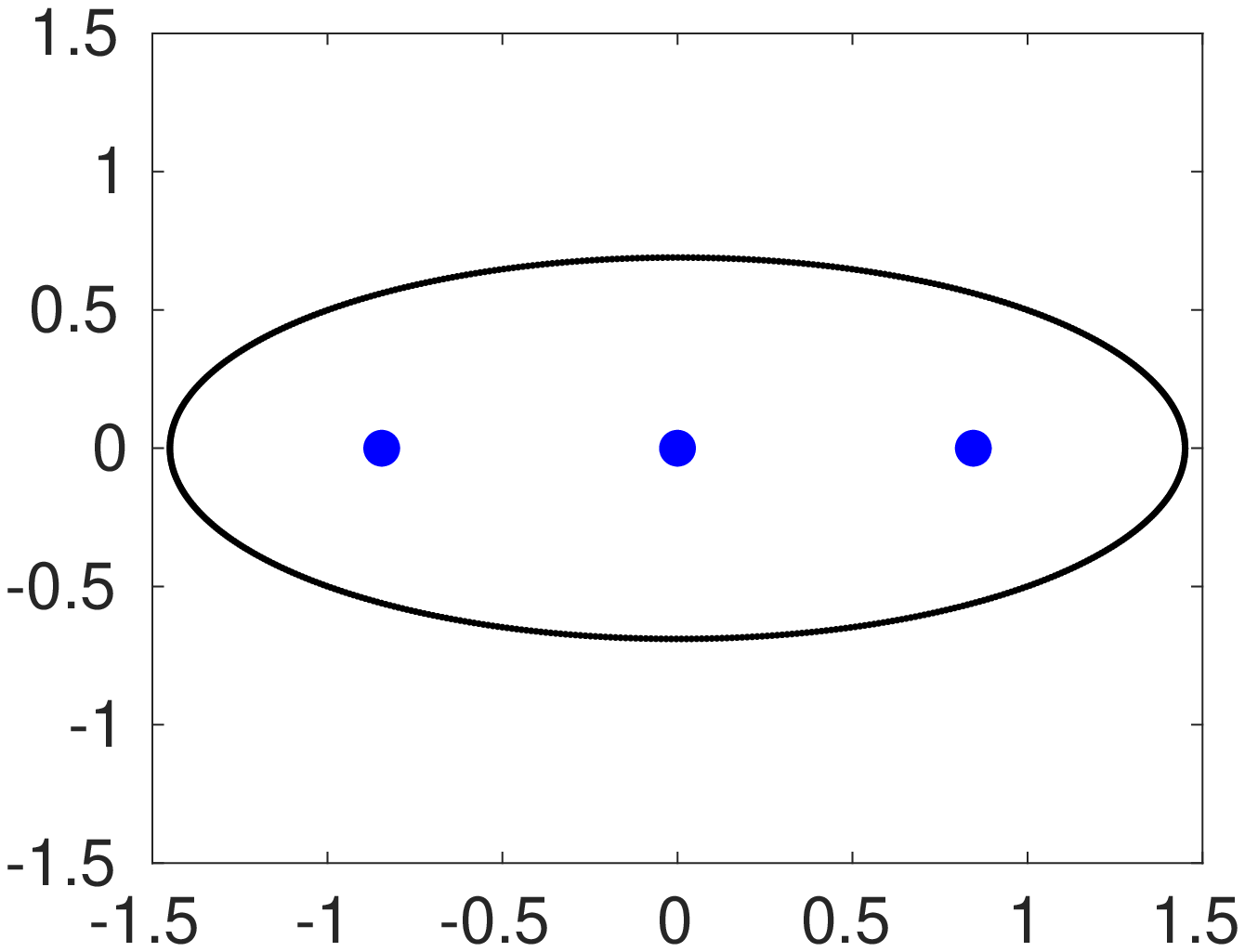}\label{fig:three_snap4}}
\caption{Optimal three-trap configurations for $D=1$ in a deforming
  ellipse of area $\pi$ with semi-major axis $a$ and a common trap
  radius $\eps=0.05$.  Left: $a=1$, $b=1$. Middle Left: $a=1.184$,
  $b\approx 0.845$. Middle Right: $a=1.351$, $b\approx 0.740$. Right:
  $a=1.450$, $b\approx 0.690$. The optimally located traps form an
  isosceles triangle as they deform from a ring pattern in the unit
  disk to a collinear pattern as $a$ increases.}
\end{center}
\label{fig:three_snap}
\end{figure}

\begin{figure}[htbp]
\begin{center}
{\includegraphics[height=4.1cm,width=0.49\textwidth]{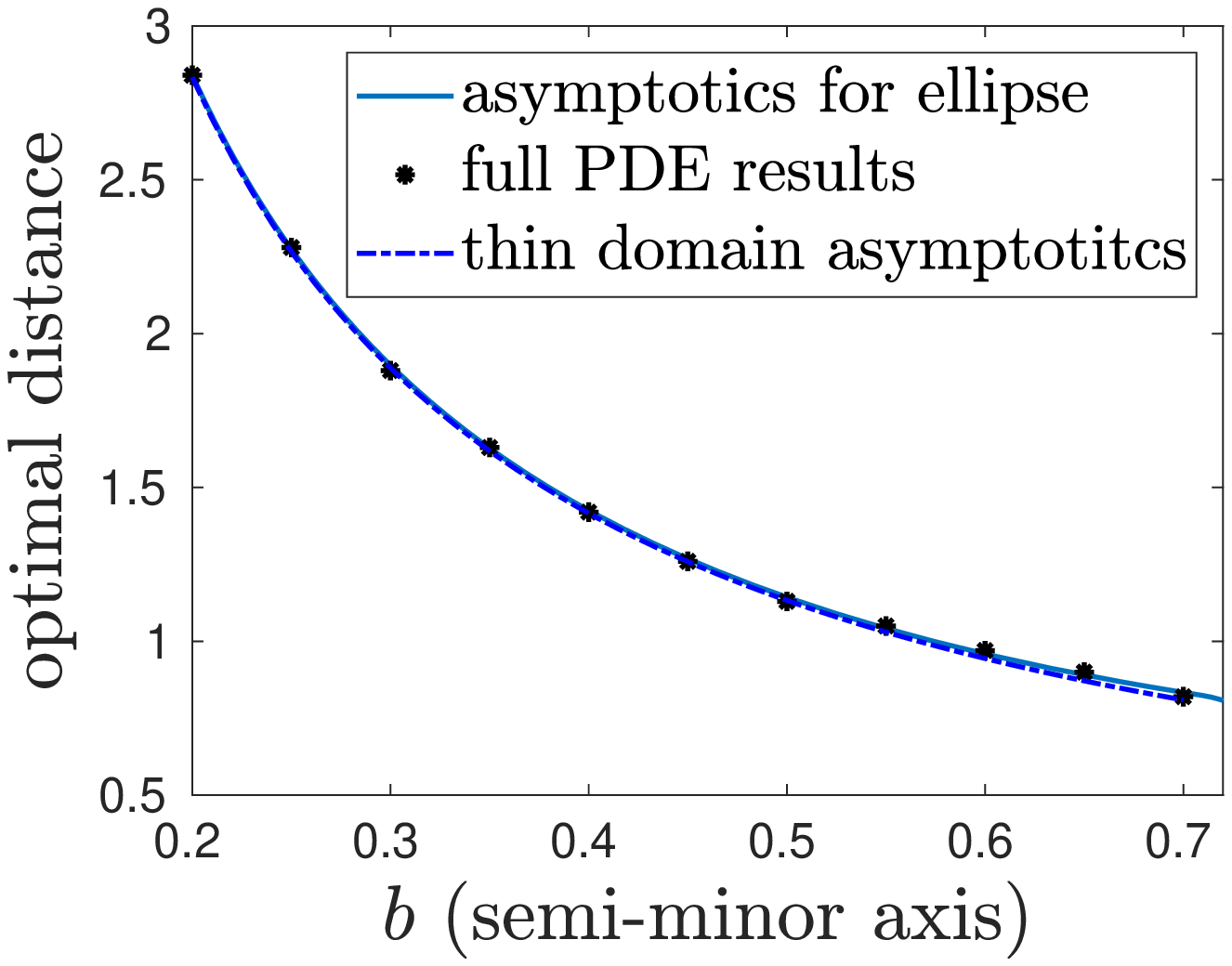}\label{fig:three_ell:x0}}
{\includegraphics[height=4.1cm,width=0.49\textwidth]{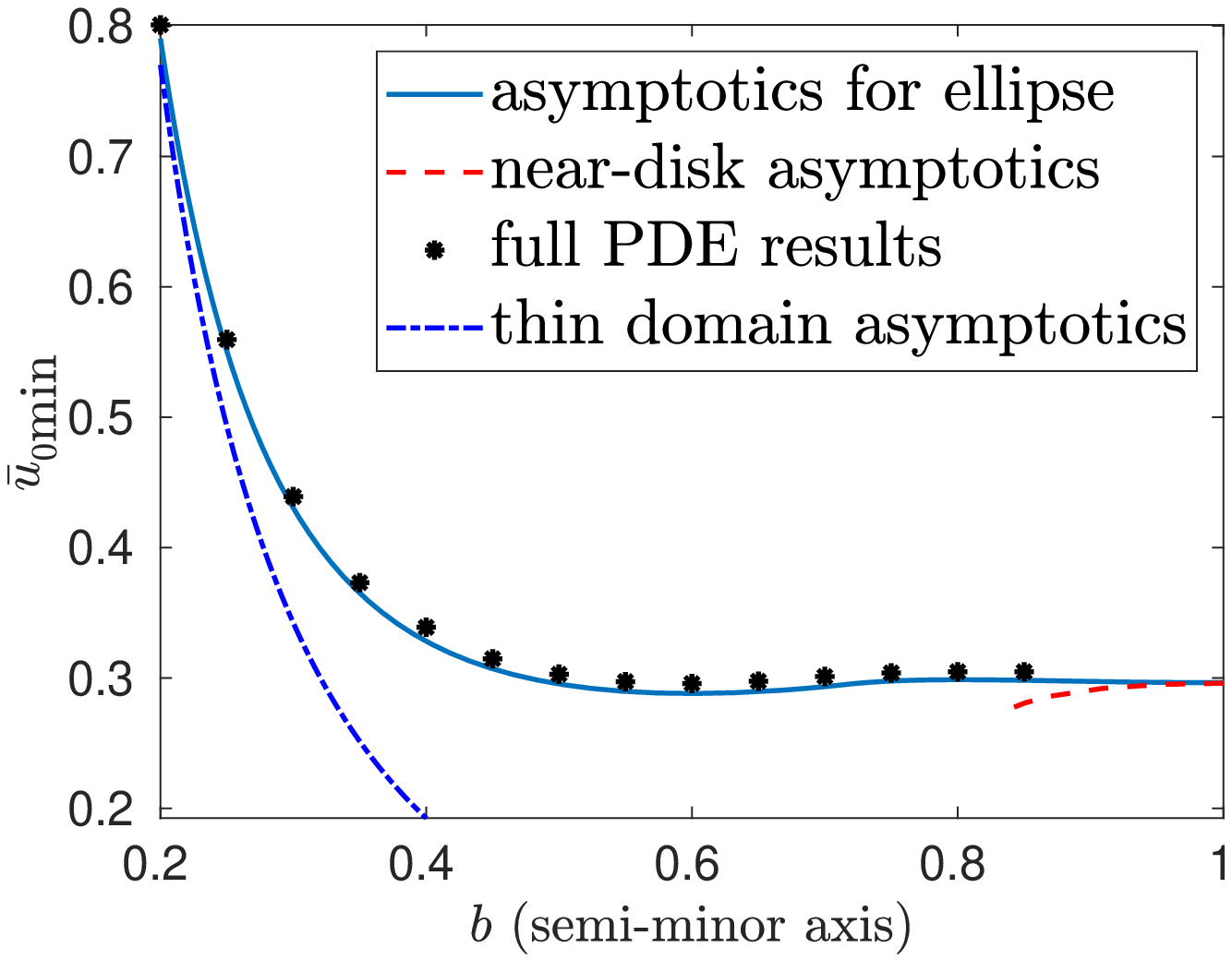}\label{fig:three_ell:u0}}
\caption{Left panel: Optimal distance from the origin for a collinear
  three-trap pattern on the major-axis of an ellipse of area $\pi$ versus the
  semi-minor axis $b$. When $b\le 0.71$ the optimal pattern has a trap
  at the center and a pair of traps symmetrically located on either
  side of the origin. Right panel: optimal average MFPT
  $\overline{u}_{0\mbox{min}}$ versus $b$.  Solid curves: hybrid asymptotic
  theory \eqref{e:u0_bar} for the ellipse coupled to the ODE relaxation
  scheme \eqref{ode:relax} to find the minimum.  Dashed line (red):
  near-disk asymptotics of \eqref{final:avemfpt}. Discrete points:
  Full PDE numerical results computed using the closest point method.
  Dashed-dotted line (blue): thin-domain asymptotics \eqref{thin:m3}.}
\end{center}
\label{fig:three_ellipse}
\end{figure}

Next, we consider the case $m=3$. To clearly illustrate how the
optimal trap configuration changes as the aspect ratio of the ellipse
is varied, we use the hybrid theory to compute the area of the
triangle formed by the three optimally located traps. The results
shown in Fig.~\ref{fig:three_area} are seen to compare favorably with
full PDE results. These results show that that the optimal
traps become colinear on the semi-major axis when $a\ge 1.45$.  In
Fig.~\ref{fig:three_snap} we show snapshots, at certain values of the
semi-major axis, of the optimal trap locations in the ellipse. In the
right panel of Fig.~\ref{fig:three_ellipse}, we show that the optimal
average MFPT from the hybrid theory compares very well with full
numerical PDE results for all $b\leq 1$, and that the thin domain
asymptotics \eqref{thin:m3} provides a good approximation when
$b\leq 0.3$. In the left panel of Fig.~\ref{fig:three_ellipse} we plot
the optimal trap locations on the semi-major axis when the trap
pattern is collinear. We observe that results for the optimal trap
locations from the hybrid theory, the thin domain asymptotics
\eqref{thin:m3}, and the full PDE simulations, essentially coincide on
the full range $0.2<b<0.7$.

\begin{figure}[htbp]
\begin{center}
\includegraphics[height=4.5cm,width=0.60\textwidth]{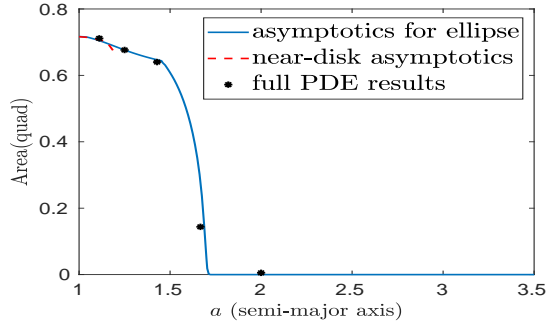}
\caption{Area of the quadrilateral formed by the four optimally
  located traps of a common radius $\eps=0.05$ with $D=1$ in a
  deforming ellipse of area $\pi$ and semi-major axis $a$. The optimal
  traps become collinear as $a$ increases. Solid curve: hybrid
  asymptotic theory \eqref{e:u0_bar} for the ellipse coupled to the
  ODE relaxation scheme \eqref{ode:relax} to find the minimum. Dashed
  line (red): near-disk asymptotics of \eqref{final:avemfpt}. Discrete
  points: full numerical PDE results computed from the closest point
  method.}
\end{center}
\label{fig:four_area}
\end{figure}

\begin{figure}[htbp]
\begin{center}
{\includegraphics[height=3.5cm,width=0.24\textwidth]{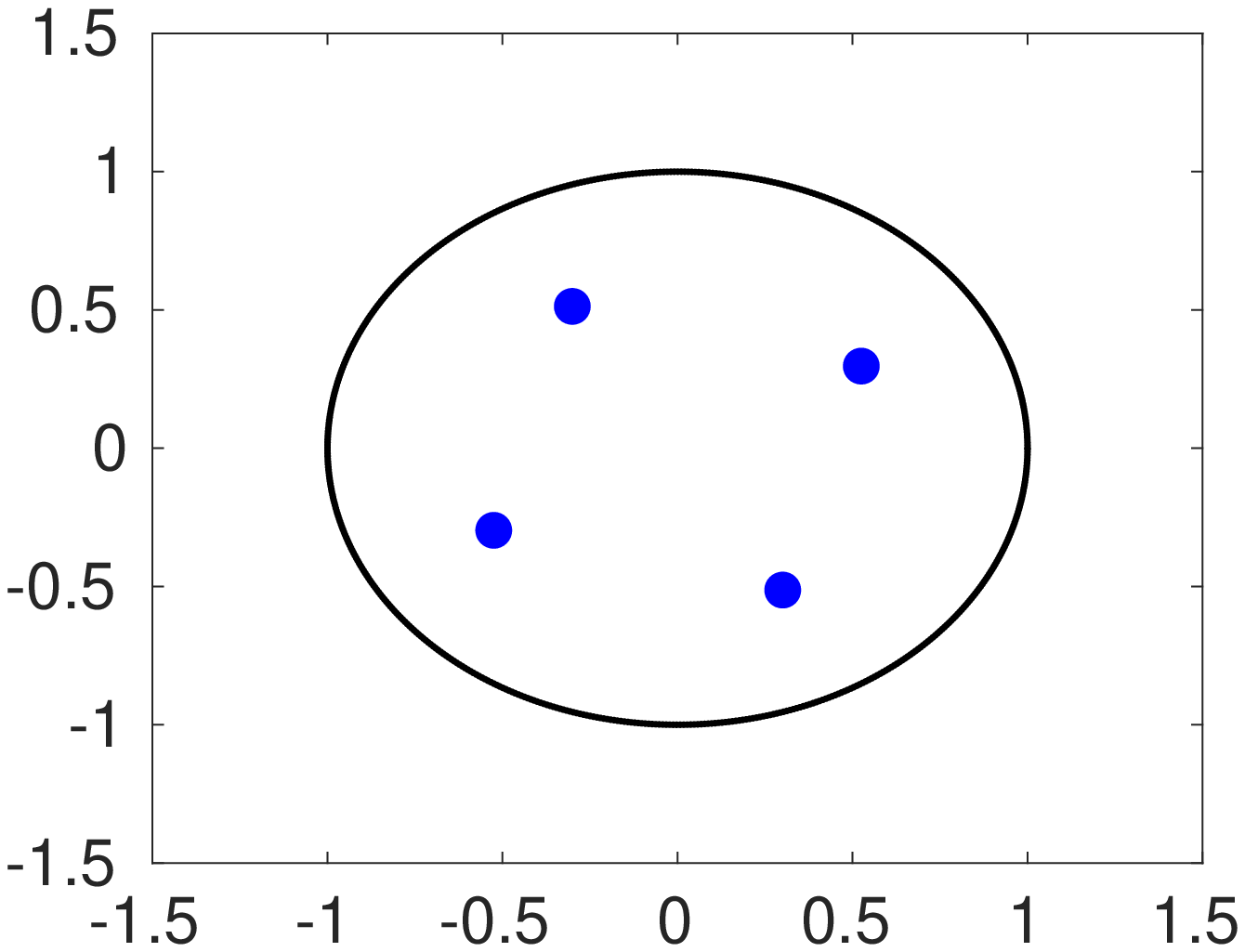}\label{fig:four_snap1}}
{\includegraphics[height=3.5cm,width=0.24\textwidth]{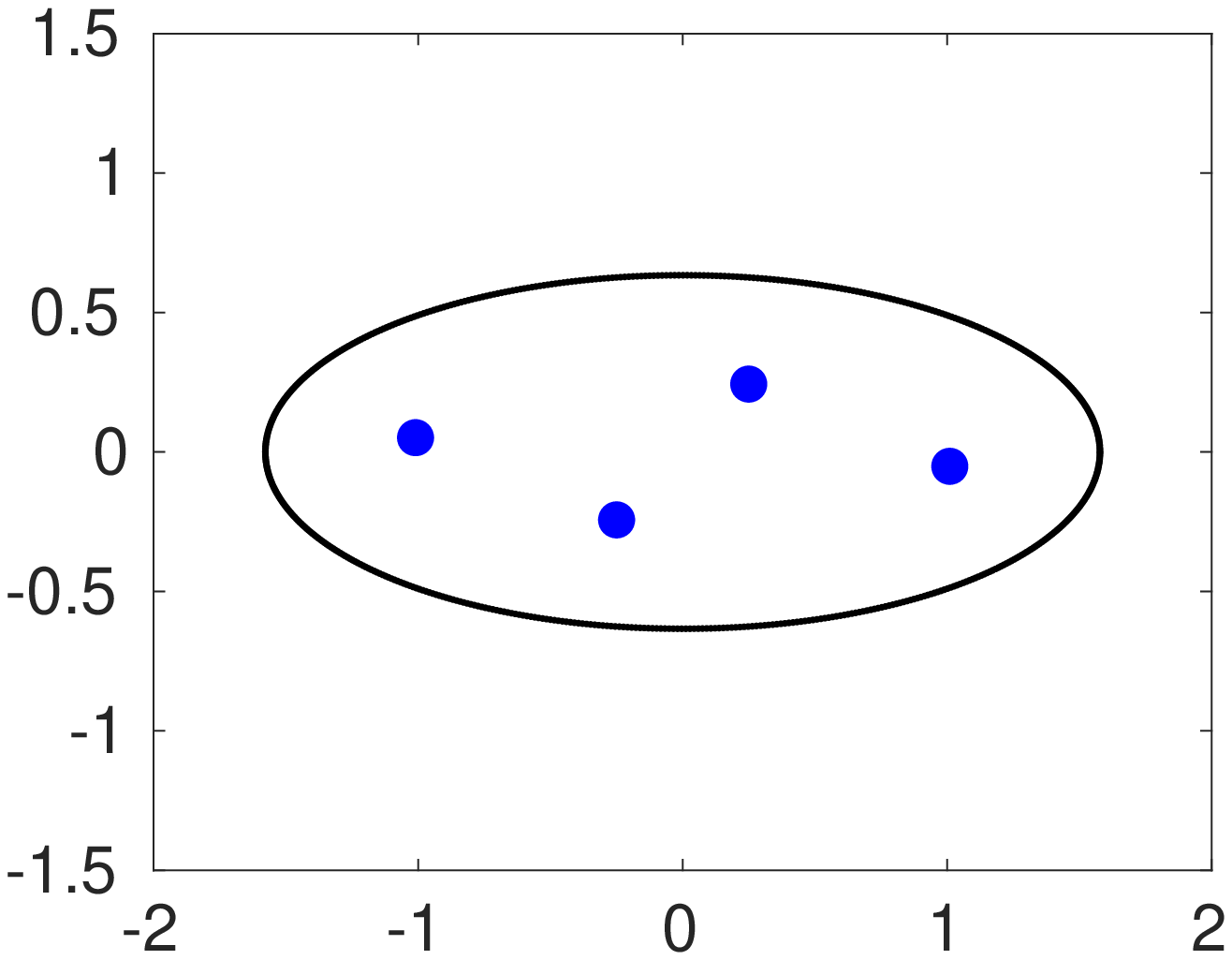}\label{fig:four_snap2}}
{\includegraphics[height=3.5cm,width=0.24\textwidth]{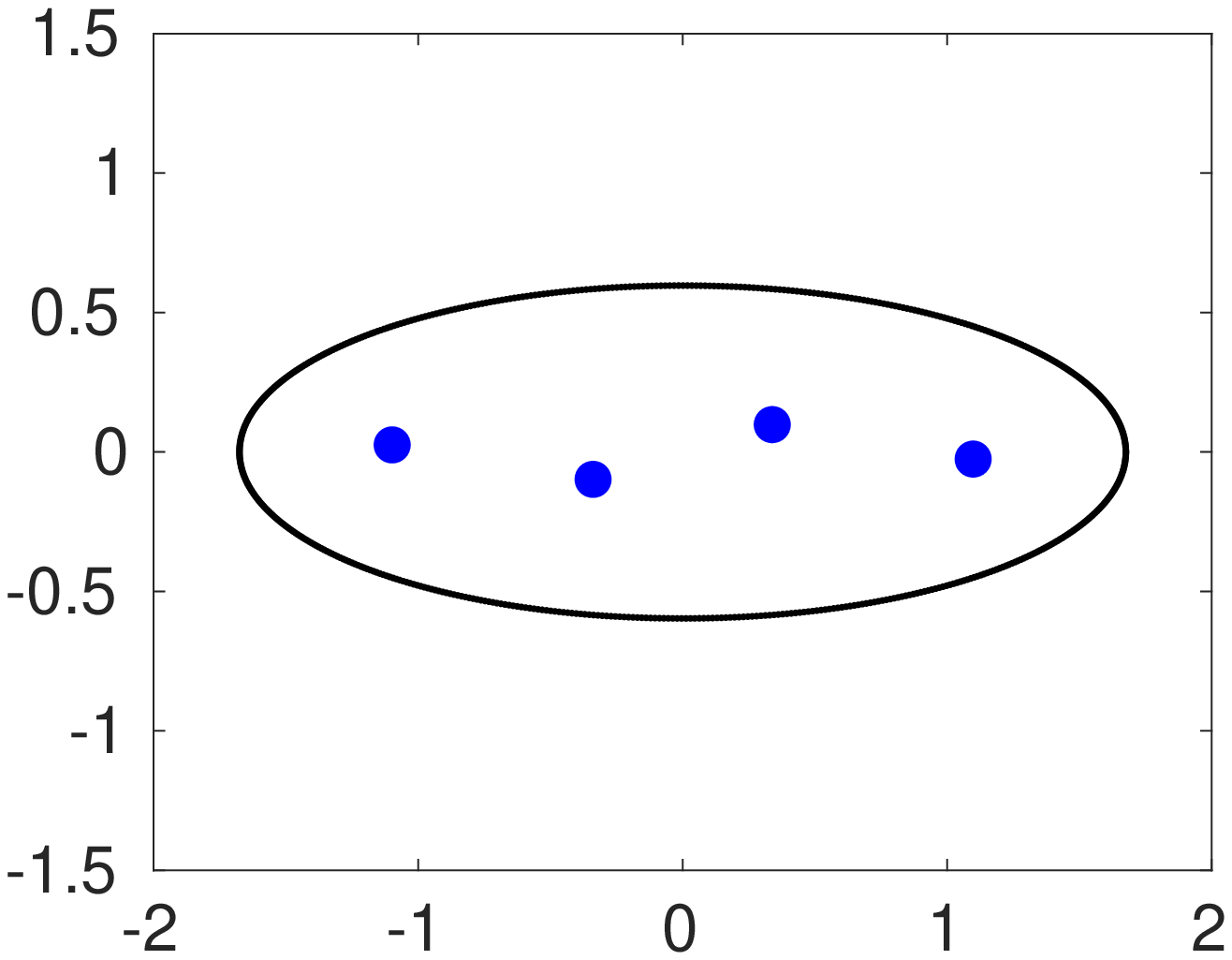}\label{fig:four_snap3}}
{\includegraphics[height=3.5cm,width=0.24\textwidth]{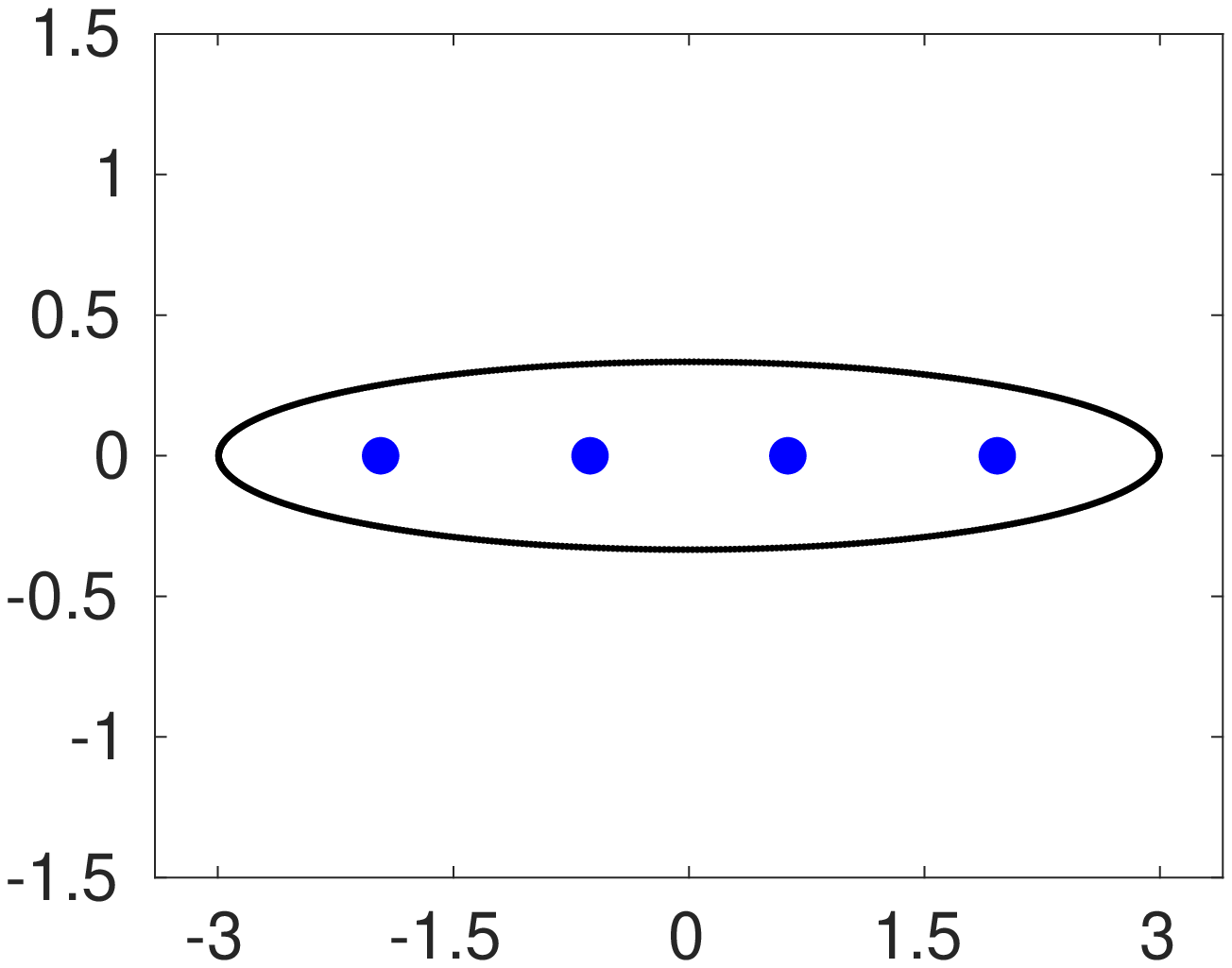}\label{fig:four_snap4}}
\caption{Optimal four-trap configurations for $D=1$ in a deforming
  ellipse of area $\pi$ with semi-major axis $a$ and a common trap
  radius $\eps=0.05$.  Left: $a=1$, $b=1$. Middle Left: $a=1.577$,
  $b\approx 0.634$. Middle Right: $a=1.675$, $b\approx 0.597$. Right:
  $a=3.0$, $b\approx 0.333$. The optimally located traps form a rectangle,
  followed by a parallelogram, as they deform from a ring pattern in
  the unit disk to a collinear pattern as $a$ increases.}
\end{center}
\label{fig:four_snap}
\end{figure}

\begin{figure}[htbp]
\begin{center}
{\includegraphics[height=4.1cm,width=0.49\textwidth]{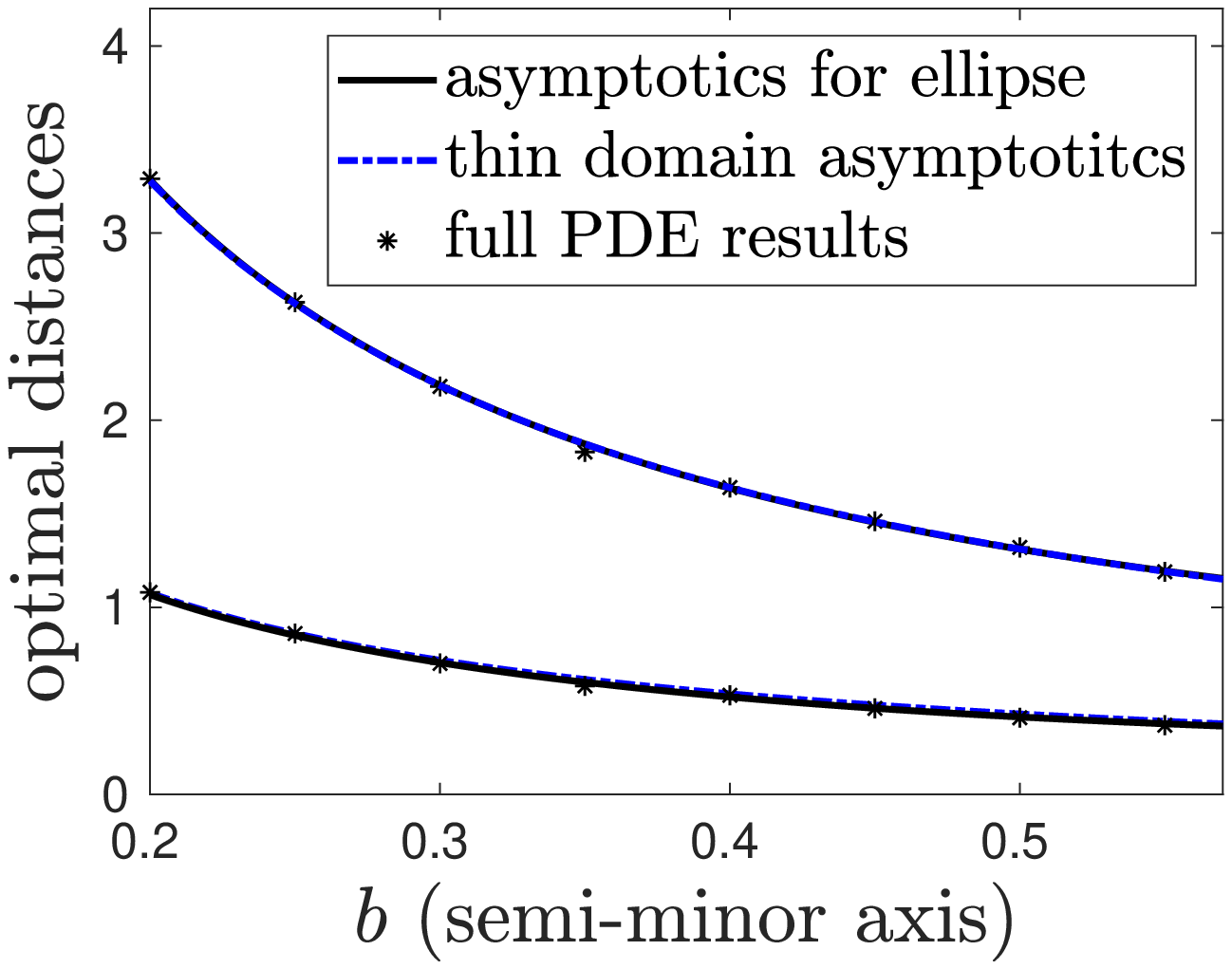}\label{fig:four_ell:x0}}
{\includegraphics[height=4.1cm,width=0.49\textwidth]{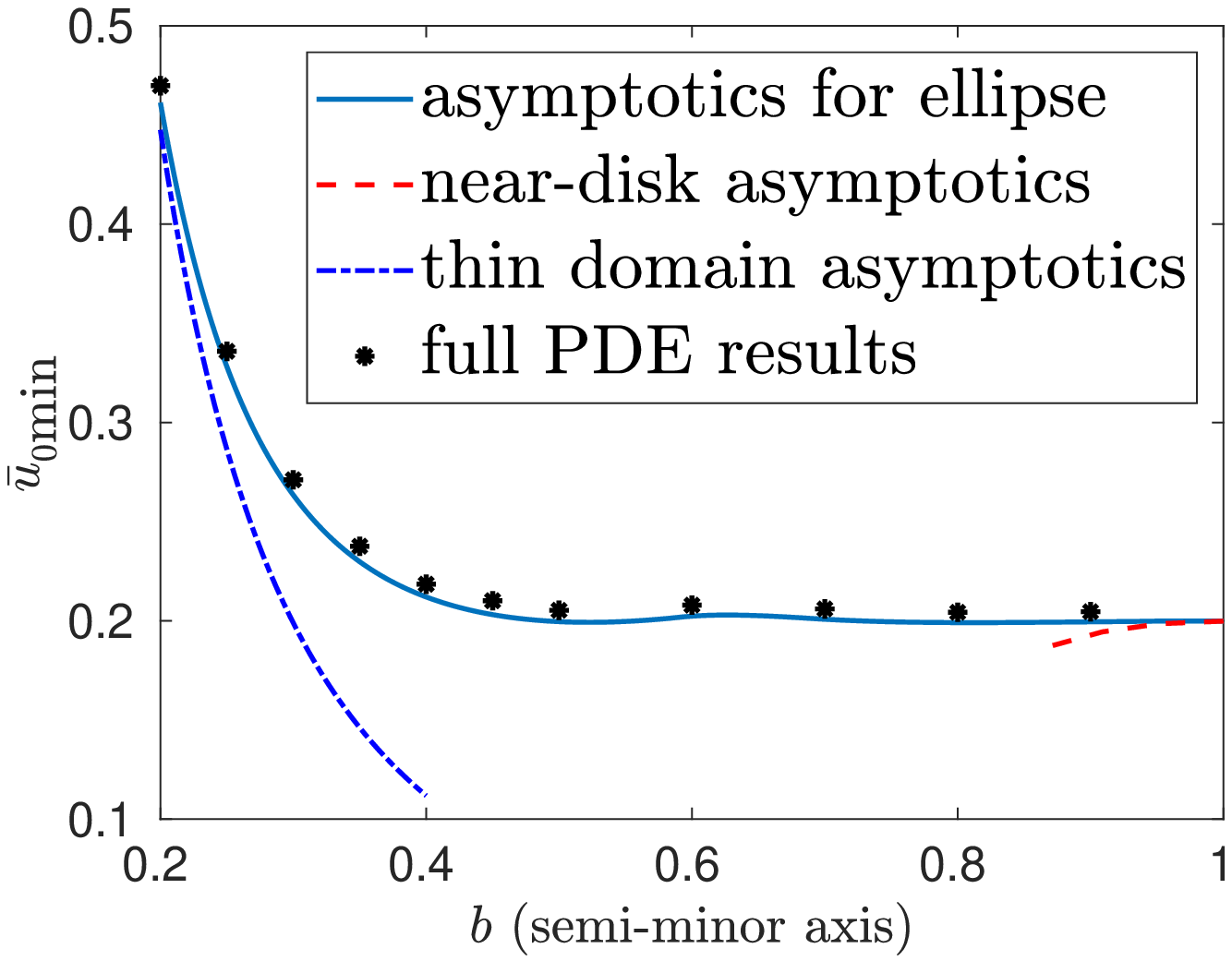}\label{fig:four_ell:u0}}
\caption{Left panel: Optimal distances from the origin for a collinear
  four-trap pattern on the major-axis of an ellipse of area $\pi$ and
  semi-minor axis $b$. When $b\le 0.57$ the optimal pattern has two
  pairs of traps symmetrically located on either side of the
  origin. Right panel: the optimal average MFPT
  $\overline{u}_{0\mbox{min}}$ versus $b$.  Solid curves: hybrid asymptotic
  theory \eqref{e:u0_bar} for the ellipse coupled to the ODE
  relaxation scheme \eqref{ode:relax} to find the minimum. Dashed line
  (red): near-disk asymptotics of \eqref{final:avemfpt}.  Discrete
  points: full numerical PDE results computed from the closest point
  method.  Dashed-dotted line (blue): thin-domain asymptotics
  \eqref{thin:m4}.}
\end{center}
\label{fig:four_ellipse}
\end{figure}

For the case of four traps, where $m=4$, in Fig.~\ref{fig:four_area}
we use the hybrid theory to plot the area of the quadrilateral formed
by the four optimally located traps versus the semi-major axis $a>1$.
The full PDE results, also shown in Fig.~\ref{fig:four_area}, compare
well with the hybrid results. This figure shows that as the aspect
ratio of the ellipse increases the traps eventually become collinear on
the semi-major axis when $a\geq 1.7$. This feature is further
illustrated by the snapshots of the optimal trap locations shown in
Fig.~\ref{fig:four_snap} at representative values of $a$. In the
right panel of Fig.~\ref{fig:four_ellipse}, we show that the hybrid and
full numerical PDE results for the optimal average MFPT agree very
closely for all $b\leq 1$, but that the thin-domain asymptotic result
\eqref{thin:m4} agrees well only when $b\leq 0.25$. However, as
similar to the three-trap case, on the range of $b$ where the trap
pattern is collinear, in the left panel of Fig.~\ref{fig:four_ellipse}
we show that the hybrid theory, the full PDE simulations, and the
thin-domain asymptotics all provide essentially indistinguishable
predictions for the optimal trap locations on the semi-major axis.

\begin{figure}[htbp]
\begin{center}
{\includegraphics[height=3.5cm,width=0.32\textwidth]{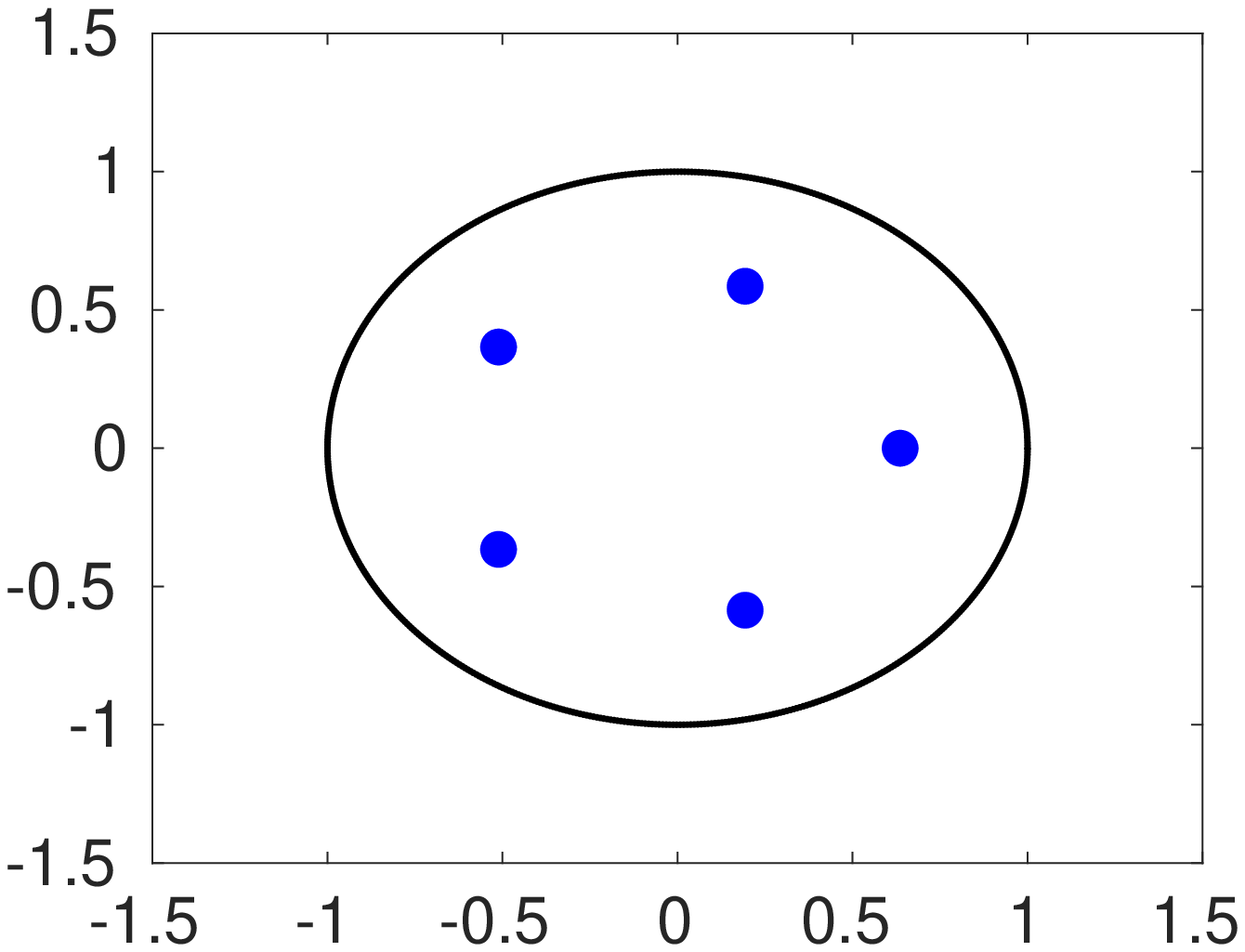}\label{fig:five_snap1}}
{\includegraphics[height=3.5cm,width=0.32\textwidth]{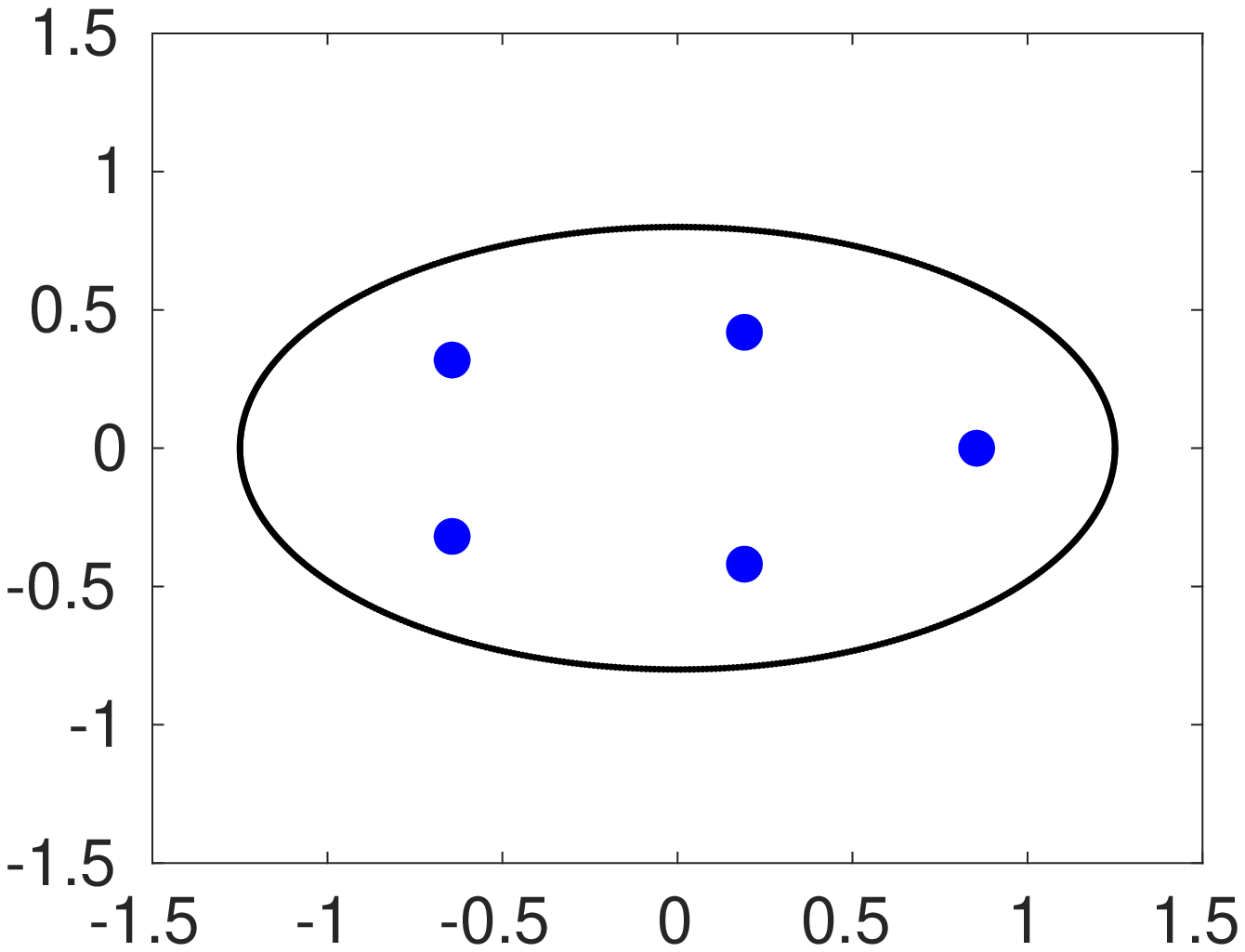}\label{fig:five_snap2}}
{\includegraphics[height=3.5cm,width=0.32\textwidth]{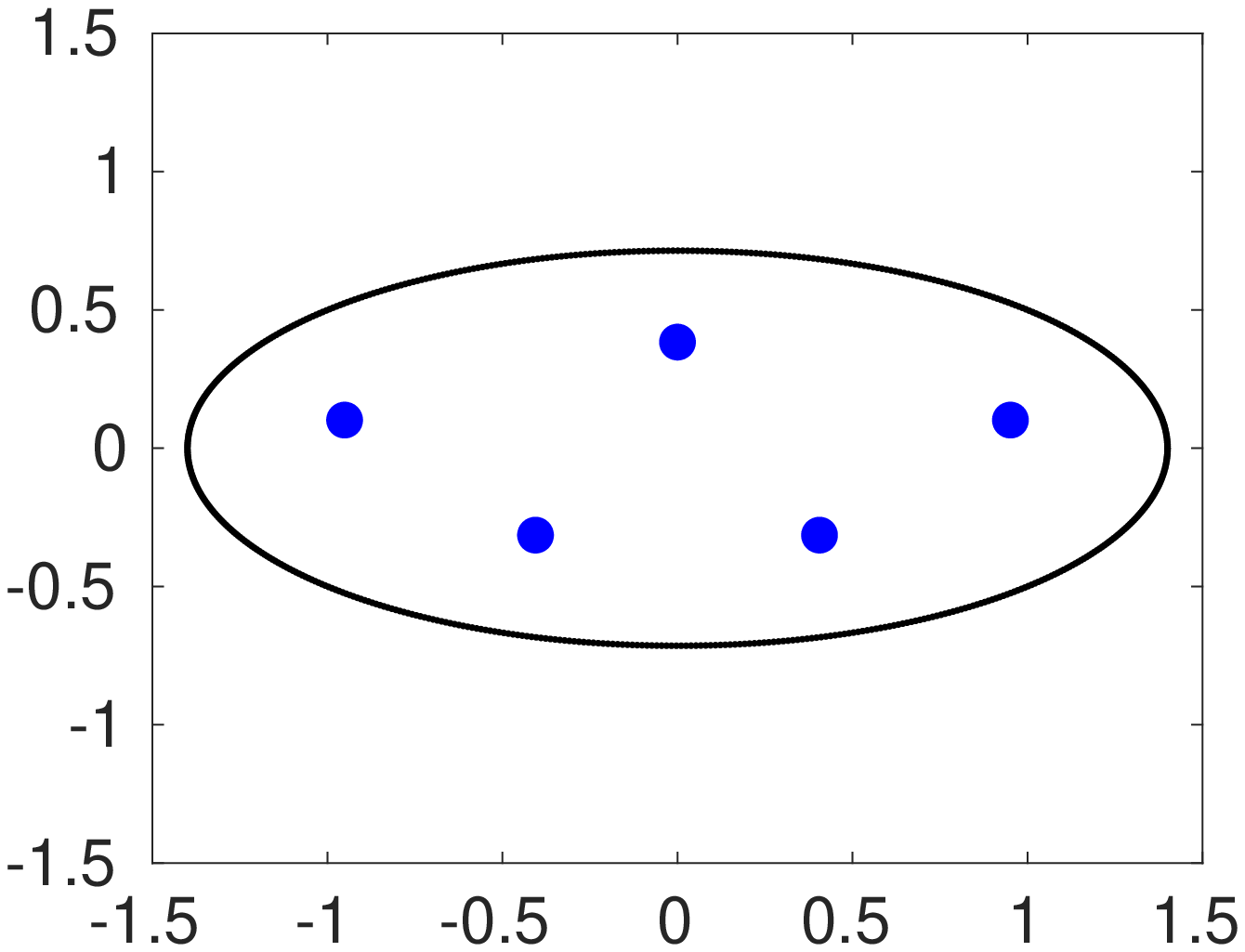}\label{fig:five_snap3}}
{\includegraphics[height=3.5cm,width=0.32\textwidth]{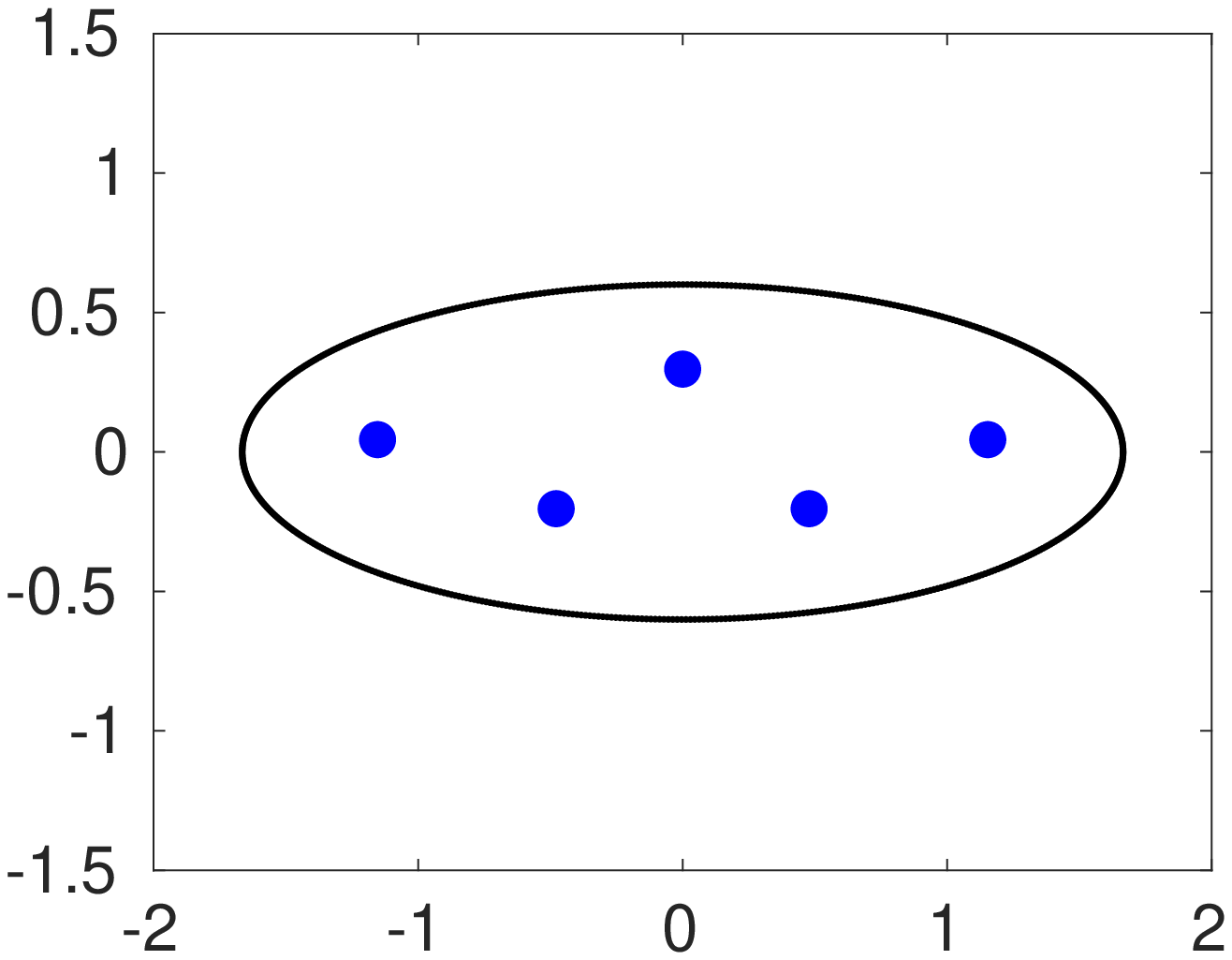}\label{fig:five_snap4}}
{\includegraphics[height=3.5cm,width=0.32\textwidth]{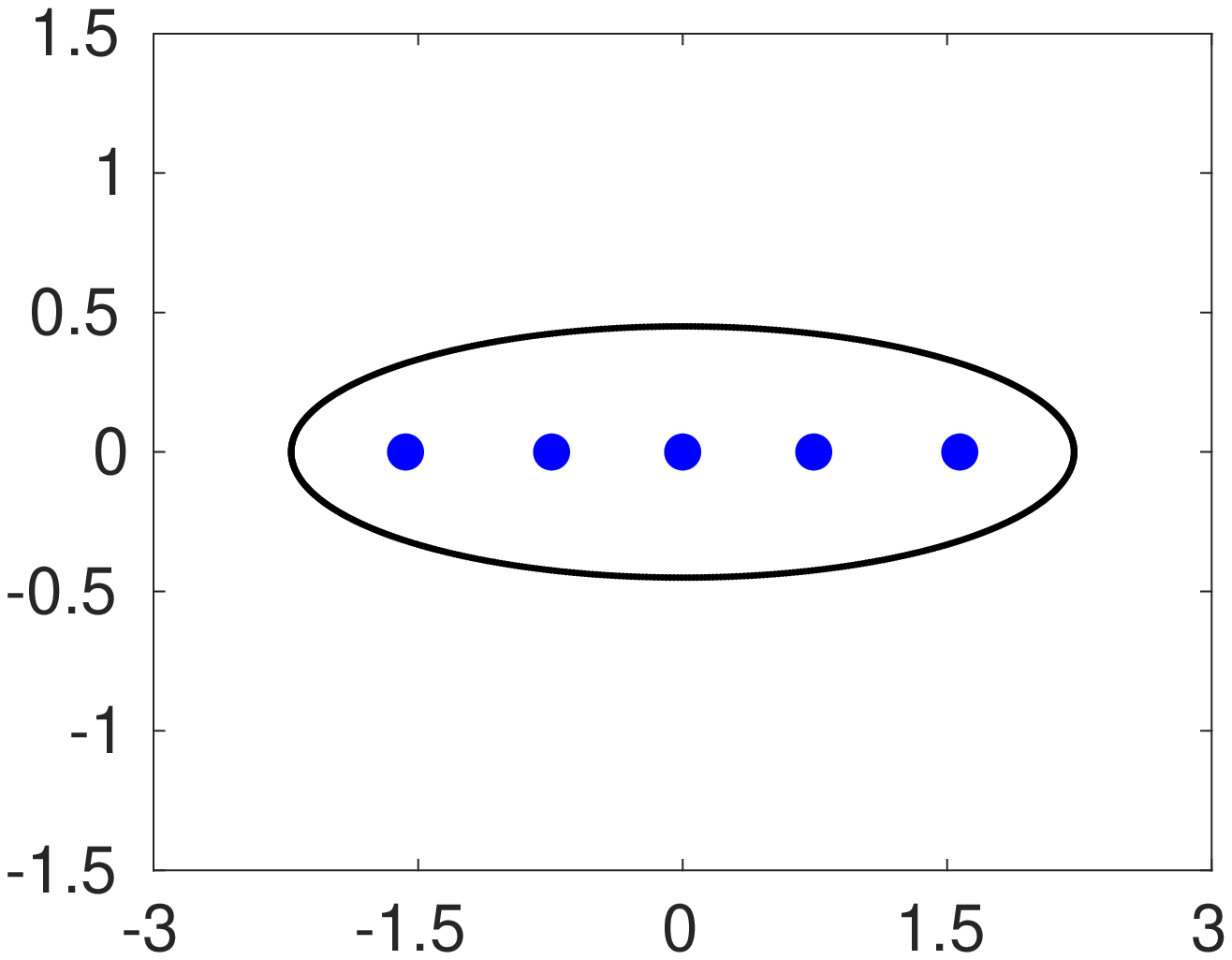}\label{fig:five_snap5}}
{\includegraphics[height=3.5cm,width=0.32\textwidth]{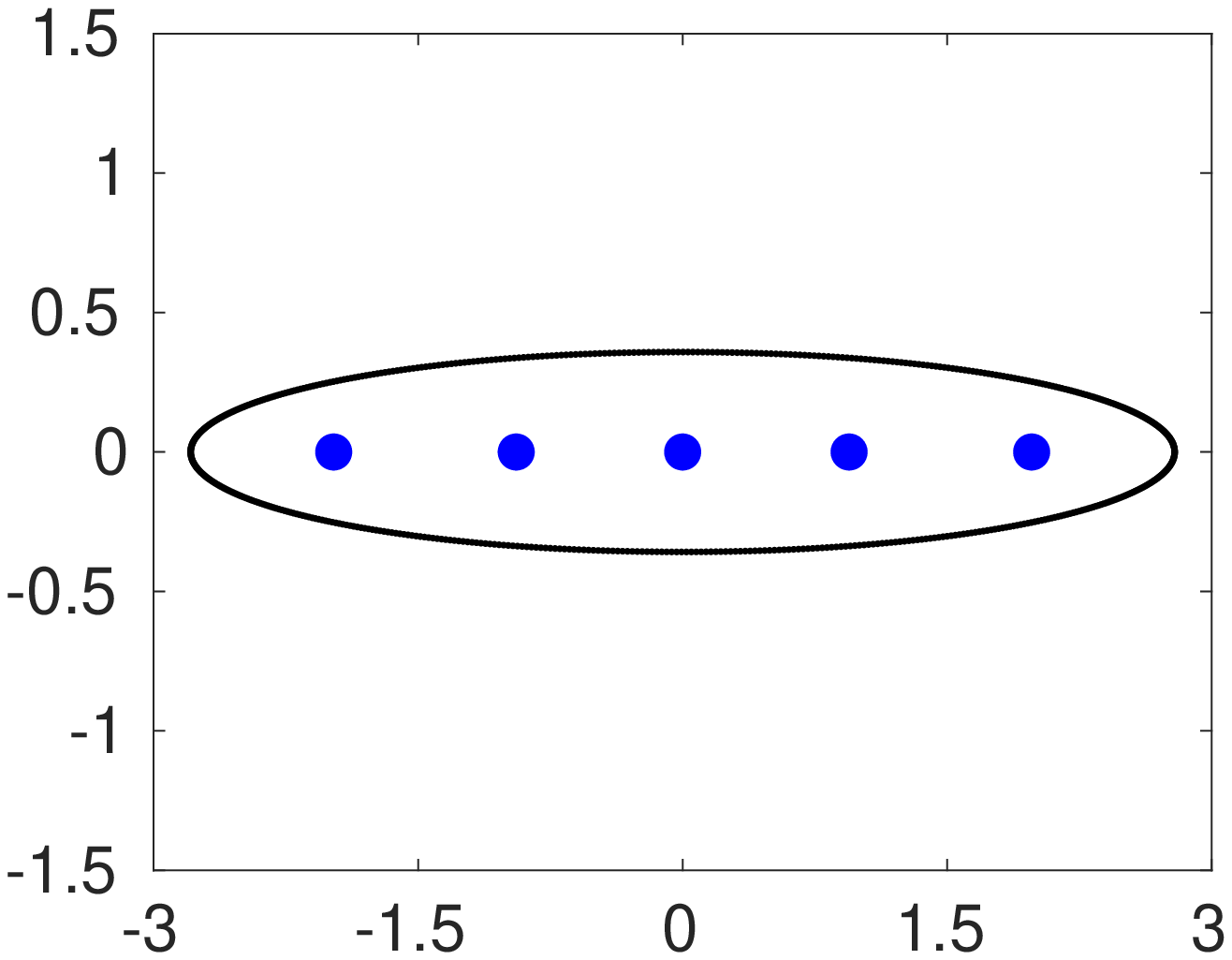}\label{fig:five_snap6}}
\caption{Optimal five-trap configurations for $D=1$ in a deforming
  ellipse of area $\pi$ with semi-major axis $a$ and a common trap
  radius $\eps=0.05$.  Top left: $a=1$, $b=1$. Top middle: $a=1.25$,
  $b=0.8$. Top right: $a=1.4$, $b\approx 0.690$.  Bottom left:
  $a=1.665$, $b\approx 0.601$. Bottom middle: $a=2.22$,
  $b\approx 0.450$. Bottom right: $a=2.79$, $b\approx 0.358$.  The
  optimal traps become collinear as $a$ increases and the edge-most
  traps become closer to the corner of the domain as $a$ increases.}
\end{center}
\label{fig:five_snap}
\end{figure}

\begin{figure}[htbp]
\begin{center}
{\includegraphics[height=4.1cm,width=0.49\textwidth]{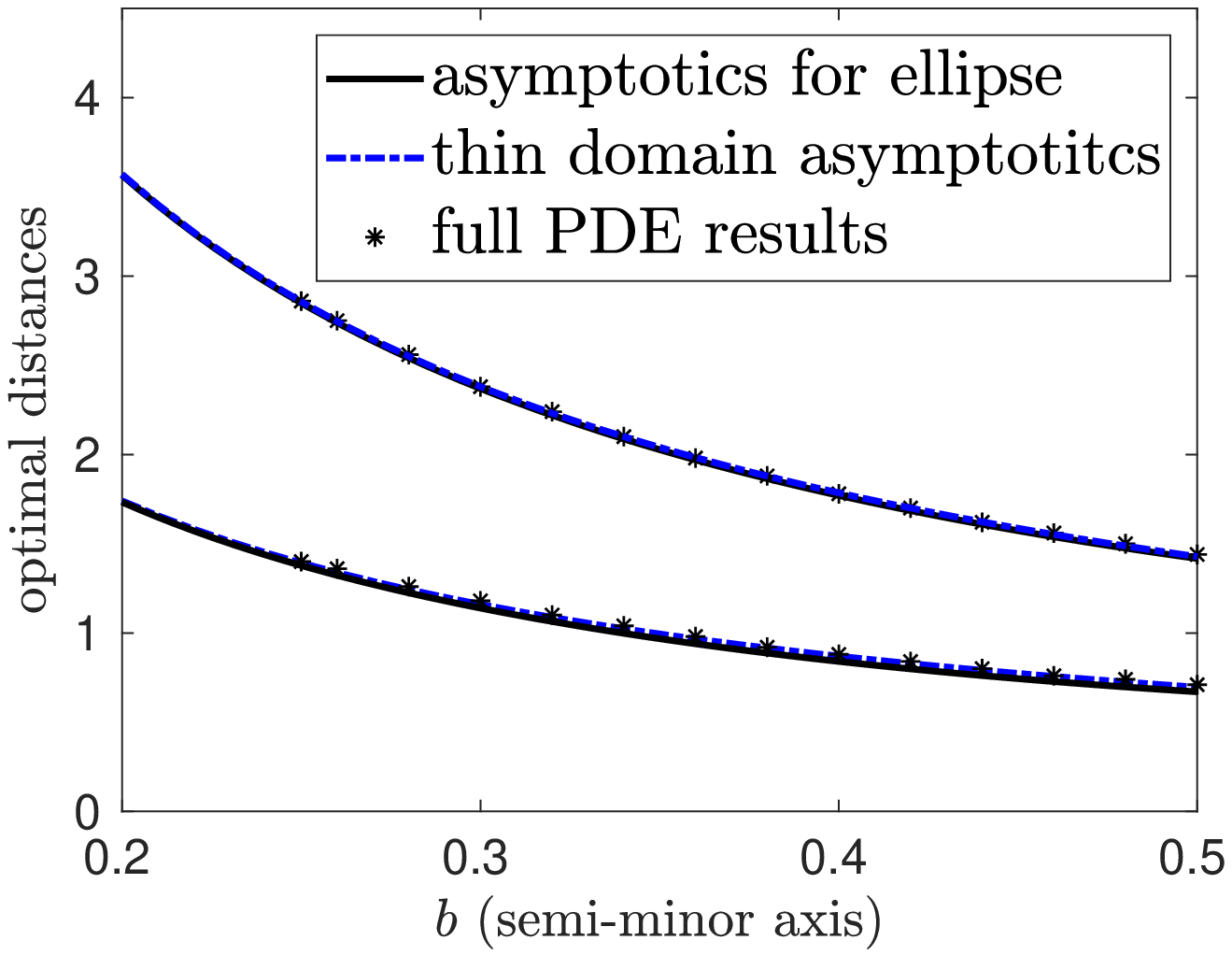}\label{fig:five_ell:x0}}
{\includegraphics[height=4.1cm,width=0.49\textwidth]{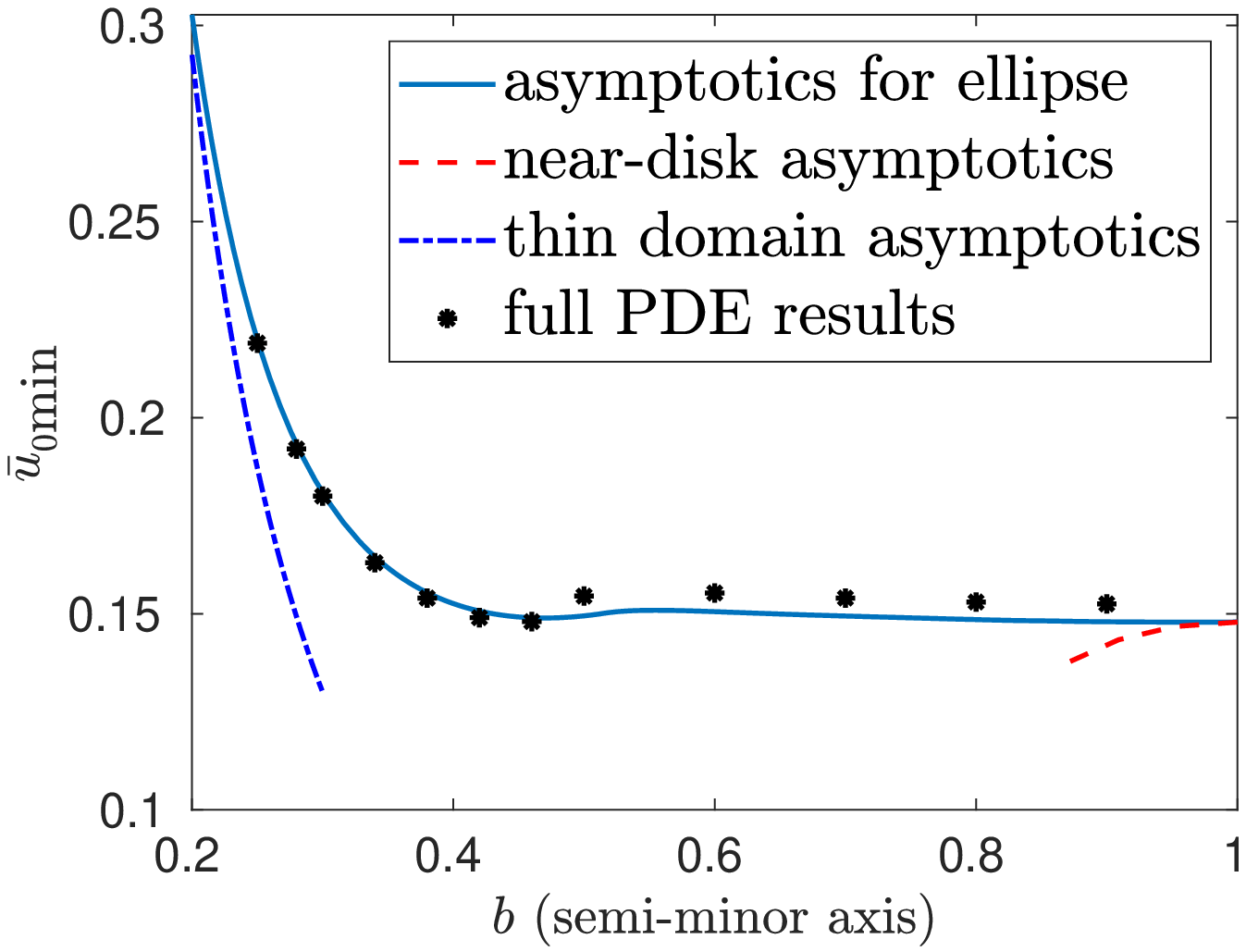}\label{fig:five_ell:u0}}
\caption{Left panel: Optimal distances from the origin for a collinear
  five-trap pattern on the major-axis of an ellipse of area $\pi$ and
  semi-minor axis $b$. When $b\le 0.51$ the optimal pattern has a trap
  at the center and two pairs of traps symmetrically located on either
  side of the origin. Right panel: The optimal average MFPT
  $\overline{u}_{0\mbox{min}}$ versus $b$.  Solid curves: hybrid asymptotic
  theory for the ellipse \eqref{e:u0_bar} coupled to the ODE
  relaxation scheme \eqref{ode:relax} to find the minimum.
  Dashed line (red): near-disk asymptotics of \eqref{final:avemfpt}.
  Discrete points: full numerical PDE results computed from the
  closest point method.  Dashed-dotted line (blue): thin-domain
  asymptotics \eqref{thin:m5}.}
\end{center}
\label{fig:five_ellipse}
\end{figure}

Finally, we show similar results for the case of five traps. In
Fig.~\ref{fig:five_snap}, we plot the optimal trap locations in the
ellipse as the semi-major axis of the ellipse is varied. This plot
shows that the optimal pattern becomes collinear when (roughly)
$a\geq 2$. In the right panel of Fig.~\ref{fig:five_ellipse}, we show a
close agreement between the hybrid and full numerical PDE results for
the optimal average MFPT.  However, as seen in
Fig.~\ref{fig:five_ellipse}, the thin-domain asymptotic result
\eqref{thin:m5} accurately predicts the optimal MFPT only for rather
small $b$. As for the four-trap case, in the left panel of
Fig.~\ref{fig:five_ellipse} we show that the hybrid theory, the full
PDE simulations, and the thin-domain asymptotics all yield similar
predictions for the optimal trap locations on the semi-major axis.

\subsection{Thin-Domain Asymptotics}\label{sec:thin}

For a long and thin ellipse, where $b=\delta\ll 1$ and
$a={1/\delta}$ but with $|\Omega|=\pi$, we now derive simple
approximations for the optimal trap locations and the optimal average
MFPT using an approach based on thin-domain asymptotics. For $m=2$ the
optimal trap locations are on the semi-major axis
(cf.~Fig.~\ref{fig:two_ellipse}), while for $3\leq m\leq 5$ the
optimal trap locations become collinear when the semi-minor axis $b$
decreases below a threshold (see Fig.~\ref{fig:three_snap},
Fig.~\ref{fig:four_snap}, and Fig.~\ref{fig:five_snap}).

As derived in Appendix \ref{app:thin}, the leading-order approximation for
the MFPT $u$ satisfying \eqref{Ellip_Model} in a thin elliptical with
$b=\delta\ll 1$ is
\begin{equation}
  u(x,y)\sim \delta^{-2}U_0(\delta x) + {\mathcal O}(\delta^{-1}) \,,
\end{equation}
where the one-dimensional profile $U_0(X)$, with $x={X/\delta}$,
satisfies the ODE
\begin{equation}\label{sec:long_u0}
  \left[\sqrt{1-X^2} \, U_0^{\prime} \right]^{\prime} = -\frac{\sqrt{1-X^2}}{D}\,,
  \quad \mbox{on} \quad |X|\leq 1 \,,
\end{equation}
with $U_0$ and $U_0^{\prime}$ bounded as $X\to \pm 1$. In terms of
$U_0(X)$, the average MFPT for the thin ellipse is estimated for
$\delta\ll 1$ as
\begin{equation}\label{thin:ave}
  \overline{u}_0 \sim \frac{1}{\pi} \int_{-1/\delta}^{1/\delta}
  \int_{-\delta \sqrt{1-\delta^2 x^2}}^{\delta \sqrt{1- \delta^2 x^2}}  u \, dx dy \sim
  \frac{4}{\pi \delta^2} \int_{0}^{1} \sqrt{1-X^2} \, U_0(X) \, dX \,.
\end{equation}

In the thin domain limit, the circular traps of a common radius $\eps$
centered on the semi-major axis are approximated by zero point
constraints for $U_0$ at locations on the interval $|X|\leq 1$. In
this way, \eqref{sec:long_u0} becomes a multi-point BVP problem, whose
solution depends on the locations of the zero point
constraints. Optimal values for the location of these constraints are
obtained by minimizing the 1-D integral in \eqref{thin:ave}
approximating $\overline{u}_0$. We now apply this approach for
$m=2,\ldots,5$ collinear traps.

For $m=2$ traps centered at $X=\pm d$ with $0<d<1$, the multi-point BVP
for $U_0(X)$ on $0<X<1$ satisfies
\begin{equation}\label{thin_m2:U0}
  \left[\sqrt{1-X^2} \, U_0^{\prime} \right]^{\prime} = -\frac{\sqrt{1-X^2}}{D}\,,
  \quad 0<X<1 \,; \qquad U_0^{\prime}(0)=0 \,, \quad U_0(d)=0 \,,
\end{equation}
with $U_0$ and $U_0^{\prime}$ bounded as $X\to \pm 1$. A particular
solution for \eqref{thin_m2:U0} is
$U_{0p}=-{[(\sin^{-1}(X))^2+X^2]/(4D)}$, while the homogeneous
solution is $U_{0H}=c_1 \sin^{-1}(X) + c_2$. By combining these
solutions, we readily calculate that
\begin{subequations}
\begin{equation}\label{thin_m2:u0_solve_1}
U_0(X) = \begin{cases}
  -\frac{1}{4D} \left[ \left(\sin^{-1}{X}\right)^2 + X^2 - \pi
    \sin^{-1} X + c_2 \right]\,, \quad d\leq X \leq 1\,, \\
  -\frac{1}{4D} \left[ \left(\sin^{-1}{X}\right)^2 + X^2 + c_1
   \right]\,, \quad 0\leq X \leq d \,,
\end{cases}
\end{equation}
where $c_1$ and $c_2$ are given by
\begin{equation}\label{thin_m2:u0_solve_2}
  c_1 =  - d^2 - \left( \sin^{-1}{d}\right)^2 \,, \qquad
  c_2 = - d^2 + \pi \sin^{-1}{d} - \left(\sin^{-1}{d}\right)^2 \,.
\end{equation}
\end{subequations}
Upon substituting \eqref{thin_m2:u0_solve_1} into \eqref{thin:ave}, we
obtain that
\begin{subequations}
\begin{equation}\label{thin_m2:ave_ell}
  \overline{u}_0 \sim -\frac{1}{\pi D\delta^2} \left[ J_0 + {\mathcal H}(d)
  \right] \,, 
\end{equation}
where the two integrals $J_0$ and ${\mathcal H}(d)$ are given by
\begin{align}
  J_0 &\equiv \int_{0}^{1} F(X) \left[ \left( \sin^{-1}{X}\right)^2 +
    X^2 - \pi\sin^{-1}(X) \right] \, dX  \approx -0.703 \,, \label{thin_ave_J}\\
  {\mathcal H}(d) &\equiv \pi \int_{0}^{d} F(X) \sin^{-1}(X) \, dX
                    + c_2 \int_{d}^{1} F(X) \, dX + c_1\int_{0}^{d}
                    F(X) \, dX \,, \label{thin_ave_H}
\end{align}
\end{subequations}
where $F(X)=\sqrt{1-X^2}$.  By performing a few quadratures, and using
\eqref{thin_m2:u0_solve_2} for $c_1$ and $c_2$, we obtain an explicit
expression for ${\mathcal H}(d)$:
\begin{equation}\label{H:m2}
  {\mathcal H}(d) = -\frac{\pi}{2} \left[\sin^{-1}(d)\right]^2 +
  \frac{\pi^2}{4} \sin^{-1}(d)  - \frac{\pi d^2}{2} \,.
\end{equation}

To estimate the optimal average MFPT we simply maximize
${\mathcal H}(d)$ in \eqref{H:m2} on $0<d<1$. We compute that
$d_{\textrm{opt}}\approx 0.406$, and correspondingly
$\overline{u}_{0\mbox{min}}=-\left(\pi D\delta^2\right)^{-1} \left[ J_0 +
  {\mathcal H}(d_{\textrm{opt}})\right]$. Then, by setting $\delta=b$
and $x_{\textrm{opt}}={d_{\textrm{opt}}/\delta}$, we obtain the
following estimate for the optimal trap location and minimum average
MFPT for $m=2$ traps in the thin domain limit:
\begin{equation}\label{thin:m2}
  x_{0 \textrm{opt}}\sim {0.406/b} \,, \qquad \overline{u}_{0\textrm{opt}} \sim
  {0.0652/( b^2 D)} \,, \quad \mbox{for} \quad b\ll 1\,.
\end{equation}
These estimates are favorably compared in Fig.~\ref{fig:two_ellipse}
with full PDE solutions computed using the closest point method
\cite{IWWC2019} and with the full asymptotic theory based on
\eqref{e:u0_bar}.

Next, suppose that $m=3$. Since there is an additional trap at the
origin, we simply replace the condition $U_0^{\prime}(0)=0$ in
\eqref{thin_m2:U0} with $U_0(0)=0$. In place of
\eqref{thin_m2:u0_solve_1},
\begin{subequations}
\begin{equation}\label{thin_m3:u0_solve_1}
U_0(X) = \begin{cases}
  -\frac{1}{4D} \left[ \left(\sin^{-1}{X}\right)^2 + X^2 - \pi
    \sin^{-1} X + c_2 \right]\,, \quad d\leq X \leq 1\,, \\
  -\frac{1}{4D} \left[ \left(\sin^{-1}{X}\right)^2 + X^2 + c_1\sin^{-1}{X}
   \right]\,, \quad 0\leq X \leq d \,,
\end{cases}
\end{equation}
where $c_1$ and $c_2$ are given by
\begin{equation}\label{thin_m3:u0_solve_2}
  c_1 =  - {\left(d^2 + \left[ \sin^{-1}(d)\right]^2 \right)/\sin^{-1}(d)}
  \,, \qquad c_2 = - d^2 + \pi \sin^{-1}(d) - \left[\sin^{-1}(d)\right]^2 \,.
\end{equation}
\end{subequations}
The average MFPT is given by \eqref{thin_m2:ave_ell}, where ${\mathcal H}(d)$
is now defined by
\begin{equation}\label{thin_ave_H_3}
  {\mathcal H}(d) \equiv c_2 \int_{d}^{1} F(X) \, dX + (c_1+\pi)
  \int_{0}^{d} F(X) \, \sin^{-1}(X) \, dX \,,
\end{equation}
with $F(X)=\sqrt{1-X^2}$. By maximizing ${\mathcal H}(d)$ on $0<d<1$,
we obtain $d_{\textrm{opt}}\approx 0.567$, so that
$\overline{u}_{0\mbox{min}}=-\left(\pi D\delta^2\right)^{-1} \left[ J_0 +
  {\mathcal H}(d_{\textrm{opt}})\right]$.  In this way, the optimal
trap location and the minimum of the average MFPT satisfies
\begin{equation}\label{thin:m3}
  x_{0 \textrm{opt}}\sim {0.567/b} \,, \qquad \overline{u}_{0\textrm{opt}} \sim
  {0.0308/( b^2 D)} \,, \quad \mbox{for} \quad b\ll 1\,.
\end{equation}
In Fig.~\ref{fig:three_ellipse} these scaling laws are seen to compare
well with full PDE solutions and with the full asymptotic theory of
\eqref{e:u0_bar}, even when $b$ is only moderately small.

Next, we consider the case $m=4$, with two symmetrically placed traps on
either side of the origin. Therefore, we solve \eqref{thin_m2:U0} with
$U_0^{\prime}(0)=0$, $U_0(d_1)=0$, and $U_0(d_2)=0$, where $0<d_1<d_2$. In
place of \eqref{thin_m2:u0_solve_1}, we get
\begin{subequations}
\begin{equation}\label{thin_m4:u0_solve_1}
U_0(X) = \begin{cases}
  -\frac{1}{4D} \left[ \left(\sin^{-1}{X}\right)^2 + X^2 - \pi
    \sin^{-1} X + c_2 \right]\,, \quad d_2\leq X \leq 1\,, \\
  -\frac{1}{4D} \left[ \left(\sin^{-1}{X}\right)^2 + X^2 + b_1 \sin^{-1}{X}
        + b_2 \right]\,, \quad d_1\leq X \leq d_2\,, \\
    -\frac{1}{4D} \left[ \left(\sin^{-1}{X}\right)^2 + X^2 + c_1
   \right]\,, \quad 0\leq X \leq d_1 \,,
\end{cases}
\end{equation}
where $c_1$ and $c_2$ are given by
\begin{equation}\label{thin_m4:u0_solve_2}
\begin{split}
  & \qquad\qquad\qquad c_1 =  - d_1^2 - \left( \sin^{-1}{d_1}\right)^2 \,, \qquad
  c_2 = - d_2^2 + \pi \sin^{-1}{d_2} - \left(\sin^{-1}{d_2}\right)^2 \,,\\
  b_1 &= \frac{\left(\sin^{-1}{d_1}\right)^2 -\left(\sin^{-1}{d_2}\right)^2 +
    d_1^2 - d_2^2}{\sin^{-1}{d_2} -\sin^{-1}{d_1}} \,, \qquad
   b_2 =-b_1 \sin^{-1} d_1 - d_1^2 - \left(\sin^{-1}{d_1}\right)^2 \,.
\end{split}
\end{equation}
\end{subequations}
The average MFPT is given by \eqref{thin_m2:ave_ell}, where
${\mathcal H}={\mathcal H}(d_1,d_2)$ is now given by
\begin{equation}\label{thin_ave_H_4}
\begin{split}
  {\mathcal H}(d_1,d_2) &\equiv c_2 \int_{d_2}^{1} F(X) \, dX + (b_1+\pi)
  \int_{d_1}^{d_2} F(X) \, \sin^{-1}(X) \, dX + b_2\int_{d_1}^{d_2}
  F(X)\, dX \\
  & \qquad + \pi \int_{0}^{d_1} F(X) \, \sin^{-1}(X) \, dX +
    c_1 \int_{0}^{d_1} F(X) \, dX \,,
\end{split}
\end{equation}
where $F(X)\equiv\sqrt{1-X^2}$.  By using a grid search to maximize
${\mathcal H}(d_1,d_2)$ on $0<d_1<d_2<1$, we obtain that
$d_{1\textrm{opt}}\approx 0.215$ and
$d_{2\textrm{opt}} \approx 0.656$. This yields that the optimal trap
locations and the minimum of the average MFPT, given by
$\overline{u}_{0\mbox{min}}=-\left(\pi D\delta^2\right)^{-1} \left[ J_0 +
  {\mathcal H}(d_{1\textrm{opt}},d_{2\textrm{opt}})\right]$, have the
scaling law
\begin{equation}\label{thin:m4}
  x_{1 \textrm{opt}}\sim {0.215/b} \,, \quad
  x_{2 \textrm{opt}}\sim {0.656/b} \,, \quad
  \overline{u}_{0\textrm{opt}} \sim
  {0.0179/( b^2 D)} \,, \quad \mbox{for} \quad b\ll 1\,.
\end{equation}
These scaling laws are shown in Fig.~\ref{fig:four_ellipse} to agree well
with the full PDE solutions and with the full asymptotic theory of
\eqref{e:u0_bar} when $b$ is small.

Finally, we consider the case $m=5$, where we need only modify the
$m=4$ analysis by adding a trap at the origin. Setting $U_0(0)=0$,
$U_0(d_1)=0$, and $U_0(d_2)=0$ we obtain that $U_0$ is again given by
\eqref{thin_m4:u0_solve_1}, except that now $c_1$ in
\eqref{thin_m4:u0_solve_1} is replaced by $c_1\sin^{-1}(X)$, with
$c_1$ as defined in \eqref{thin_m3:u0_solve_2}. The average MFPT
satisfies \eqref{thin_m2:ave_ell}, where in place of
\eqref{thin_ave_H_4} we obtain that ${\mathcal H}(d_1,d_2)$ is given
by
\begin{equation}\label{thin_ave_H_5}
\begin{split}
  {\mathcal H}(d_1,d_2) &\equiv c_2 \int_{d_1}^{1} F(X) \, dX +
  (b_1+\pi) \int_{d_1}^{d_2} F(X) \, \sin^{-1}(X) \, dX \\
  & \qquad + b_2\int_{d_1}^{d_2} F(X)\, dX + (c_1+\pi) \int_{0}^{d_1}
  F(X) \, \sin^{-1}{X} \, dX \,,
\end{split}
\end{equation}
with $F(X)=\sqrt{1-X^2}$.  A grid search yields that
${\mathcal H}(d_1,d_2)$ is maximized on $0<d_1<d_2<1$ when
$d_{1\textrm{opt}}\approx 0.348$ and
$d_{2\textrm{opt}} \approx 0.714$. In this way, the corresponding
optimal trap locations and minimum average MFPT have the scaling law
\begin{equation}\label{thin:m5}
  x_{1 \textrm{opt}}\sim {0.348/b} \,, \quad
  x_{2 \textrm{opt}}\sim {0.714/b} \,, \quad
  \overline{u}_{0\textrm{opt}} \sim {0.0117/( b^2 D)} \,, \quad \mbox{for}
  \quad b\ll 1\,.
\end{equation}
Fig.~\ref{fig:five_ellipse} shows that \eqref{thin:m5} compares well
with the full PDE solutions and with the full asymptotic theory of
\eqref{e:u0_bar} when $b$ is small.

\section{An Explicit Neumann Green's Function for the
  Ellipse}\label{sec:g_ell}

We derive the {\em new explicit formula} \eqref{cell:finz_g} for the
Neumann Green's function and its regular part in \eqref{cell:R0} in
terms of rapidly converging infinite series. This Green's function
$G(\x;\x_0)$ for the ellipse
$\Omega \equiv \lbrace{ \x = (x,y) \, \vert \, {x^2/a^2}+{y^2/b^2}\leq
  1\rbrace}$ is the unique solution to
\begin{subequations}\label{ell:g}
\begin{align}
  \Delta G  &= \frac{1}{|\Omega|} - \delta(\x - \x_0)\, \quad \x \in \Omega\,;
 \qquad \partial_n G =0\,, \,\,\, \x \in \partial \Omega\,;
                \label{ell:g_a}\\
  G  \sim -\frac{1}{2\pi}& \log{|\x - \x_0|} + R_e + o(1) \quad \text{as}
  \quad\x \to \x_0\,; \qquad \int_{\Omega} G \, \text{d}\x=0\,,
                           \label{ell:g_b}
\end{align}
\end{subequations}
where $|\Omega|=\pi a b$ is the area of $\Omega$ and $R_e$ is the
regular part of the Green's function. Here $\partial_n G$ is the
outward normal derivative to the boundary of the ellipse. To remove the
$|\Omega|^{-1}$ term in \eqref{ell:g_a}, we introduce $N(\x;\x_0)$ defined
by
\begin{equation}\label{ell:g_to_n}
  G(\x;\x_0) = \frac{1}{4|\Omega|} (x^2+y^2) + N(\x;\x_0) \,.
\end{equation}
We readily derive that $N(\x;\x_0)$ satisfies
\begin{subequations}\label{ell:n}
\begin{align}
  \Delta N  &= - \delta(\x - \x_0)\quad \x \in \Omega\,;
  \quad \partial_n N = -\frac{1}{2|\Omega|\sqrt{{x^2/a^4} + {y^2/b^4}}}
   \,,\,\,\, \x \in \partial \Omega\,; \label{ell:n_a}\\
  \int_{\Omega} N \, \text{d}\x &= - \frac{1}{4 |\Omega|} \int_{\Omega}
 (x^2 + y^2) \, \text{d}\x = - \frac{1}{4 |\Omega|} \left( \frac{|\Omega|}{4}
   (a^2+ b^2) \right) = -\frac{1}{16}(a^2 + b^2)\,. \label{ell:n_b}
\end{align}
\end{subequations}

We assume that $a>b$, so that the semi-major axis is on the
$x$-axis.  To solve \eqref{ell:n} we introduce the elliptic cylindrical
coordinates $(\xi,\eta)$ defined by \eqref{ell:coord} and its inverse
mapping \eqref{ell:inverse_mapping}.  We set
$\mc{N}(\xi,\eta)\equiv N(x(\xi,\eta),y(\xi,\eta))$ and seek to convert
\eqref{ell:n} to a problem for $\mc{N}$ defined in a rectangular
domain. It is well-known that
\begin{equation}\label{cell:0}
  N_{xx}+N_{yy}= \frac{1}{f^2(\cosh^2\xi - \cos^2\eta)} \left(
    \mc{N}_{\xi\xi} + \mc{N}_{\eta\eta} \right) \,.
\end{equation}
Moreover, by computing the scale factors
$h_{\xi}=\sqrt{x_{\xi}^2 + y_{\xi}^2}$ and
$h_{\eta}=\sqrt{x_{\eta}^2 + y_{\eta}^2}$ of the transformation, we
obtain that
\begin{equation}\label{cell:1}
  \delta(x-x_0) \delta(y-y_0)=\frac{1}{h_{\eta} h_{\xi}}
  \delta(\xi-\xi_0) \delta(\eta-\eta_0) =\frac{1}{f^2(\cosh^2\xi -
    \cos^2\eta)} \delta(\xi-\xi_0) \delta(\eta-\eta_0)\,,
\end{equation}
where we used $h_\xi=h_\eta=f \sqrt{\cosh^2 \xi_0 -\cos^2\eta_0}$.
By using \eqref{cell:0} and \eqref{cell:1}, we obtain that the PDE in
\eqref{ell:n_a} transforms to
\begin{equation}\label{cell_t:pde}
  \mc{N}_{\xi\xi}+ \mc{N}_{\eta\eta} = -\delta(\xi-\xi_0)\delta(\eta-\eta_0)
  \,, \quad \mbox{in} \quad 0\leq \eta \leq 2\pi \,, \,\, 0\leq \xi\leq u_b\,.
\end{equation}

To determine how the normal derivative in \eqref{ell:n_a} transforms, we
calculate
\begin{equation}
  \begin{pmatrix} N_x \\ N_y \end{pmatrix} = \frac{1}{x_\xi y_\eta-x_\eta y_\xi}
    \begin{pmatrix} y_\eta & -y_\xi \\ -x_\eta & x_\xi \end{pmatrix} 
    \begin{pmatrix} \mc{N}_\xi \\ \mc{N}_\eta \end{pmatrix} \,,
\end{equation}
where from \eqref{ell:coord_1} we calculate
\begin{equation}\label{cell:xder}
 x_\xi=f\sinh\xi\cos\eta = y_\eta \,, \qquad x_\eta=-f\cosh\xi\sin\eta =-y_\xi\,.
\end{equation}
Now using $x=a\cos\eta$ and $y=b\sin\eta$ on $\partial\Omega$, we calculate on
$\partial\Omega$ that
\begin{equation}\label{cell:4}
  \partial_n N = \nabla N \cdot \frac{(x/a^2\,, y/b^2)}
  {\sqrt{x^2/a^4+y^2/b^4}} = \frac{ \left(\frac{1}{a}\cos\eta\,\,,
      \frac{1}{b}\sin\eta\right)}{ \sqrt{x^2/a^4+y^2/b^4}
   \left( x_\xi y_\eta-x_\eta y_\xi\right)} 
    \begin{pmatrix} y_\eta & -y_\xi \\ -x_\eta & x_\xi \end{pmatrix} 
    \begin{pmatrix} \mc{N}_\xi \\ \mc{N}_\eta \end{pmatrix}.
\end{equation}
By using \eqref{cell:xder}, we calculate on $\partial\Omega$ that
$x_\xi y_\eta-x_\eta y_\xi = b^2 \cos^2\eta + a^2 \sin^2\eta$. With this
expression, we obtain after some algebra that \eqref{cell:4} becomes
\begin{equation}\label{cell:6}
  \partial_n N = \frac{1}{ab \sqrt{x^2/a^4+y^2/b^4}} \, \mc{N}_u \,, \quad
  \mbox{on} \quad \xi =\xi_b \,.
\end{equation}
By combining \eqref{cell:6} and \eqref{ell:n_a}, we obtain
$ \mc{N}_\xi = - {1/(2\pi)}$ on $\xi =\xi_b$.

Next, we discuss the other boundary conditions in the transformed
plane.  We require that $\mc{N}$ and $\mc{N}_{\eta}$ are $2\pi$
periodic in $\eta$. The boundary condition imposed on $\eta=0$, which
corresponds to the line segment $y=0$ and $|x|\leq f=\sqrt{a^2-b^2}$
between the two foci, is chosen to ensure that $N$ and the normal
derivative $N_y$ are continuous across this segment. Recall from
\eqref{ell:xy_to_eta} that the top of this segment $y=0^{+}$ and
$|x|\leq f$ corresponds to $0\leq \eta\leq \pi$, while the bottom of
this segment $y=0^{-}$ and $|x|\leq f$ corresponds to
$\pi\leq \eta\leq 2\pi$. To ensure that $N$ is continuous across this
segment, we require that $\mc{N}(\xi,\eta)$ satisfies
$\mc{N}(0,\eta)=\mc{N}(0,2\pi-\eta)$ for any $0\leq \eta\leq
\pi$. Moreover, since $\mc{N}_\xi=N_y f\sin\eta$ on $\xi=0$, and
$\sin(2\pi-\eta)=-\sin(\eta)$, we must have
$\mc{N}_{\xi}(0,\eta)=\mc{N}_{\xi}(0,2\pi-\eta)$ on
$0\leq \eta\leq \pi$.

Finally, we examine the normalization condition in \eqref{ell:n_b} by using
\begin{equation}\label{cell:7}
  \int_{\Omega} N(x,y) \, dx \, dy = \int_{0}^{\xi_b}\int_{0}^{2\pi}
  \mc{N}(\xi,\eta) \,\, \Big{\vert}  \mbox{det}
  \begin{pmatrix} x_\xi & x_\eta \\ y_\xi & y_\eta
  \end{pmatrix} \Big{\vert} \, d\xi\, d\eta\,.
\end{equation}
Since $x_\xi y_\eta - x_\eta y_\xi=f^2\left(\cosh^2\xi-\cos^2\eta\right)$,
we obtain from \eqref{cell:7} that \eqref{ell:n_b} becomes
\begin{equation}\label{cell_t:norm}
  \int_{0}^{\xi_b}\int_{0}^{2\pi} \mc{N}(\xi,\eta) \left[\cosh^2\xi-\cos^2\eta
  \right]\, d\xi \, d\eta = -\frac{1}{16f^2}{(a^2+b^2)} = -
  \frac{(a^2+b^2)}{16(a^2-b^2)} \,.
\end{equation}
In summary, from  \eqref{cell_t:pde}, \eqref{cell_t:norm}, and the
condition on $\xi=\xi_b$, $\mc{N}(\xi,\eta)$ satisfies
\begin{subequations}\label{cell:n}
\begin{gather}
  \Delta \mc{N}  = - \delta(\xi-\xi_0)\delta(\eta -\eta_0)\,\quad
    0\leq \xi\leq \xi_b \,, \,\, 0\leq \eta\leq \pi \,, \label{cell:n_pde}\\
  \partial_\xi \mc{N}= -\frac{1}{2\pi} \,, \quad \mbox{on}\,\,\xi=\xi_b
  \,; \qquad \mc{N}\,, \,\, \mc{N}_\eta \quad 2\pi \,\, \mbox{periodic in }
  \eta \,, \label{cell:n_bnd1}\\
  \mc{N}(0,\eta)=\mc{N}(0,2\pi-\eta) \,, \quad
  \mc{N}_{\xi}(0,\eta)=-\mc{N}_{\xi}(0,2\pi-\eta) \,, \quad \mbox{for} \quad
  0\leq \eta\leq \pi \,, \label{cell:n_bnd2}\\ 
  \int_{0}^{\xi_b}\int_{0}^{2\pi} \mc{N}(\xi,\eta) \left[\cosh^2\xi-\cos^2\eta
  \right]\, d\xi \, d\eta = -\frac{(a^2+b^2)}{16(a^2-b^2)} \,.
  \label{cell:n_int}
\end{gather}
\end{subequations}

The solution to \eqref{cell:n} is expanded in terms of the eigenfunctions
in the $\eta$ direction:
\begin{equation}\label{ncell:eig_ex}
  \mc{N}(\xi,\eta) = \mc{A}_0(\xi) + \sum_{k=1}^{\infty} \mc{A}_k(\xi)
  \cos(k\eta) + \sum_{k=1}^{\infty} \mc{B}_k(\xi) \sin(k\eta) \,.
\end{equation}
The boundary condition \eqref{cell:n_bnd1} is satisfied with
$\mc{A}_0^{\prime}(\xi_b)=-{1/(2\pi)}$ and
$\mc{A}_k^{\prime}(\xi_b)=\mc{B}_k^{\prime}(\xi_b)=0$, for $k\geq
1$. To satisfy $\mc{N}(0,\eta)=\mc{N}(0,2\pi-\eta)$, we require
$\mc{B}_k(0)=0$ for $k\geq 1$. Finally, to satisfy
$\mc{N}_{\xi}(0,\eta)=-\mc{N}_{\xi}(0,2\pi-\eta)$, we require that
$\mc{A}_0^{\prime}(0)=0$ and $\mc{A}_k^{\prime}(0)=0$ for $k\geq
1$. In the usual way, we can derive ODE boundary value problems for
$\mc{A}_0$, $\mc{A}_k$, and $\mc{B}_k$. We obtain that
\begin{subequations}\label{ncell:odes}
\begin{equation}\label{ncell:ode_0}
    \mc{A}_0^{\prime\prime} = - \frac{1}{2\pi}\delta(\xi-\xi_0) \,, \quad
    0\leq \xi\leq \xi_b \,; \qquad \mc{A}_0^{\prime}(0)=0 \,, \,\,\,
    \mc{A}_0^{\prime}(\xi_b)=-\frac{1}{2\pi} \,,
\end{equation}
while on $0\leq\xi\leq \xi_b$, and for each $k=1,2,\ldots$, we have
\begin{gather}
  \mc{A}_k^{\prime\prime} - k^2 \mc{A}_k = -\frac{1}{\pi} \cos(k\eta_0)
  \delta(\xi-\xi_0) \,; \qquad  \mc{A}_k^{\prime}(0)=0 \,, \,\,\,
  \mc{A}_k^{\prime}(\xi_b)=0 \,, \label{ncell:ode_ak} \\
  \mc{B}_k^{\prime\prime} - k^2 \mc{B}_k = -\frac{1}{\pi} \sin(k\eta_0)
  \delta(\xi-\xi_0) \,; \qquad  \mc{B}_k(0)=0 \,, \,\,\,
  \mc{B}_k^{\prime}(\xi_b)=0 \,. \label{ncell:ode_bk}
\end{gather}
\end{subequations}
We observe from \eqref{ncell:ode_0} that $\mc{A}_0$ is specified only
up to an arbitrary constant.

We determine this constant from the normalization condition
\eqref{cell:n_int}. By substituting \eqref{ncell:eig_ex} into
\eqref{cell:n_int}, we readily derive the identity that
\begin{equation}\label{cell:n_int_1}
  \int_{0}^{\xi_b} \mc{A}_0(\xi) \cosh(2\xi) \, d\xi - \frac{1}{2}
  \int_{0}^{\xi_b} \mc{A}_2(\xi) \, d\xi = - \frac{1}{16\pi} \left(
    \frac{a^2+ b^2}{a^2-b^2} \right) \,.
\end{equation}
We will use \eqref{cell:n_int_1} to derive a point constraint on
$\mc{A}_0(\xi_b)$.  To do so, we define $\phi(\xi)=\cosh(2\xi)$, which
satisfies $\phi^{\prime\prime}-4\phi=0$ and $\phi^{\prime}(0)=0$. We
integrate by parts and use $\mc{A}_0^{\prime}(0)=0$ and
$\mc{A}_0^{\prime}(\xi_b)=-{1/(2\pi)}$ to get
\begin{equation}\label{cell:n_int_2}
  \begin{split}
    4\int_{0}^{\xi_b} \mc{A}_0\phi \, d\xi = \int_{0}^{\xi_b} \mc{A}_0
    \phi^{\prime\prime} \, d\xi &= \left(\phi^{\prime}\mc{A}_0 -
      \phi\mc{A}^{\prime}_0 \right)\vert_{0}^{\xi_b}  +
    \int_{0}^{\xi_b} \phi \mc{A}_0^{\prime\prime}\, d\xi \,, \\
    & = \phi^{\prime}(\xi_b) \mc{A}_0(\xi_b) + \frac{1}{2\pi}
    \left[ \phi(\xi_b) - \phi(\xi_0)\right] \,.
  \end{split}
\end{equation}
Next, set $k=2$ in \eqref{ncell:ode_ak} and integrate over
$0<\xi<\xi_b$. Using the no-flux boundary conditions we get
$\int_{0}^{\xi_b} \mc{A}_2 \, d\xi={\cos(2\eta_0)/(4\pi)}$. We substitute
this result, together with \eqref{cell:n_int_2}, into
\eqref{cell:n_int_1} and solve the resulting equation for $\mc{A}_0(\xi_b)$
to get
\begin{equation}\label{ncell:a00_t}
  \mc{A}_0(\xi_b) = \frac{1}{4\pi\sinh(2\xi_b)} \left[
    \cosh(2\xi_0)+\cos(2\eta_0) -\cosh(2\xi_b) -\frac{1}{2}
  \left( \frac{a^2 + b^2}{a^2-b^2} \right) \right]\,.
\end{equation}
To simplify this expression we use $\tanh\xi_b={b/a}$ to calculate
$\sinh(2\xi_b)={2ab/(a^2-b^2)}$ and $\coth(2\xi_b)={(a^2+b^2)/(2ab)}$, while
from \eqref{ell:coord_1} we get
\begin{equation*}
  x_0^2 + y_0^2 = f^2 \left[\cosh^2\xi_0 -\sin^2\eta_0\right] =
  \frac{ (a^2-b^2)}{2} \left[\cosh(2\xi_0) + \cos(2\eta_0)\right] \,.
\end{equation*}
Upon substituting these results into \eqref{ncell:a00_t}, we conclude that
\begin{equation}\label{ncell:a00}
  \mc{A}_0(\xi_b) = -\frac{3}{16|\Omega|} (a^2 + b^2) + \frac{1}{4|\Omega|}
  \left( x_0^2 + y_0^2 \right) \,,
\end{equation}
where $|\Omega|=\pi a b$ is the area of the ellipse. With this explicit
value for $\mc{A}_0(\xi_b)$,  the normalization condition
\eqref{cell:n_int}, or equivalently the constraint
$\int_{\Omega} G \, \text{d}\x=0$, is satisfied.

Next, we solve the ODEs \eqref{ncell:odes} for $\mc{A}_0$, $\mc{A}_k$,
and $\mc{B}_k$, for $k\geq 1$, to obtain
\begin{subequations}\label{ncell:sol_odes}
\begin{gather}
  \mc{A}_0(\xi) = \frac{1}{2\pi} \left(\xi_b-\xiM\right) + \mc{A}_0(\xi_b) \,,
  \quad \mc{A}_k(\xi) = 
  \frac{ \cos(k\eta_0)}{k\pi \sinh(k\xi_b)} \cosh(k\xim)\cosh\left(k(\xiM-\xi_b)
  \right) \,, \label{ncell:sol_ak} \\
  \mc{B}_k(\xi) = \frac{ \sin(k\eta_0)}{k\pi \cosh(k\xi_b)}
  \sinh(k\xim)\cosh\left(k(\xiM-\xi_b) \right) \,, \label{ncell:sol_bk}
\end{gather}
\end{subequations}
where we have defined $\xiM\equiv\max(\xi_0,\xi)$ and $\xim\equiv
\min(\xi_0,\xi)$.

To determine an explicit expression for
$G(\x;\x_0)={|\x|^2/(4|\Omega|)} + \mc{N}(\xi,\eta)$, as given in
\eqref{ell:g_to_n}, we substitute \eqref{ncell:a00} and
\eqref{ncell:sol_odes} into the eigenfunction expansion
\eqref{ncell:eig_ex} for $\mc{N}$. In this way, we get
\begin{subequations}\label{cell:gexp}
\begin{equation}\label{cell:gexp_t1}
    G(\x;\x_0) = \frac{1}{4|\Omega|} \left(|\x|^2 + |\x_0|^2\right) -
    \frac{3}{16|\Omega|}(a^2 + b^2) + \frac{1}{2\pi} \left(\xi_b-\xiM\right)
    + \mc{S} \,,
\end{equation}
where the infinite sum $\mc{S}$ is defined by
\begin{equation}\label{cell:gexp_sum_t}
  \begin{split}
  \mc{S} & \equiv \sum_{k=1}^{\infty}
  \frac{ \cos(k\eta_0)\cos(k\eta)} {\pi k \sinh(k\xi_b)}
  \cosh(k\xim)\cosh\left(k(\xiM-\xi_b)\right) \\
  & \qquad + \sum_{k=1}^{\infty}\frac{ \sin(k\eta_0)\sin(k\eta)}
  {\pi k \cosh(k\xi_b)} \sinh(k\xim)\cosh\left(k(\xiM-\xi_b)\right) \,.
   \end{split}
 \end{equation}
\end{subequations}
Next, from the product to sum formulas for $\cos(A)\cos(B)$ and
$\sin(A)\sin(B)$ we get
\begin{equation}\label{cell:gexp_sum_t2}
  \begin{split}
    \mc{S}  &= \frac{1}{2\pi} \sum_{k=1}^{\infty} \frac{\cosh\left(k(\xiM-\xi_b)
      \right)}{k} \left[ \frac{\cosh(k\xim)}{\sinh(k\xi_b)} +
      \frac{\sin(k\xim)}{\cosh(k\xi_b)} \right] \cos\left(k(\eta-\eta_0\right)
      \\
      & \qquad + 
        \frac{1}{2\pi} \sum_{k=1}^{\infty} \frac{\cosh\left(k(\xiM-\xi_b)
      \right)}{k} \left[ \frac{\cosh(k\xim)}{\sinh(k\xi_b)} -
   \frac{\sin(k\xim)}{\cosh(k\xi_b)} \right] \cos\left(k(\eta+\eta_0\right)\,.
\end{split}
\end{equation}
Then, by using product to sum formulas for $\cosh(A)\cosh(B)$, the
identity $\sinh(2A)=2\sinh(A)\cosh(A)$, $\xiM+\xim=\xi+\xi_0$,
and $\xiM-\xim=|\xi-\xi_0|$, some algebra yields that
\begin{equation}\label{cell:sinf}
  \begin{split}
    \mc{S} &= \frac{1}{2\pi} \mbox{Re} \left(\sum_{k=1}^{\infty}
      \frac{\left[\cosh\left(k(\xi+\xi_0)\right) +
          \cosh\left(k(|\xi-\xi_0|-2\xi_b)\right)\right]}{k\sinh(2k\xi_b)}
      e^{ik(\eta-\eta_0)} \right) \\
      & \quad + \frac{1}{2\pi} \mbox{Re} \left(\sum_{k=1}^{\infty}
      \frac{\left[\cosh\left(k(\xi+\xi_0-2\xi_b)\right) +
          \cosh\left(k|\xi-\xi_0|\right)\right]}{k\sinh(2k\xi_b)}
      e^{ik(\eta +\eta_0)} \right)\,.
  \end{split}
\end{equation}

The next step in the analysis is to convert the hyperbolic functions in
\eqref{cell:sinf} into pure exponentials. A simple calculation yields that
\begin{subequations}\label{cell:s_exp_t}
\begin{equation}\label{cell:s_exp_t1}
  \mc{S} = \frac{1}{2\pi} \mbox{Re} \left(  \sum_{k=1}^{\infty} \frac{\mc{H}_1}{k}
    e^{ik(\eta-\eta_0)} + \sum_{k=1}^{\infty} \frac{\mc{H}_2}{k}
    e^{ik(\eta+\eta_0)} \right)\,,
\end{equation}
where $\mc{H}_1$ and $\mc{H}_2$ are defined by
\begin{equation}
\begin{split}
  \mc{H}_1 &\equiv \frac{1}{1-e^{-4k\xi_b}}
  \left[ e^{k(\xi+\xi_0-2\xi_b)} + e^{-k(\xi+\xi_0+2\xi_b)} +
  e^{k(|\xi-\xi_0|-4\xi_b)} + e^{-k|\xi-\xi_0|}\right] \,, \\
  \mc{H}_2 &\equiv \frac{1}{1-e^{-4k\xi_b}}
  \left[ e^{k(\xi+\xi_0-4\xi_b)} + e^{k(|\xi-\xi_0|-2\xi_b)} +
  e^{-k(|\xi-\xi_0|+2\xi_b)} + e^{-k(\xi+\xi_0)}\right] \,.
\end{split}
\end{equation}
\end{subequations}
Then, for any $q$ with $0<q<1$ and integer $k\geq 1$, we use the
identity $\sum_{n=0}^{\infty} \left(q^k\right)^n = \frac{1}{1-q^k}$
for the choice $q=e^{-4\xi_b}$, which converts $\mc{H}_1$ and
$\mc{H}_2$ into infinite sums. This leads to a doubly-infinite sum
representation for $\mc{S}$ in \eqref{cell:s_exp_t1} given by
\begin{equation}\label{cell:sum_z}
  \mc{S} = \frac{1}{2\pi} \mbox{Re}\left(
    \sum_{k=1}^{\infty} \sum_{n=0}^{\infty} \frac{\left(q^n\right)^k}{k}
    \left( z_1^k + z_2^k + z_3^k + z_4^k + z_5^k + z_6^k + z_7^k + z_8^k
      \right)\right) \,,
\end{equation}
where the complex constants $z_1,\ldots,z_8$ are defined by
\eqref{cell:def_z}.  From these formulae, we readily observe that
$|z_j|<1$ on $0\leq\xi\leq \xi_b$ for any
$(\xi,\eta)\neq (\xi_0,\eta_0)$. Since $0<q<1$, we can then switch the
order of the sums in \eqref{cell:sum_z} when
$(\xi,\eta)\neq (\xi_0,\eta_0)$ and use the identity
$\mbox{Re}\left(\sum_{k=1}^{\infty} k^{-1} \omega^k\right)=
-\log|1-\omega|$, where $|1-\omega|$ denotes modulus. In this way,
upon setting $\omega_j=q^nz_j$ for $j=1,\ldots,8$, we obtain a compact
representation for $\mc{S}$. Finally, by using this result in
\eqref{cell:gexp} we obtain for $(\xi,\eta)\neq (\xi_0,\eta_0)$, or
equivalently $(x,y)\neq (x_0,y_0)$, the result given explicitly in
\eqref{cell:finz_g} of \S~\ref{sec:ellipse}.

Next, to determine the regular part of the Neumann Green's function we
must identify the singular term in \eqref{cell:finz_g1} at
$(\xi,\eta)=(\xi_0,\eta_0)$. Since $z_1=1$, while $|z_j|<1$ for
$j=2,\ldots,8$, at $(\xi,\eta)=(\xi_0,\eta_0)$, the singular
contribution arises only from the $n=0$ term in
$\sum_{n=0}^{\infty} \log |1-\beta^{2n} z_1|$. As such, we add and
subtract the fundamental singularity $-{\log|\x-\x_0|/(2\pi)}$ in
\eqref{cell:finz_g1} to get
\begin{subequations}
\begin{equation}
  G(\x;\x_0) = -\frac{1}{2\pi} \log|\x-\x_0| + R(\x;\x_0)\,,\label{cell:G_and_R}
\end{equation}
\begin{equation}\label{cell:R}
  \begin{split}
    R(\x;\x_0) &= \frac{1}{4|\Omega|} \left( |\x|^2 + |\x_0|^2\right)
    -\frac{3(a^2+b^2)}{16|\Omega|} - \frac{1}{4\pi} \log\beta -
    \frac{1}{2\pi}\xiM +
    \frac{1}{2\pi} \log\left( \frac{|\x-\x_0|}{|1-z_1|}\right) \\
    &\qquad -\frac{1}{2\pi}\sum_{n=1}^{\infty}\log|1-\beta^{2n} z_1|
    -\frac{1}{2\pi} \sum_{n=0}^{\infty}\log\left(
    \displaystyle \prod_{j=2}^{8} |1-\beta^{2n} z_j| \right)  \,.
  \end{split}
\end{equation}
\end{subequations}

To identify $\lim_{\x\to\x_0}R(\x;\x_0)=R_{e}$, we must find
$\lim_{\x\to\x_0} \log\left( {|\x-\x_0|/|1-z_1|}\right)$.  To do so,
we use a Taylor approximation on \eqref{ell:coord_1} to derive at
$(\xi,\eta)=(\xi_0,\eta_0)$ that
\begin{equation}\label{cell:loc_jac}
  \begin{pmatrix} \xi-\xi_0 \\ \eta-\eta_0 \end{pmatrix} =
  \frac{1}{(x_\xi y_\eta-x_\eta y_\xi)}
    \begin{pmatrix} y_\eta & -x_\eta \\ -y_\xi & x_\xi \end{pmatrix} 
    \begin{pmatrix} x-x_0 \\ y-y_0 \end{pmatrix} \,.
\end{equation}
By calculating the partial derivatives in \eqref{cell:loc_jac} using
\eqref{cell:xder}, and then noting from \eqref{cell:def_z} that
$|1-z_1|^2\sim (\xi-\xi_0)^2+(\eta-\eta_0)^2$ as
$(\xi,\eta)\to (\xi_0,\eta_0)$, we readily derive that
\begin{equation}\label{cell:local}
  \lim_{\x\to\x_0} \log\left( \frac{|\x-\x_0|}{|1-z_1|}\right) =
  \frac{1}{2} \log{(a^2-b^2)} + \frac{1}{2} \log\left(\cosh^2\xi_0
    -\cos^2\eta_0\right) \,.
\end{equation}

Finally, we substitute \eqref{cell:local} into \eqref{cell:R} and let
$\x\to\x_0$. This yields the formula for the regular part of the
Neumann Green's function as given in \eqref{cell:R0} of
\S~\ref{sec:ellipse}. In Appendix \ref{ell:disk} we show that the
Neumann Green's function \eqref{cell:finz_g} for the ellipse reduces
to the expression given in \eqref{gr:gmrm} for the unit disk when
$a\to b=1$.

\section{Discussion}\label{sec:discussion}

Here we discuss the relationship between our problem of optimal trap
patterns and a related optimization problem for the fundamental
Neumann eigenvalue $\lambda_0$ of the Laplacian in a bounded 2-D
domain $\Omega$ containing $m$ small circular absorbing traps of a
common radius $\eps$. That is, $\lambda_0$ is the lowest eigenvalue of
\begin{equation}\label{eig:low}
\begin{split}
  \Delta u + \lambda u &= 0\,,  \quad x \in \Omega\setminus
  \cup_{j=1}^{m}\Omega_{\varepsilon j} \,; \qquad \partial_n u  = 0 \,,
  \quad x\in \partial \Omega \,, \\
  u &= 0\,,  \quad x\in \partial \Omega_{\varepsilon j}\,,
\quad j = 1, \ldots,m\,.
\end{split}
\end{equation}
Here $\Omega_{\varepsilon j}$ is a circular disk of radius $\eps\ll 1$
centered at $\x_j\in \Omega$.  In the limit $\eps\to 0$, a two-term
asymptotic expansion for $\lambda_0$ in powers of
$\nu\equiv {-1/\log\eps}$ is
(see \cite[Corollary~2.3]{KTW2005} and Appendix~\ref{app:eig_low})
\begin{equation}\label{eig:2term}
\lambda_{0} \sim \frac{2\pi m \nu}{|\Omega|} - 
\frac{4\pi^2 \nu^2}{|\Omega|} p(\x_1,\ldots,\x_m) + O(\nu^3) \,,
\quad \mbox{with} \quad p(\x_1,\ldots,\x_m)\equiv \v{e}^T \mc{G} \v{e},
\end{equation}
where $\v{e}\equiv (1,\ldots,1)^T$ and $\mc{G}$ is the Neumann Green's
matrix. To relate this result for $\lambda_0$ with that for the
average MFPT $\overline{u}_0$ satisfying \eqref{e:u0_bar}, we let
$\nu\ll 1$ in \eqref{e:u0_bar} and calculate that
$\mathcal{A}\sim {|\Omega| \v{e}/(2\pi D m)} + {\mathcal O}(\nu)$.
>From \eqref{e:u0_bar}, we conclude that
\begin{equation}\label{disc:u0_bar}
  \overline{u}_0 = \frac{|\Omega|}{2\pi D \nu m} \left( 1 +
    \frac{2\pi \nu}{m} p(\x_1,\ldots,\x_m) + {\mathcal O}(\nu^2)\right)\,,
\end{equation}
where $p(\x_1,\ldots,x_m)$ is defined in \eqref{eig:2term}. By
comparing \eqref{disc:u0_bar} and \eqref{eig:2term} we conclude, up to
terms of ${\mathcal O}(\nu^2)$, that the trap configurations that
provide local minima for the average MFPT also provide local maxima
for the first Neumann eigenvalue for \eqref{eig:low}. Qualitatively,
this implies that, up to terms of order ${\mathcal O}(\nu^2)$, the
trap configuration that maximizes the rate at which a Brownian
particle is captured also provides the best configuration to minimize
the average mean first capture time of the particle. In this way, our
optimal trap configurations for the average MFPT for the ellipse
identified in \S~\ref{ell:ex} also correspond to trap patterns that
maximize $\lambda_0$ up to terms of order ${\mathcal
  O}(\nu^2)$. Moreover, we remark that for the special case of a
ring-pattern of traps, the first two-terms in \eqref{disc:u0_bar} provide
an exact solution of \eqref{e:u0_bar}. As such, for these special
patterns, the trap configuration that maximizes the
${\mathcal O}(\nu^2)$ term in $\lambda_0$ provides the optimal trap
locations that minimize the average MFPT to {\em all orders in $\nu$.}

Finally, we discuss two possible extensions of this study. Firstly, in
near-disk domains and in the ellipse it would be worthwhile to use a
more refined gradient descent procedure such as in \cite{ridgway} and
\cite{gilbert} to numerically identify globally optimum trap
configurations for a much larger number of identical traps than
considered herein. One key challenge in upscaling the optimization
procedure to a larger number of traps is that the energy landscape can
be rather flat or else have many local minima, and so identifying the
true optimum pattern is delicate. Locally optimum trap patterns with
very similar minimum values for the average MFPT already occurs in
certain near-disk domains at a rather small number of traps (see
Fig.~\ref{fig:neardisk_cos4} and Fig.~\ref{fig:neardisk_odd}). One
advantage of our asymptotic theory leading to \eqref{final:avemfpt}
for the near-disk and \eqref{e:u0_bar} for the ellipse, is that it can
be implemented numerically with very high precision. As a result,
small differences in the average MFPT between two distinct locally
optimal trap patterns are not due to discretization errors arising
from either numerical quadratures or evaluations of the Neumann
Green's function. As such, combining our hybrid theory with a refined
global optimization procedure should lead to the reliable
identification of globally optimal trap configurations for these
domains.

Another open direction is to investigate whether there are
computationally useful analytical representations for the Neumann
Green's function in an arbitrary bounded 2-D domain. In this
direction, in \cite[Theorem~4.1]{KW2003} an explicit analytical result
for the gradient of the regular part of the Neumann Green's function
was derived in terms of the mapping function for a general class of
mappings of the unit disk. It is worthwhile to study whether
this analysis can be extended to provide a simple and accurate
approach to compute the Neumann Green's matrix for an arbitrary
domain. This matrix could then be used in the linear algebraic system
\eqref{e:u0_bar} to calculate the average MFPT, and a gradient descent
scheme implemented to identify optimal patterns.

\section{Acknowledgements}\label{sec:ak}
Colin Macdonald and Michael Ward were supported by NSERC
Discovery grants. Tony Wong was partially supported by a UBC Four-Year
Graduate Fellowship.

\begin{appendix}
\renewcommand{\theequation}{\Alph{section}.\arabic{equation}}
\setcounter{equation}{0}

\section{Derivation of the Thin Domain ODE}\label{app:thin}
In the asymptotic limit of a long thin domain, we use a perturbation
approach on the MFPT PDE \eqref{Ellip_Model} for $u(x,y)$ in order to
derive the limiting problem \eqref{sec:long_u0}. We introduce the
stretched variables $X$ and $Y$ by $X = \delta x, Y = {y/\delta}$ and
$d = {x_0/\delta}$, and set $U(X,Y)=u({X/\delta},Y\delta)$. Then the
PDE in \eqref{Ellip_Model} becomes
$\delta^4 \partial_{XX} U + \partial_{YY} U = -{\delta^2/D}$. By
expanding $U = \delta^{-2} U_0 + U_1 + \delta^2 U_2 + \ldots$ in this
PDE, we collect powers of $\delta$ to get
\begin{equation}
{\mathcal O}(\delta^{-2})\,:\,\,\partial_{YY} U_0 = 0\,; \quad
{\mathcal O}(1)\,: \,\,\partial_{YY} U_1 = 0\,; \quad
{\mathcal O}(\delta^2)\,: \,\,\partial_{YY} U_2 = -\frac{1}{D} -
\partial_{XX} U_0 \,. \label{app:pde_order}
\end{equation}

On the boundary $y = \pm\delta F(\delta x)$, or equivalently
$Y = \pm F(X)$, where $F(X)=\sqrt{1-X^2}$, the unit outward normal is
$\hat{\mathbf{n}} = {\mathbf{n}/|\mathbf{n}|}$, where
$\mathbf{n} \equiv (-\delta^2 F^{\prime}(X),\pm1)$. The condition for
the vanishing of the outward normal derivative in
\eqref{Ellip_Model} becomes
\begin{equation*}
  \partial_n u = \hat{\mathbf{n}} \cdot (\partial_x u, \partial_y u) =
  \frac{1}{|\mathbf{n}|}(-\delta^2F^{\prime}, \pm 1) \cdot
  (\delta\partial_X U, \delta^{-1}\partial_Y U) = 0\,, \,\,\, \mbox{on}
  \,\,\, Y = \pm F(X) \,.
\end{equation*}
This is equivalent to the condition that
$\partial_Y U = \pm \delta^4 F^{\prime}(X) \partial_X U$ on
$Y = \pm F(X)$. Upon substituting  $U = \delta^{-2} U_0 + U_1 + \delta^2 U_2 +
\ldots$ into this expression, and equating powers of $\delta$, we obtain on
$Y=\pm F(X)$ that
\begin{equation}
{\mathcal O}(\delta^{-2})\,: \,\,\partial_Y U_0 =0\,; \quad
{\mathcal O}(1)\,; \,\,\partial_Y U_1 = 0\,; \quad
{\mathcal O}(\delta^2)\,; \,\, \quad\partial_Y U_2 = \pm F^{\prime}(X)
\partial_X U_0 \,. \label{eqn:three_traps_bc_different_order}
\end{equation}
From (\ref{app:pde_order}) and
(\ref{eqn:three_traps_bc_different_order}) we conclude that
$U_0 = U_0(X)$ and $U_1 = U_1(X)$.  Assuming that the trap radius
$\varepsilon$ is comparable to the domain width $\delta$, we will
approximate the zero Dirichlet boundary condition on the three traps
as zero point constraints for $U_0$.

The ODE for $U_0(X)$ is derived from a solvability condition on
the ${\mathcal O}(\delta^2)$ problem:
\begin{equation}\label{long:u2}
  \partial_{YY} U_2 = -\frac{1}{D} - U_0^{\prime\prime}\,, \,\,\,
  \mbox{in}\,\,\, \Omega\setminus\Omega_a\,; \quad \partial_Y U_2 =
  \pm F^{\prime}(X) U_0^{\prime}\,, \,\,\, \mbox{on} \,\,\,
   Y = \pm F(X)\,, \,\, |X|<1 \,.
\end{equation}
We multiply this problem for $U_2$ by $U_0$ and integrate in $Y$ over
$|Y|<F(X)$. Upon using Lagrange's identity and the boundary
conditions in \eqref{long:u2} we get
\begin{equation}
\begin{aligned}
  \int_{-F(X)}^{F(X)} \left(U_0 \partial_{YY} U_2 - U_2 \partial_{YY} U_0\right)\,
  dY &= \left[ U_0 \partial_Y U_2 - U_2 \partial_Y U_0  \right]
  \Big{\vert}_{-F(X)}^{F(X)}= 2U_0 F^{\prime}(X) U_0^{\prime} \,, \\
  \int_{-F(X)}^{F(X)} U_0 \left( -\frac{1}{D} - U_{0}^{\prime\prime} \right) \, dY
  &= -2F(X)U_0\left(\frac{1}{D} + U_0^{\prime\prime}\right) = 2U_0 F^{\prime}(X)
  U_0^{\prime}\,. 
\end{aligned}
\end{equation}
Thus, $U_0(X)$ satisfies the ODE
$\left[F(X)U_0^{\prime}\right]^{\prime}= -{F(X)/D}$, with
$F(X)=\sqrt{1-X^2}$, as given in \eqref{sec:long_u0} of
\S~\ref{sec:thin}. This gives the leading-order asymptotics
$u\sim \delta^{-2}U_0(X)$.

\section{Limiting Case of the Unit Disk}\label{ell:disk}
We now show how to recover the well-known Neumann Green's function and
its regular part for the unit disk by letting $a\to b =1$ in
\eqref{cell:finz_g} and \eqref{cell:R0}, respectively. In the limit
$\beta\equiv {(a-b)/(a+b)}\to 0$ only the $n=0$ terms in the
infinite sums in \eqref{cell:finz_g} and \eqref{cell:R0} are
non-vanishing. In addition, as $\beta\to 0$, we obtain from
\eqref{ell:coord} that $|\x|^2\sim {f^2 e^{2\xi}/4}$ and
$|\x_0|^2\sim {f^2 e^{2\xi_0}/4}$, and $\xi_b=-\log{f} + \log(a+b)\to
-\log{f}+\log{2}$, where $f\equiv \sqrt{a^2-b^2}$. This yields that
\begin{equation}\label{lim:iden}
  \xi + \xi_0 - 2 \xi_b \sim \log\left(\frac{2|\x|}{f}\right) +
  \log\left(\frac{2|\x_0|}{f}\right) -2\log{2} + 2\log{f} = \log\left(
    |\x||\x_0|\right) \,.
\end{equation}
As such, only the $z_1$ and $z_4$ terms in the infinite sums
in \eqref{cell:finz_g1} with $n=0$ persist as $a\to b=1$, and so
\eqref{cell:finz_g1} reduces in this limit to
\begin{equation}\label{lim:g1}
  G(\x;\x_0) \sim \frac{1}{4|\Omega|} \left( |\x|^2 + |\x_0|^2\right)
  -\frac{3}{8|\Omega|} + \frac{1}{2\pi}\left(\xi_b-\xiM\right) -
  \frac{1}{2\pi}\log|1-z_1| -  \frac{1}{2\pi}\log|1-z_4| \,,
\end{equation}
where $|\Omega|=\pi$ and $\xiM\equiv\max(\xi_0,\xi)$. Since
$\eta\to\theta$ and $\eta_0\to\theta_0$, where $\theta$ and $\theta_0$
are the polar angles for $\x$ and $\x_0$, we get from
\eqref{cell:def_z} that $z_4\to |\x||\x_0|e^{i(\theta-\theta_0)}$ as
$a\to b=1$. We then calculate that
\begin{equation}\label{lim:z4}
    -\frac{1}{2\pi} \log|1-z_4|=  -\frac{1}{4\pi} \log|1-z_4|^2=
  -\frac{1}{4\pi} \log\left( 1 -2 |\x||\x_0| \cos(\theta-\theta_0)
    + |\x|^2|\x_0|^2 \right) \,.
\end{equation}

Next, with regards to the $z_1$ term we calculate for $a\to b=1$ that
\begin{equation}\label{lim:z1_1}
  |\xi-\xi_0| = \begin{cases}
 \xi-\xi_0 \sim \log\left( \frac{|\x|}{|\x_0|} \right)\,, &\quad \mbox{if}\,\,
    0<|\x_0|<|\x|\,, \\
    -(\xi-\xi_0) \sim \log\left( \frac{|\x_0|}{|\x|} \right)\,, &\quad \mbox{if}
    \,\, 0<|\x|<|\x_0| \,.
    \end{cases}
\end{equation}
From \eqref{cell:def_z} this yields for $a\to b=1$ that
\begin{equation}\label{lim:z1_2}
  z_1=e^{-|\xi-\xi_0|+i(\eta-\eta_0)} \sim \begin{cases}
    \frac{|\x_0|}{|\x|} e^{i(\theta-\theta_0)} \,, &\quad \mbox{if} \,\,
    0<|\x_0|<|\x| \,, \\
    \frac{|\x|}{|\x_0|} e^{i(\theta-\theta_0)} \,, &\quad \mbox{if} \,\,
    0<|\x|<|\x_0| \,.
    \end{cases}
\end{equation}
By using \eqref{lim:z1_2}, we calculate for $a\to b=1$ that
\begin{equation}\label{lim:z1}
    -\frac{1}{4\pi} \log|1-z_1|^2 = -\frac{1}{2\pi}\log|\x-\x_0|
   + \begin{cases}
     \frac{1}{4\pi} \log|\x|^2 \,,& \,\, \mbox{if } 0<|\x_0|<|\x|\,, \\
     \frac{1}{4\pi} \log|\x_0|^2 \,, & \,\, \mbox{if } 0<|\x|<|\x_0|\,. 
   \end{cases}
\end{equation}
Next, we estimate the remaining term in \eqref{lim:g1} as $a\to b=1$ using
\begin{equation}\label{lim:umin}
  \frac{1}{2\pi} \left(\xi_b-\xiM\right) =\frac{1}{2\pi} \begin{cases}
    \xi_b-\xi \sim -\frac{1}{2\pi}\log|\x|\,, & \quad \mbox{if }\,\,
        |\x|>|\x_0|>0 \,,\\
    \xi_b-\xi_0 \sim -\frac{1}{2\pi}\log|\x_0|\,, & \quad \mbox{if }\,\,
    0<|\x|<|\x_0| \,. \end{cases}
\end{equation}
Finally, by using \eqref{lim:z4}, \eqref{lim:z1}, and \eqref{lim:umin}
into \eqref{lim:g1}, we obtain for $a\to b=1$ that
\begin{equation}\label{lim:gfinal}
  \begin{split}
    G(\x;\x_0) &\sim -\frac{1}{2\pi}\log|\x-\x_0| 
 -\frac{1}{4\pi} \log\left( 1 -2 |\x||\x_0| \cos(\theta-\theta_0)
   + |\x|^2 |\x_0|^2 \right)\\
 & \qquad + \frac{1}{4|\Omega|}\left( |\x|^2 + |\x_0|^2\right)
 -\frac{3}{8|\Omega|},
\end{split}
\end{equation}
where $|\Omega|=\pi$. This result agrees with that in \eqref{gr:gm}
for the Neumann Green's function in the unit disk. Similarly, we can
show that the regular part $R_e$ for the ellipse given in
\eqref{cell:R0} tends as $a\to b=1$ to that given in \eqref{gr:rm} for
the unit disk. 

\section{Asymptotics of the Fundamental Neumann
  Eigenvalue}\label{app:eig_low}
For $\nu\ll 1$, it was shown in \cite{KTW2005}, by using a matched
asymptotic expansion analysis in the limit of small trap radii similar
to that leading to \eqref{e:u0_bar}, that the fundamental Neumann
eigenvalue $\lambda_0$ for \eqref{eig:low} is the smallest positive
root of
\begin{equation}\label{a:eig_helm}
  {\mathcal K}(\lambda) \equiv \mbox{det}\left(I + 2\pi \nu {\mathcal G}_H
  \right)=0\,.
\end{equation}
Here $\nu=-{1/\log\eps}$ and ${\mathcal G}_H$ is the Helmholtz Green's
matrix with matrix entries
\begin{align}\label{a:GreenMAtrix}
  (\mathcal{G})_{Hjj} = R_{Hj} \,\,\, \text{for} \,\,\, i = j
  \quad \text{and} \quad (\mathcal{G})_{Hij} = (\mathcal{G})_{Hji}  =
  G_{H}(\x_i ; \x_j) \,\,\, \text{for} \,\,\, i \neq j \,,
\end{align}
where the Helmholtz Green's function $G_{H}(\x;\x_j)$ and its regular
part $R_{Hj}$ satisfy
\begin{subequations}\label{a:GreenFunctionProb}
\begin{align}
  \Delta G_H  +\lambda G_H &= -\delta(\x-\x_j) \,,\quad \x \in \Omega\,;
 \qquad \partial_n G_H =0\,, \,\,\, \x \in \partial \Omega\,;
                \label{a:GreenFunctionProb_A}\\
  G_H  \sim -\frac{1}{2\pi}& \log{|\x - \x_j|} + R_{Hj} + o(1) \,,
      \quad \text{as} \quad\x \to \x_j\,. \label{a:GreenFunctionProb_B}
\end{align}
\end{subequations}
For $0<\lambda\ll 1$, we estimate ${\mathcal G}_H$ by expanding
$G_H={A/\lambda} + G +{\mathcal O}(\lambda)$, for some $A$ to be
found. From \eqref{a:GreenFunctionProb}, we derive in terms
of the Neumann Green's matrix ${\mathcal G}$ that
\begin{equation}\label{a:gh_exp}
  {\mathcal G}_H = -\frac{m}{\lambda |\Omega|} E + {\mathcal G} +
  {\mathcal O}(\lambda) \,, \qquad \mbox{with} \quad E \equiv \frac{1}{m}
  \v{e} \v{e}^T \,,
\end{equation}
for $0<\lambda \ll 1$.  From \eqref{a:gh_exp} and \eqref{a:eig_helm},
the fundamental Neumann eigenvalue $\lambda_0$ is the smallest
$\lambda>0$ for which there is a nontrivial solution $\v{c}\neq \v{0}$ to
\begin{equation}\label{a:sing_mat}
  \left(I -\frac{2\pi \nu m}{\lambda |\Omega|} E + 2\pi \nu {\mathcal G}
    + {\mathcal O}(\nu) \right) \v{c} =0 \,.
\end{equation}
Since this occurs when $\lambda={\mathcal O}(\nu)$, we define
$\lambda_c>0$ by   $\lambda = {2\pi \nu m \lambda_c/|\Omega|}$,
so that \eqref{a:sing_mat} can be written in equivalent form as
\begin{equation}\label{a:math_neum}
  E \v{c} = \lambda_c \left( I + 2\pi \nu {\mathcal G} + {\mathcal O}(\nu^2)
  \right)\v{c} \,, \qquad \mbox{where} \quad
  \lambda = \frac{2\pi \nu m} {|\Omega|} \lambda_c\,.
\end{equation}

Since $E\v{e}=\v{e}$, while $E\v{q}=0$ for any $\v{q}\in \Real^{m-1}$
with $\v{e}^T\v{q}=0$, we conclude for $\nu\ll 1$ that the only
non-zero eigenvalue of \eqref{a:math_neum} satisfies $\lambda_c\sim 1$
with $\v{c}\sim \v{e}$. To determine the correction to this
leading-order result, in \eqref{a:math_neum} we expand
$\lambda_c=1+\nu \lambda_{c1}+\cdots$ and
$\v{c}=\v{e}+\nu \v{c}_{1}+ \cdots$. From collecting
${\mathcal O}(\nu)$ terms in \eqref{a:math_neum}, we get
\begin{equation}\label{a:mat_solve}
  \left(I - E \right) \v{c}_{1} = -2\pi {\mathcal G}\v{e} - \lambda_{c1}\v{e}
  \,.
\end{equation}
Since $I-E$ is symmetric with the 1-D nullspace $\v{e}$, the solvability
condition for \eqref{a:mat_solve} is that $-2\pi\v{e}^T {\mathcal G}\v{e}-
\lambda_{c1} \v{e}^T\v{e}=0$. Since $\v{e}^T\v{e}=m$, this yields the two-term
expansion
\begin{equation}\label{a:lambda_c1}
  \lambda_{c}= 1+\nu \lambda_{c1}+\ldots \,, \qquad \mbox{where}
  \quad \lambda_{c1}  = -\frac{2\pi}{m} \v{e}^T {\mathcal G}\v{e}\,.
\end{equation}
Finally, using $\lambda = {2\pi \nu m\lambda_c/|\Omega|}$, we
obtain the two-term expansion as given in \eqref{eig:2term}.

\end{appendix}

\bibliographystyle{plain}
\bibliography{Reference_new}

\end{document}